\documentclass[a4,fleqn,usenatbib]{mnras}

\DeclareRobustCommand{\VAN}[3]{#2}
\let\VANthebibliography\thebibliography
\def\thebibliography{\DeclareRobustCommand{\VAN}[3]{##3}\VANthebibliography}

\usepackage[T1]{fontenc}
\usepackage{ae,aecompl}
\usepackage{times}
\usepackage{amssymb}
\usepackage{amsmath}
\usepackage{upgreek}
\usepackage{color}
\usepackage{graphicx}
\usepackage{pdflscape}

\title[Detection of hot cores]{ATOMS: ALMA Three-millimeter Observations of Massive Star-forming
regions - VIII. A search for hot cores by using C$_2$H$_5$CN, CH$_3$OCHO and CH$_3$OH lines}

\author[S.-L. Qin et al.]{
Sheng-Li Qin,$^{1}$\thanks{E-mail: qin@ynu.edu.cn} 
Tie Liu,$^{2}$\thanks{E-mail:liutie@shao.ac.cn} 
Xunchuan Liu,$^{2}$  
Paul F. Goldsmith,$^{3}$ 
Di Li,$^{4,5,6}$ 
Qizhou Zhang,$^{7}$ 
Hong-Li Liu,$^{1}$ 
\newauthor 
Yuefang Wu,$^{8}$ 
Leonardo Bronfman,$^{9}$ 
Mika Juvela,$^{10}$ 
Chang Won Lee,$^{11,12}$
Guido Garay,$^{9}$ 
Yong Zhang,$^{13}$ 
\newauthor 
Jinhua He,$^{14,15,9}$ 
Shih-Ying Hsu,$^{16}$ 
Zhi-Qiang Shen,$^{2}$ 
Jeong-Eun Lee,$^{17}$
Ke Wang,$^{18}$ 
Ningyu Tang,$^{19}$
\newauthor
Mengyao Tang,$^{20}$ 
Chao Zhang,$^{21}$ 
Yinghua Yue,$^{1}$  
Qiaowei Xue,$^{1}$ 
Shang-Huo Li,$^{11}$ 
Yaping Peng,$^{22}$ 
Somnath Dutta,$^{16}$
\newauthor 
Ge Jixing,$^{15}$ 
Fengwei Xu,$^{8}$ 
Longfei Chen,$^{4}$ 
Tapas Baug,$^{23}$
Lokesh dewanggan,$^{24}$
and
Anandmayee Tej$^{25}$ 
\\
Affiliations are listed at the end of the paper}
\date{Accepted XXX. Received YYY; in original form ZZZ}

\pubyear{2022}

\begin{document}
\label{firstpage}
\pagerange{\pageref{firstpage}--\pageref{lastpage}}
\maketitle

\begin{abstract}
Hot cores characterized by rich lines of complex organic molecules are
considered as ideal sites for investigating the physical and chemical
environments of massive star formation. We present a search for
hot cores by using typical nitrogen- and oxygen-bearing complex
organic molecules (C$_2$H$_5$CN, CH$_3$OCHO and CH$_3$OH), based
on ALMA Three-millimeter Observations of Massive Star-forming
regions (ATOMS).  The angular resolutions and line sensitivities of the ALMA observations are  better than 2 arcsec and 10 mJy/beam, respectively. 
A total of 60 hot cores are identified with 45
being newly detected, in which the complex organic molecules have
high gas temperatures  ($>$ 100 K) and small source sizes ($<$ 0.1 pc). So far this is the
largest sample of hot cores observed with similar angular
resolution and spectral coverage. The observations have also shown
nitrogen and oxygen differentiation in both line emission and gas
distribution in 29 hot cores.  Column densities of CH$_3$OH and
CH$_3$OCHO increase as rotation temperatures rise. The column
density of CH$_3$OCHO correlates tightly with that of CH$_3$OH.
The pathways for production of different species are discussed.
Based on the spatial position difference between hot cores and
UC~H{\sc ii} regions, we conclude that 24 hot cores are externally
heated while the other hot cores are internally heated. The
observations presented here will potentially help establish a hot core
template for studying massive star formation and astrochemistry.

\end{abstract}



\section{Introduction}
Massive stars play an important role in shaping structure and
evolution of galaxies, but also can affect star and planet
formation. They also are dominant sources of heavy elements and UV
radiation (see Zinnecker \& Yorke 2007 and references therein).
How massive stars form is not yet well understood. During
massive star formation processes, the hot core phase is
particularly important since hot cores may trace the physical and
chemical environments where massive stars are born. When clouds
collapse to form massive stars, the material
surrounding these objects are heated up, leading to formation
of hot cores with rich chemistry.

Hot cores are characterized by rich line emission from complex
organic molecules (COMs) with  higher gas temperature ($>$ 100 K) and smaller
source size ($<$0.1 pc) (e.g, Kurtz et al. 2000; Cesaroni 2005). Of the detected
250 molecular species\footnote
{https://cdms.astro.uni-koeln.de/classic/molecules} (see also
McGuire 2021), most of COMs, as well as simple molecules are
frequently observed in hot cores. Different molecules are used for
probing different physical components at different scales
(Tychoniec  et al. 2021; van Dishoeck \& Blake 1998; J{\o}gensen
et al. 2021). Especially COMs are thought to play an important
role in prebiotic chemistry which may be linked to the origin of
life (e.g., Herbst \& van Dishoeck 2009; Ceccarelli et al. 2017).
Therefore observations toward hot cores are crucial in the study
of massive star formation and astrochemistry.

Hot cores have been observed towards individual cases and large
sample of sources by single dish telescopes (Belloche et al. 2013;
Bisschop et al. 2007; Fontani et al. 2007; Gibb et al. 2000a;
Schilke et al. 1997, 2001, 2006; Halfen et al. 2013; Coletta et al.
2020; Widicus Weaver et al. 2017; Suzuki et al. 2016, 2018;
Crockett et al. 2014; Xie et al. 2021; Hern\'{a}ndez-Hern\'{a}ndez et al. 2019;
Neill et al. 2014; M\"oller et al. 2021; Ospina-Zamudio et al. 2018; Li et al. 2020; Bergin et
al. 2010; van der Tak et al. 2000). Due to small source sizes of
hot cores, single dish observations suffer from beam dilution and
sample the emission from both hot cores and surrounding cold
envelopes. The millimeter/submillimeter interferometric arrays
(e.g., SMA, NOEMA, and ALMA) offer larger bandwidth, higher
spatial resolution and increased sensitivity, which have promoted
the hot core observations greatly, but mainly focused on
individual sources or small samples (Law et al. 2021; van der Walt et al. 2021; Beuther et al. 2009; Qin et
al. 2008, 2010, 2015; Rong et al. 2016; Rivilla et al. 2017;
B{\o}gelund et al. 2019; Guzm\'an et al. 2018; Luo et al. 2009; Csengeri et al.
2019; Sakai et al. 2013, 2018; Zapata et al. 2011; Mottram et al.
2020; Fuente et al. 2021; Belloche et al. 2016, 2019; Brogan et al. 2016; Bonfand et
al. 2017; Xue et al. 2019; wong \& An 2018;  S\'anchez-Monge et al. 2010, 2014;
Hern\'{a}ndez-Hern\'{a}ndez et al. 2014; Peng et al. 2019; Liu et al. 2001, 2002; Remijan et al. 2003, 2004; Palau
et al. 2017; Wu et al. 2014; Taniguchi et al. 2020; Rathborne et al. 2011; Orozco-Aguilera et al. 2017;
 Ahmadi et al. 2018; Mottram et al. 2020;
Pagani et al. 2017; Tercero et al. 2018).
Since these observations were made with different spatial
resolution and spectral setup, it is difficult to make a comparison.
More recently, a large sample including 18 well-known high-mass
star-forming regions was observed by the NOEMA with same spectral
setup, suggesting that  most molecules are destroyed in evolved
cores having less emission lines (Gieser et al. 2021). Of the
detected hot cores, a few show chemical differentiation between nitrogen-
and oxygen-bearing COMs (e.g., Wyrowski et al. 1999;  Qin et al. 2010, 2015;
Remijan et al. 2004; Kalenskii \& Johansson 2010; 
Suzuki et al. 2018; Lee et al. 2019; Csengeri et al. 2019; Allen et al. 2017;
Jim\'enez-Serra et al. 2012; Gieser et al. 2019; Mills et al.
2018). Chemical differentiation and heating mechanism are still
longstanding problems due to a lack of large sample and systematic
observations.

 ATOMS project has observed 146 massive clumps at 3 mm band with ALMA (Liu et
al. 2020a, 2020b). 90 out of a total of 453 compact dense cores have been
considered as hot core candidates based
on number of emission lines (Liu et al. 2021). Hot cores are rich
in COM lines. Especially two typical nitrogen- and oxygen-bearing molecules 
C$_2$H$_5$CN and CH$_3$OCHO are frequently detected in hot cores. Spectral 
windows (SPWS) 7 and 8 were tuned to observe  C$_2$H$_5$CN
and CH$_3$OCHO with more than 3 transitions, which are suitable for identifying 
hot cores.  Comparing with the submillimeter waveband, the lines at 3 mm have much less
blending problem. Aiming to build up a large sample of hot cores,
we conduct a survey using C$_2$H$_5$CN and CH$_3$OCHO lines based
on the ATOMS continuum and line data at 3 mm band in this work.
The two molecules can also be used for investigating chemical
differentiation among these hot cores. The observations and data
reduction are briefly described in Section 2, and the
observational results including hot core identification and
parameter calculation are given in Section 3. We discuss the
heating mechanism and nitrogen and oxygen differentiation in Section 4. The
main results and conclusions are summarized in Section 5.

\section{Observations}

The sample selection and basic observational parameters are
described in Liu et al. (2020a). In brief, the ALMA band 3
observations were made toward 146 massive clumps from September to
middle November 2019 with both the Atacama Compact 7-m Array (ACA)
and the 12-m array (C43-2 or C43-3 configurations). The correlator
setup was tuned to include eight SPWs with 6
higher spectral resolutions of $\sim$ 0.2--0.4 km s$^{-1}$ windows
(SPWs 1--6) in the lower sideband and  SPWs 7 and 8 having lower
spectral resolutions in the upper side band. SPWs 7 and 8 have a
broad bandwidth of 1875 MHz corresponding to a spectral resolution
of $\sim$1.6 km s$^{-1}$ which were used for continuum imaging and
line survey purposes. The frequencies of  SPWs 7 and 8 range from
97536 to 99442 MHz, and from 99470 to 101390 MHz respectively,
covering many of COM lines. In addition to the COM lines,
H40$\alpha$ line at 99023 MHz is included in SPW 7 used for H{\sc
ii} region identification. SiO (2-1) at 86847 MHz are tuned in SPW
4 for tracing shocked gas.  Since hot cores have smaller source
size and COM lines suffer from less missing flux problem, we only
use the 12 m array data for identifications of COM lines and hot
cores in this work. Data reduction was done using the CASA
software package version 5.6 (McMullin et al. 2007). The resultant
continuum image and line cubes with the 12 m array data for the 146
clumps have angular resolutions of $\sim$ 1.2--1.9 arcsec, and
maximum recoverable angular scales of $\sim$ 14.5--20.3 arcsec.
The mean 1 $\sigma$ noise level is better than 10 mJy beam$^{-1}$
per channel for lines, and 0.4 mJy beam$^{-1}$ for continuum.
 Taking angular resolution of 1.9 arcsec and 3 $\sigma$  level
, the position accuracy of line images due to the noise is estimated to be better than 0.3 arcsec by using formula 
$\Delta \theta$=0.45 $\frac{\theta_{\rm FWHM}}{\rm {S/N}}$ (Reid et al. 1988).





\section{Results}

\subsection {Line and hot core identifications}
As stated before, 453 compact dense cores were found in 146
massive clumps (Liu et al. 2021). We have inspected the line
emission of the 453 cores one by one and extracted spectra at the
line-rich positions. In total 60 line-rich cores are  considered as hot
core candidates.  The results are in good agreement with those
by Liu et al. (2021) where 54 cores were found in a statistic
manner to have significant detection of at least 20 COM
transitions  with line intensities larger than 3 $\sigma$ level. Then we use the eXtended CASA Line Analysis
Software Suite
(XCLASS,\footnote{https://xclass.astro.uni-koeln.de/} M\"oller et
al. 2017) for further line identification and parameter
calculation. The XCLASS accesses the Cologne Database for
Molecular Spectroscopy (CDMS; M\"uller et al. 2001, 2005) and Jet
Propulsion Laboratory  molecular databases (JPL; Pickett et al.
1998). Assuming the molecular gas satisfies local thermodynamical
equilibrium condition (LTE), the XCLASS solves radiative transfer
equation and produces synthetic spectra for specific molecular
transitions by taking source size, beam filling factor, line profile, line blending, excitation, and opacity into account. 
In the XCLASS modelling, the input parameters are
source size, beam size, line velocity width, velocity offset,
rotation temperature and column density (M\"oller et al. 2017). In
our case, we take deconvolved angular sizes of the continuum
sources as source sizes  which are listed in Table 1. The velocity offsets are determined when
referred to commonly detected CH$_3$OH line at 100.6389 GHz.
Therefore we set  rotation temperatures, column densities
and velocity widths as free parameters to simulate the observed
spectra. To obtain optimized rotation temperature and column density
parameters, we employ Modeling and Analysis Generic Interface for
eXternal numerical codes (MAGIX; M\"oller et al. 2013) for further
calculation. Note that we only consider line-rich species of
C$_2$H$_5$CN, CH$_3$OCHO and CH$_3$OH molecule in this work.

The frequency setup covers many of C$_2$H$_5$CN and CH$_3$OCHO
transitions in SPWs 7 and 8. The observed C$_2$H$_5$CN and
CH$_3$OCHO transitions span upper level energies of 30--139 K and
22--58 K, respectively. Therefore we do not expect very hot gas
components to be detected. Only a few CH$_3$OH lines are tuned in
our observations with upper level energies ranging from 17 to 724
K. More than 3 transitions of C$_2$H$_5$CN and CH$_3$OCHO are
detected in the 59 line-rich cores. The line transitions of C$_2$H$_5$CN and CH$_3$OCHO in IRAS 09018-4816
are too weak to derive rotation temperature, but 3 CH$_3$OH transitions are detected. Rotation temperatures and
column densities can be derived simultaneously. The XCLASS
calculations suggested that all the 59 line-rich cores have gas
temperature derived from C$_2$H$_5$CN and CH$_3$OCHO higher than
100 K. IRAS 09018-4816 has gas temperature above 100 K based on CH3OH line data.  Considering the distances (Liu et al. 2021a) and source sizes, the 60 cores have source size smaller than 0.1 pc.
We then confirm the 60 cores as hot cores, and give their
coordinates, rotation temperatures and column densities in Table
1. We have checked the literature and found that 15 of these hot cores have
been reported before (I18089-1732, Beuther et al. 2004;
I18032-2032/G9.62+0.19, Liu et al. 2015;  Dall'Olio et al. 2019; I18507+0110/G34.43+0.24,
Beltr\'an et al. 2009; Fu \& Li 2017; Calcutt et al. 2014;
I17441-2822/Sgr B2(M), S\'anchez-Monge et al. 2017;
IRAS18056-1952/G10.47+0.03, Rolffs et al. 2011;
I18507+0121/G34.26+0.15 , Sakai et al. 2013; Mookerjea et al.
2007; I16060-5146, I16065-5158, I12326-6245, I14498-5856,
I15254-5621, Araya et al. 2005; Dedes et al. 2011; I16547-4247,
I17233-3606, I18182-1433, Hern\'{a}ndez-Hern\'{a}ndez et al. 2014;
I17175-3544/NGC6334I, Brogan et al. 2016). Thus, 45 hot cores are
newly detected in our work. This is currently the largest hot core
sample observed with similar angular resolution and spectral
coverage. Forty-one cores have more than 3 CH$_3$OH transitions
detected. Note that this is a lower limit since the frequency setup
 of C43-2 and C43-3 configurations has a 48 MHz difference
which leads to two CH$_3$OH transitions  not covered in C43-3
observations. For CH$_3$OH with less than 3 transitions detected,
column densities are estimated by assuming gas temperature equal
to the rotation temperatures of oxygen-bearing molecule
CH$_3$OCHO, or equal to the rotation temperatures of C$_2$H$_5$CN
in case that CH$_3$OCHO is not detected.
\begin{figure*}

\includegraphics[width=15cm,height=20cm]{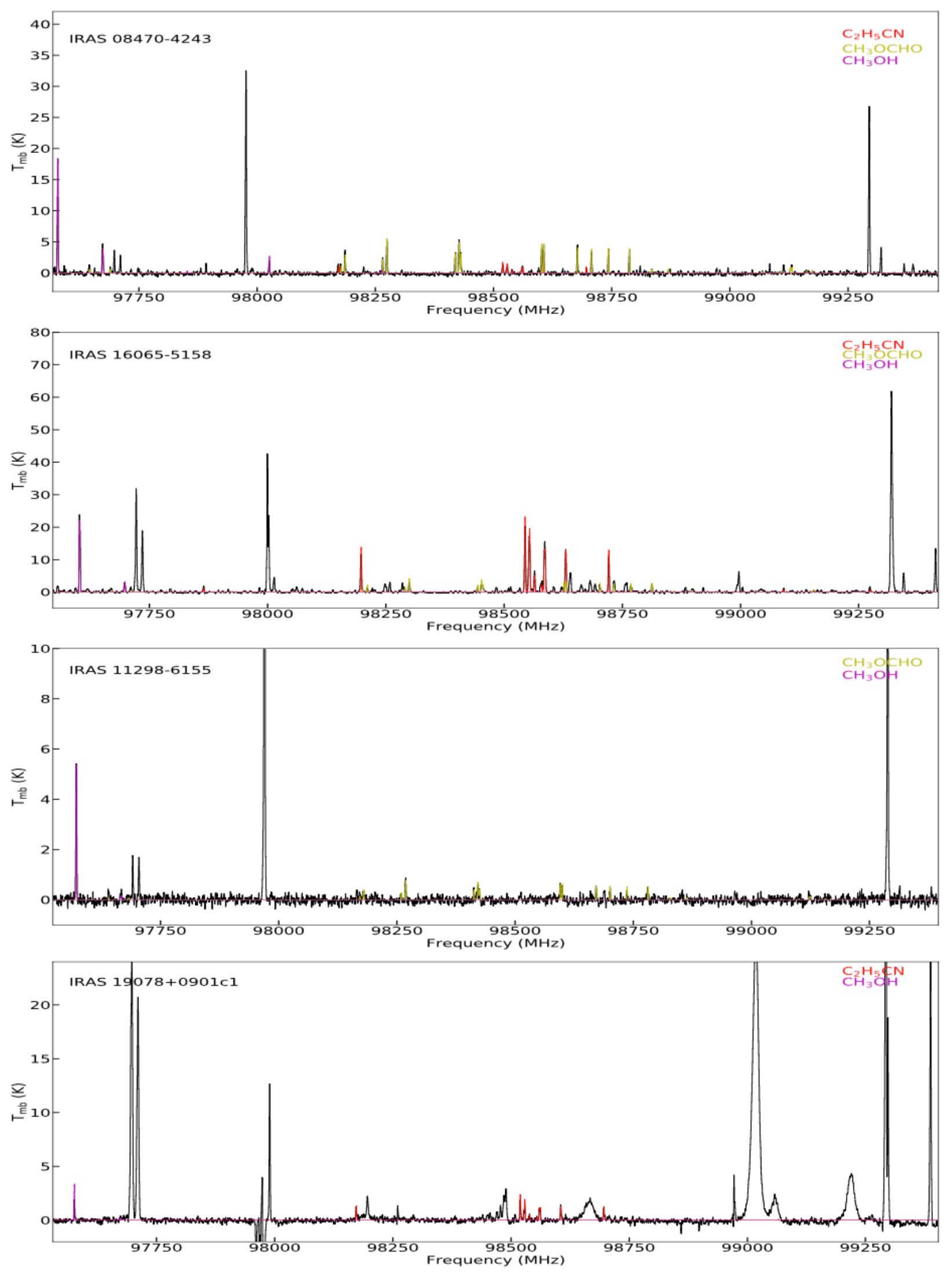}

\caption{Sample spectra in SPW 7  for four typical hot cores. The
observed spectra are shown as black curves and the XCLASS modelled
spectra are coded in color.}

\end{figure*}
\begin{figure*}
 \includegraphics[width=15cm,height=20cm]{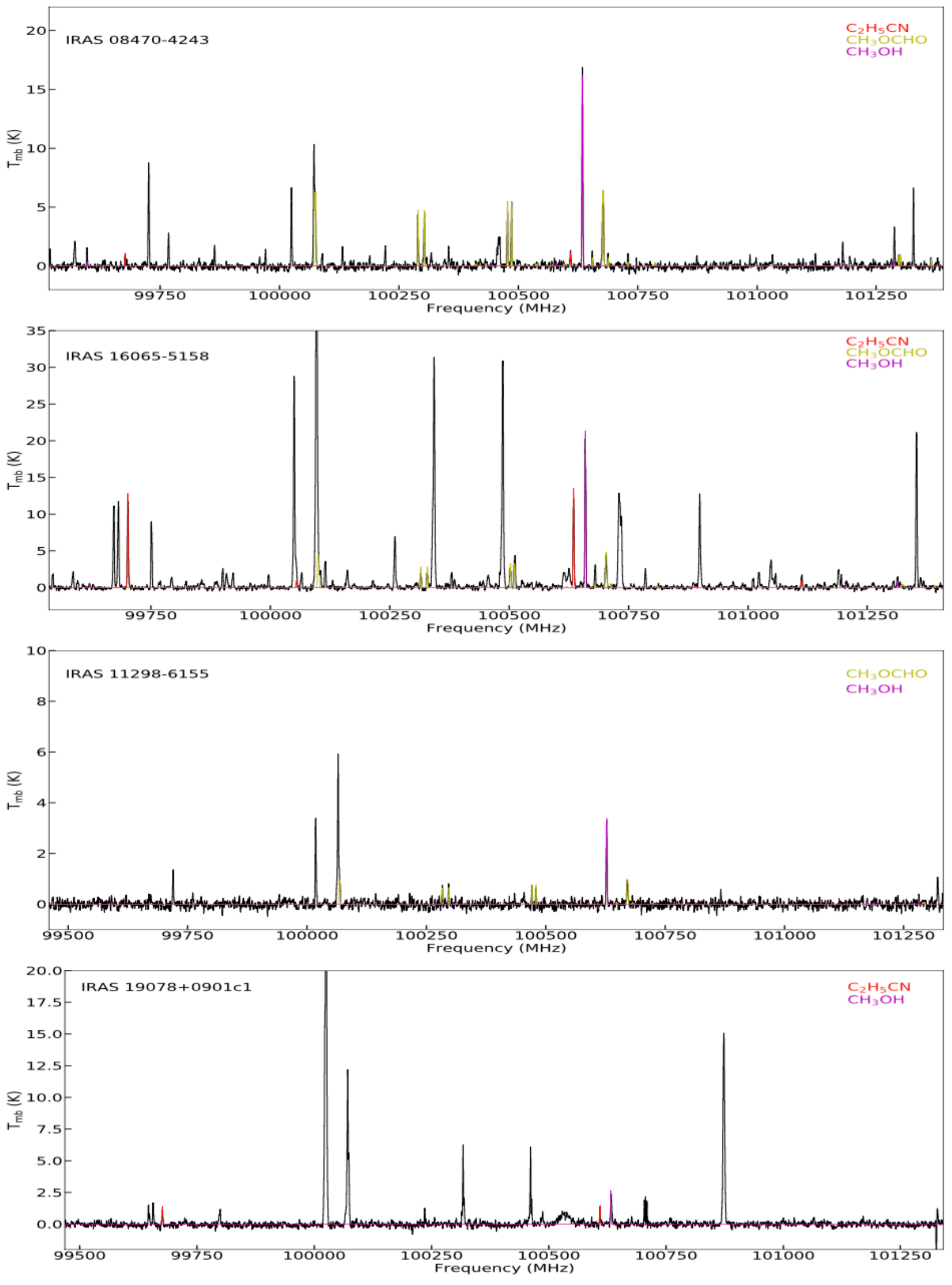}

 \contcaption{Sample spectra in SPW 8  for four typical hot cores.
 The observed spectra are shown as black curves and the XCLASS modelled spectra are coded in color. The spectra of all the sources are availble in the supplementary material.}

\end{figure*}

Figure 1 presents  sample spectra toward a few hot cores with
modeled molecular spectra (C$_2$H$_5$CN, CH$_3$OCHO and CH$_3$OH)
overlaid on the observed ones. Clearly lots of C$_2$H$_5$CN and
CH$_3$OCHO lines are observed and large number of lines are
emitted from C$_2$H$_5$CN and CH$_3$OCHO species. The spectra
emitted from the hot cores also show different emission features.
From Figure 1 and the figures in the supplementary material, one can see that
line peak intensities of C$_2$H$_5$CN are larger than those of
CH$_3$OCHO in 19 cores while the opposite is seen in 5 cores. It
is worth mentioning that  C$_2$H$_5$CN
line emission is absent in 4 cores while CH$_3$OCHO is not detected in 9
cores. The differences in line emissions from these cores may
imply differences in physical and chemical environments among
these sources.

\subsection {Line images}
Line images of various molecules can provide valuable information
on their spatial distributions and possible chemical routes. We
choose three molecular line transitions of C$_2$H$_5$CN,
CH$_3$OCHO and CH$_3$OH at 98523, 98792 and 100639 MHz for the
images, respectively. The sample images are shown in Figure 2, and
other images are presented in the supplementary material. Overall,
most of the C$_2$H$_5$CN, CH$_3$OCHO and CH$_3$OH emissions are
associated with the continuum images, though there are position
offsets between peaks of the line and continuum images. All the
line images show compact source structure and the three line
emissions mainly distribute over small regions, indicating hot
core properties. It can be seen that the emission peaks of
C$_2$H$_5$CN are coincident with those of CH$_3$OCHO in 28 cores.
Considering position accuracy of 0.3 arcsec in line images, 29
cores clearly show nitrogen and oxygen separation, i.e., the
emissions of C$_2$H$_5$CN and CH$_3$OCHO peak at different 
positions. So far only a few hot cores exhibit nitrogen and oxygen
differentiation in space. Our observations provide the largest hot
core sample with positional separation between nitrogen- and
oxygen-bearing molecules.
\begin{figure*}

\includegraphics[width=16cm,height=14cm]{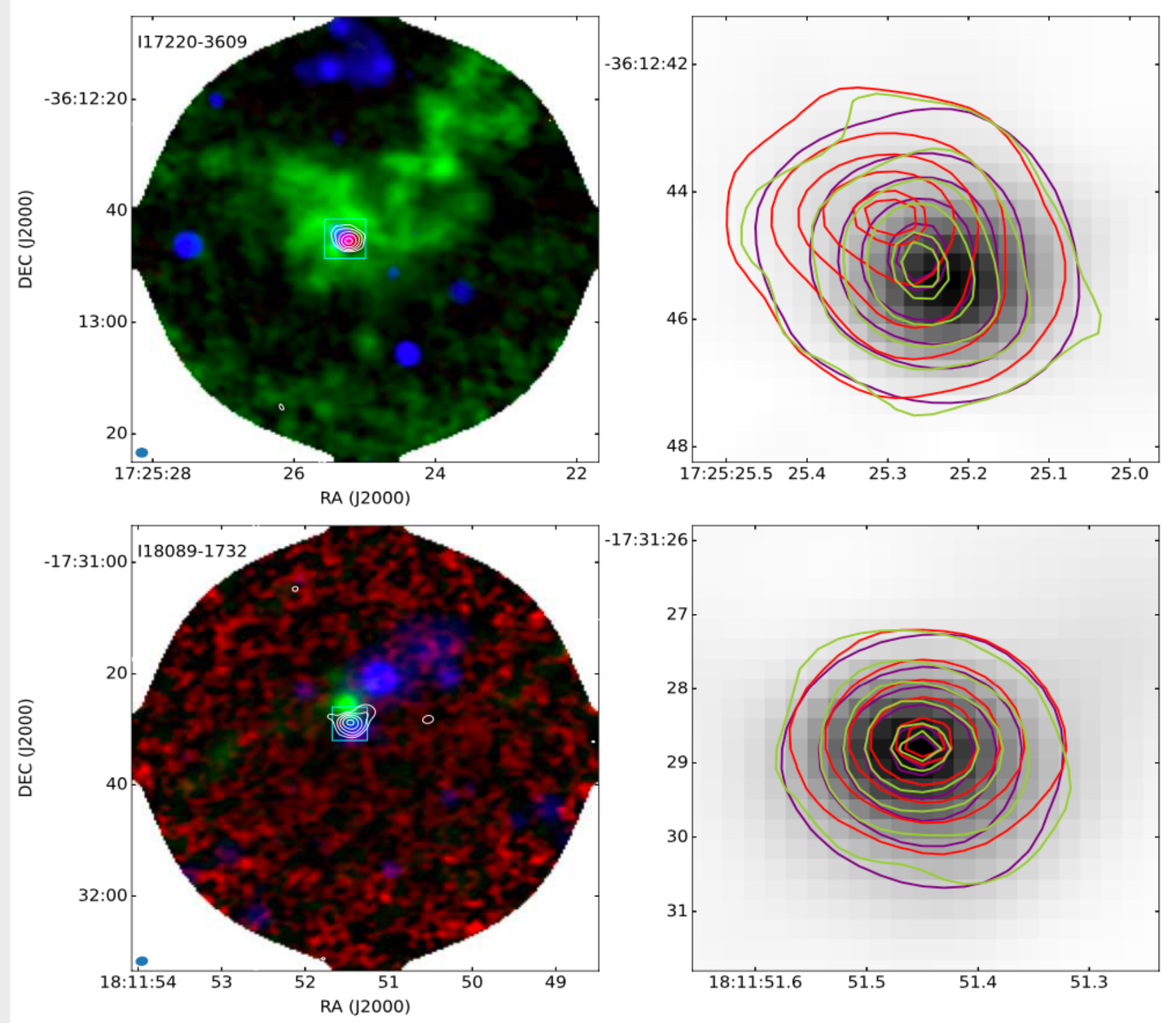}

 \caption{Sample images of the continuum and organic
molecular lines. In the left panels, the background shows the
three-color image composed by H40$\alpha$ (red), SiO (green) and
Spitzer 4.5 $\micron$ (blue), and the white contours represent the
3 mm continuum; the green rectangles mark the imaging regions of the right panels. In the right panel, the background shows the 3 mm
continuum. The red, cygn and yellow contours represent the
integrated intensities of C$_2$H$_5$CN, CH$_3$OH and CH$_3$OCHO,
respectively. The contour levels are 10 to 90 percent (stepped by
20 percent) of the peak values. The innermost contour has a level
of 95 percent of the peak value. The images of all the sources are availble in the supplementary material.}

\end{figure*}

\subsection {Rotation temperatures, column densities and abundances relative to CH$_3$OH}
From Table 1, one can see that the rotation temperatures of
C$_2$H$_5$CN, CH$_3$OCHO and CH$_3$OH range from 100--285 K,
100--230 K and 100--290 K respectively. Most of the cores have gas
temperatures of 100--200 K for the three species. When compared
with C$_2$H$_5$CN and CH$_3$OCHO, CH$_3$OH has the largest column
densities ranging from 8.6$\times$10$^{16}$ to 1.5
$\times$10$^{19}$ cm$^{-2}$ which is 2--3 orders of magnitude
higher than those of C$_2$H$_5$CN  and 1--2 orders of magnitude
higher than those of CH$_3$OCHO. Figure 3 shows the relationships
between rotation temperatures and column densities for the three
molecules in our sample. Column densities of CH$_3$OH increase as
their rotation temperatures rise. A linear fitting to the data
gives log(N(CH$_3$OH))=0.022T$_{\rm rot}$+14 with
correlation coefficient of 0.86. A similar trend is seen for
CH$_3$OCHO with the relation
log(N(CH$_3$OCHO))=0.017T$_{\rm rot}$+14 and correlation
coefficient of 0.68.  The linear fitting to C$_2$H$_5$CN data gives a correlation coefficent of 0.29, suggesting that column densities of C$_2$H$_5$CN are
not sensitive to its rotation temperatures, though a weak
increasing trend is seen. The results appear to indicate that the
two oxygen-bearing molecules are chemically-related species or the
two species reside in the same astrophysical environments.

\begin{figure}

\includegraphics[width=8cm,height=15cm]{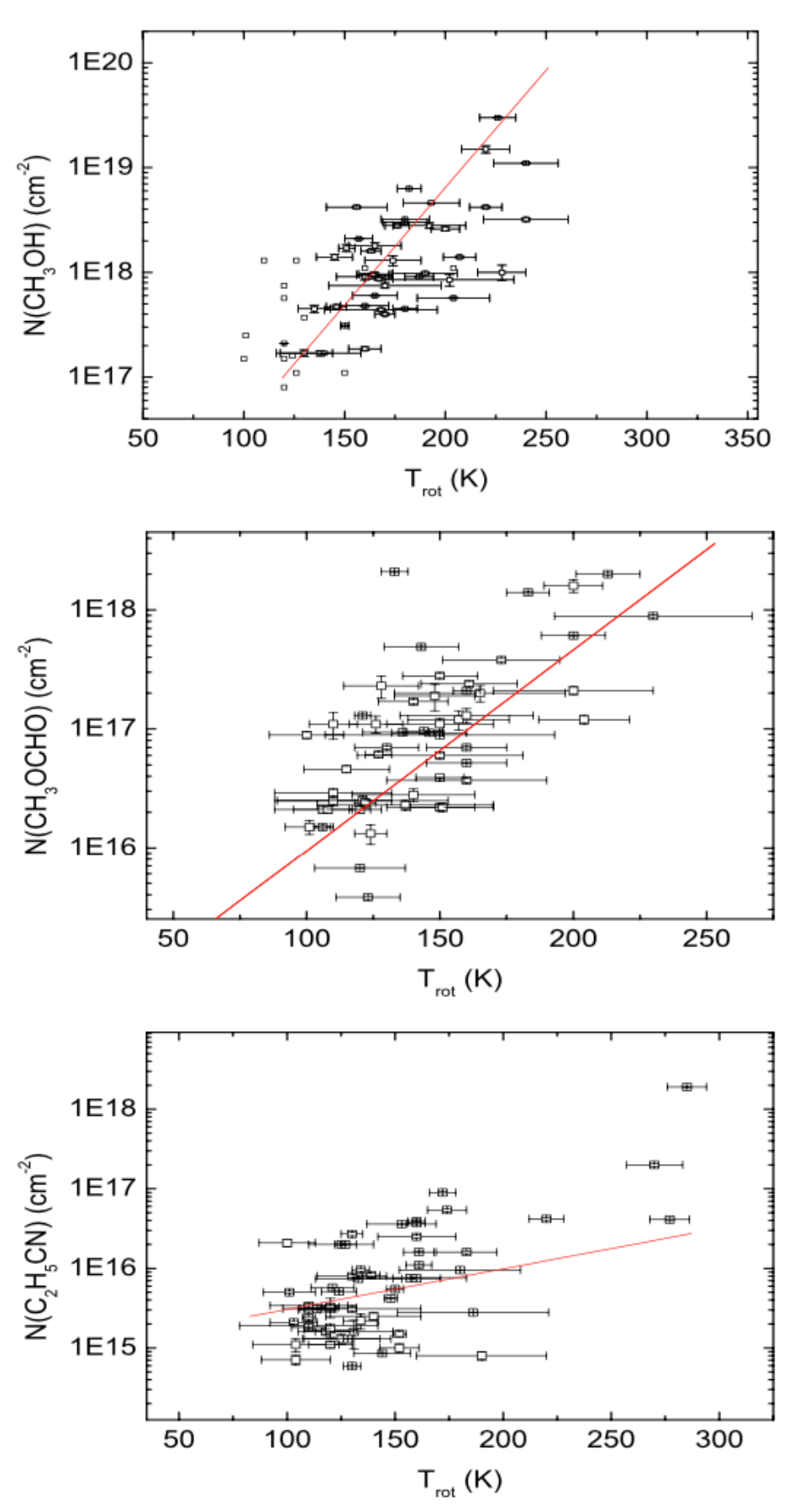}

 \caption{Relationships between column densities and rotation temperatures for CH$_3$OH, CH$_3$OCHO and C$_2$H$_5$CN.
 The open squares indicate the derived column densities and rotation temperatures for the hot cores.
 The  bars indicate the 1$\sigma$ errors. The data points without error bars in the upper panel are for the sources
 without three CH$_3$OH line detected.  The linear least-squares fit is shown as red solid  line.}

\end{figure}
Figure 4 presents the relationships of column densities among the
three species. Overall, column densities of both C$_2$H$_5$CN and
CH$_3$OCHO have positive correlation with that of  CH$_3$OH. The
linear fittings give
log(N(CH$_3$OCHO))=1.1log(N(CH$_3$OH))--2.5 and
log(N(C$_2$H$_5$CN))=1.2log(N(CH$_3$OH))--5.1 with
correlation coefficients of 0.95 and 0.82 for CH$_3$OCHO and
C$_2$H$_5$CN, respectively. The models suggested that
oxygen-bearing molecules CH$_3$OCHO and CH$_3$OH and
nitrogen-bearing molecule C$_2$H$_5$CN have different forming
pathways (Charnley et al. 1992; Rodgers \& Charnley 2001).
Probably the column densities of the three molecules depend on the
initial cloud environments. As shown in Figure 3, the column
densities of the oxygen-bearing molecules of CH$_3$OCHO and
CH$_3$OH correlate well with the rotation temperatures while the
nitrogen-bearing molecule C$_2$H$_5$CN does not follow the same trend.
A possible explanation is that excitation of CH$_3$OCHO and
CH$_3$OH depends on both temperature and density while excitation
of C$_2$H$_5$CN is mainly related to its column density.

Among the detected COMs so far, CH$_3$OH has the highest gas phase
abundance relative to H$_2$. The infrared observations have shown
that CH$_3$OH is the most abundant molecule relevant to water ice
(e.g., Dartois et al. 1999; Gibb et al. 2000b; Ehrenfreund \&
Charnley 2000). Grain surface chemical models also suggested that
formation of many of other COMs (especially the oxygen-bearing
molecules) is related to CH$_3$OH. From the abundance criteria,
the CH$_3$OH can be taken as one of the reference molecules to compare
with other sources. The relative abundances defined by f =
N(x)/N(CH$_3$OH) are listed in Table 1, where x is the specific
molecule CH$_3$OCHO or C$_2$H$_5$CN. The relative abundances range
from 8.4$\times$10$^{-4}$ to 4$\times$10$^{-2}$ for C$_2$H$_5$CN
and from 2.2$\times$10$^{-2}$ to 2.8 $\times$10$^{-1}$ for
 CH$_3$OCHO, respectively, which are in agreement with previous results in hot
cores (B{\o}elund et al. 2019; Mookerjea et al. 2007; Allen et al.
2017; Guzm\'an et al. 2018; Law et al. 2021; Feng et al. 2016;
Bonfand et al. 2017, 2019; Molet et al. 2019; Qin et al. 2010).

\begin{figure}

\includegraphics[width=8cm,height=10cm]{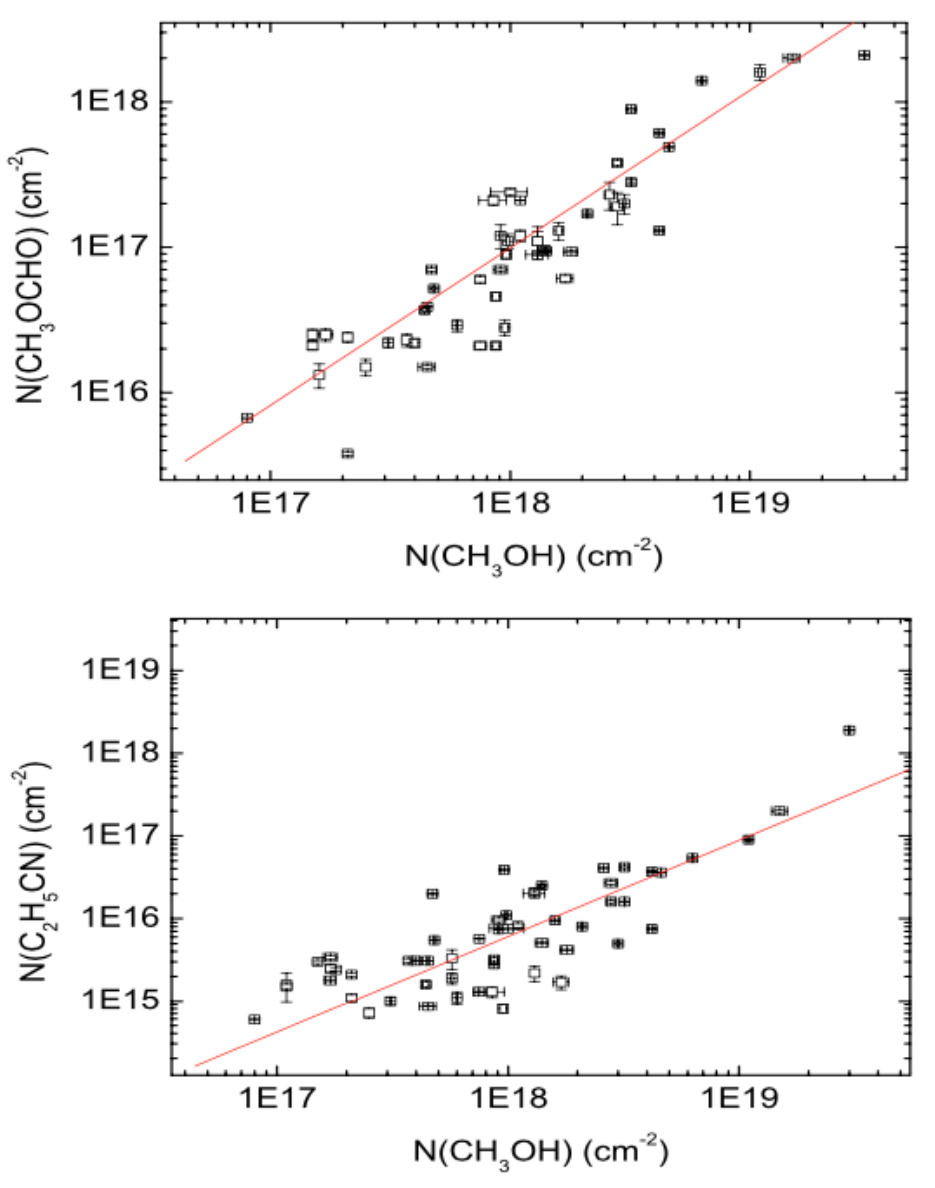}

 \caption{Correlations of column densities of CH$_3$OH, CH$_3$OCHO and C$_2$H$_5$CN.
 The open squares indicate the column densities and the bars indicate the 1$\sigma$ errors.
 The linear least-squares fit is shown as  red solid line.}

\end{figure}

\section{Discussion}

\subsection {Heating mechanism}

Observationally the evolutionary sequence of high-mass star
formation starts from massive pre-stellar cores, to high-mass
protostellar objects (HMPOs), hot cores, HC H{\sc ii} and UC H{\sc
ii} regions where massive stars are already formed (e.g. Beuther
et al. 2007; Menten et al. 2005). During the processes of massive
star formation, the gravitational energy is converted into thermal
energy to form the HMPOs. The radiation from HMPOs, HC H{\sc ii}
and UC H{\sc ii} regions can  heat up their immediate environments
to form  hot cores. Therefore hot cores are observed to be
associated with dense and hot regions near the young high-mass
protostellar objects and the UC H{\sc ii} regions.

In this section, we want to differentiate internal and
external heating mechanisms by comparing the relative position
of line images of COMs and UC H{\sc ii} regions. Radio continuum
at centimeter wavelength and radio recombination lines are
characteristics of the UC H{\sc ii} regions associated with newly
formed high-mass young stars. ATOMS covers the H40$\alpha$ line
transition which can be used for identifying UC H{\sc ii} regions.
Figure 2 presents sample images of the continuum, organic
molecular, H40$\alpha$ and SiO lines. From the left panels of Figure 2
and the figures in the supplementary material, 28 UC H{\sc ii} regions are
identified based on H40$\alpha$ line images. When comparing the
continuum and H40$\alpha$ images with the CH$_3$OH line images, we
find that the morphologies of CH$_3$OH line images are similar to
those of continuum images, and that the CH$_3$OH emissions peak at the peak
positions of the continuum, for most sources without associated
UC H{\sc ii} regions. We note that although IRAS 08303--4303, IRAS
13484--6100, IRAS 16484--4603, IRAS 17158--3901, IRAS 17233-3606,
and IRAS 18182--1433 have no  UC H{\sc ii} region counterparts,
their CH$_3$OH emission peaks are located offset from those of
continuum emissions. Probably shocks traced by the 4.5 $\micron$ and SiO push
the molecular materials away. For the sources having
H40$\alpha$ line emission, peaks of CH$_3$OH are offset from the
continuum images except for IRAS 18056--1952. Previous molecular
line images were observed to be not always consistent with the
continuum emission in cases where continuum emission
includes contributions from dust and free-free emission of UC
H{\sc ii} regions. Also no internal energy sources were
identified, suggesting that these hot cores are externally heated
(e.g., Wyrowski et al. 1999; De Buizer et al. 2003; Mookerjea et
al. 2007). Then we simply classify the observed hot cores in our
work into two categories: externally heated hot cores and
internally heated hot cores. Twenty four hot cores are thought to be heated
by adjacent UC H{\sc ii} regions and the other sources are
classified as internally heated hot cores, suggesting that some hot
cores are not simply linked to HMPOs and precursors to UC H{\sc
ii} regions in evolutionary sequence during massive star formation
(Law et al. 2021), but they only trace physical and chemical
environments of hot, dense and line-rich regions. Note that we consider IRAS 18056-1952 
and I18032-2032c1 as internally heated since both line and continuum images of the two sources peak at 
 UC H{\sc ii} region positions.  

Figure 5 shows the cumulative distributions of the rotational
temperatures and column densities of  C$_2$H$_5$CN and CH$_3$OCHO
of the two groups of hot cores, the internally heated ones and the
externally heated ones, respectively. No statistical difference of
the rotational temperatures of C$_2$H$_5$CN between the two groups
can be found. The rotational temperatures of  CH$_3$OCHO tend to
be higher in the externally heated hot cores than in the
internally heated hot cores, although this difference  is not very much
significant with  a p-value of 0.14 given by the Kolmogorov-Smirnov test. The externally and internally heated cores 
show no obvious difference (with a p-value of 0.25) in the cumulative
distributions of the column densities of CH$_3$OCHO.  While significant difference (with
a p-value of 0.027) can be seen between the cumulative
distributions of the column densities of C$_2$H$_5$CN ($N_{\rm C_2H_5CN}$). The values of $N_{\rm C_2H_5CN}$ in the
externally heated hot cores are a factor of two higher than those
in the internally heated hot cores. Of course, one may expect that
higher temperature component from highly excited lines are
shifting towards the UC H{\sc ii} regions for the externally
heated hot cores while higher temperature components will be
located at the dusty continuum peaks for internally heated hot
cores. Future high angular resolution and high frequency
observations covering high energy line transitions can verify the
occurrence and characteristics of externally and internally heated
mechanisms in detail.

\begin{figure*}

 \includegraphics[width=18cm,height=14cm]{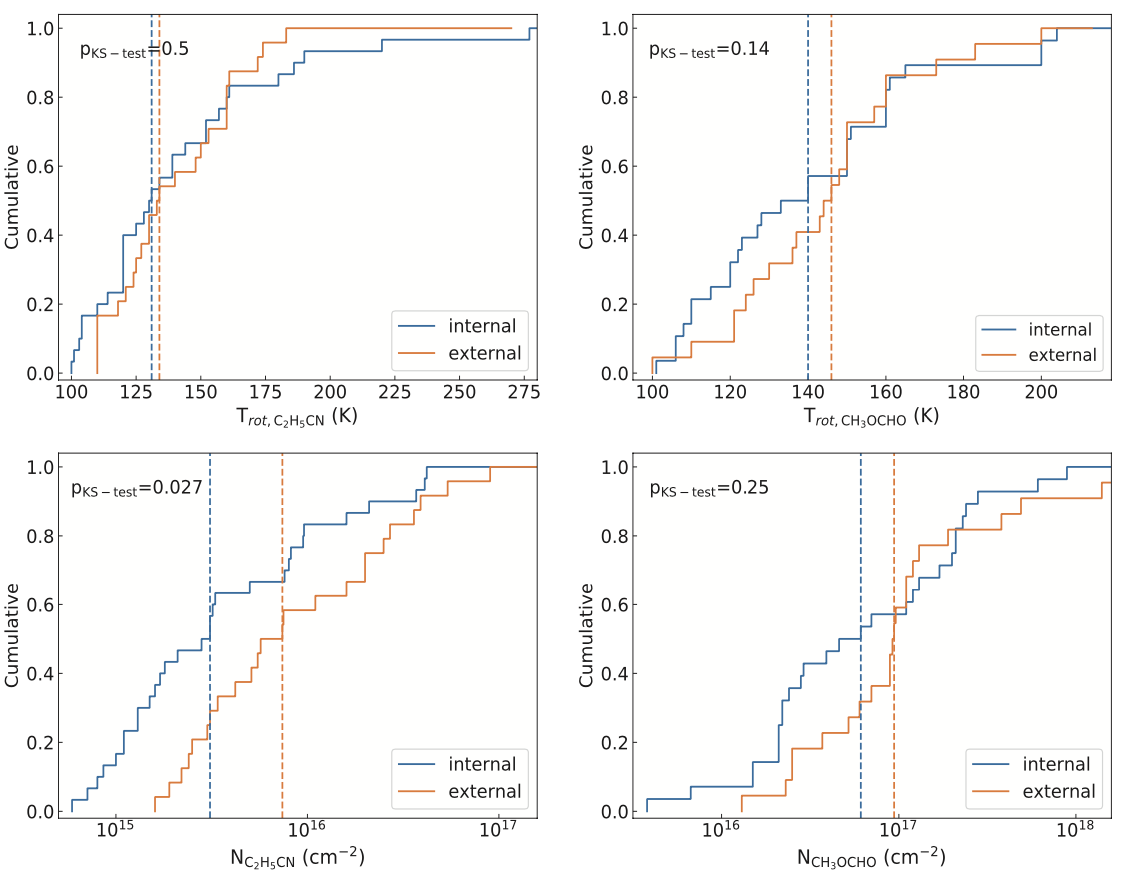}

 \caption{The cumulative distributions of the rotational temperatures (upper) and column densities (lower)
of  C$_2$H$_5$CN (left) and  CH$_3$OCHO (right). The parameters of
the internally heated ones and the externally heated ones are
shown in blue and orange colors, respectively. The vertical dashed
lines represent the corresponding median values. The p-value of
the KS-test between the cumulative distributions of the two groups
of hot cores is shown in the upper-left corner of each panel.}

\end{figure*}

\subsection {Nitrogen and oxygen differentiation}
Nitrogen and oxygen differentiation has been suggested in some hot
cores (Blake et al. 1987; Wyrowski et al. 1999; Fayolle et al.
2015; Qin et al. 2015, 2010), showing different rotation
temperatures and abundances, and nitrogen- and oxygen-bearing COMs peaking at different spatial
positions. Observations have shown that many hot cores have
excitation temperatures of nitrogen-bearing COMs higher than those
of oxygen-bearing COMs (van't Hoff et al. 2020, and references
therein). Derived abundances of oxygen-bearing and
nitrogen-bearing COMs have positive correlations with the other
oxygen-bearing and nitrogen-bearing COMs respectively (e.g.,
Bisschop et al. 2007; Bergner et al. 2017; Suzuki et al. 2018).
The observations have also shown that nitrogen-bearing COMs peak
at protostar position while oxygen-bearing COMs are peaked offset
from the central source (Qin et al., 2010; Fayolle et al. 2015).

 From Table 1, rotation temperatures of C$_2$H$_5$CN in most hot cores are generally
 lower than those of CH$_3$OH, which is inconsistent with previous observations. As stated in Section 3.1,
a possible explanation is that our 3 mm observations of
C$_2$H$_5$CN lines have low upper level energies of 30--139 K, and
then the hot components of the cores are not sampled.  The LTE
calculations find CH$_3$OCHO and CH$_3$OH molecules to have larger
column density than that of the nitrogen-bearing molecule
 C$_2$H$_5$CN. Column density of CH$_3$OH correlates well with
 that of CH$_3$OCHO, but also has a positive correlation with
 C$_2$H$_5$CN which also conflicts with previous results as stated
 above. Probably our targeted lines of the C$_2$H$_5$CN and  CH$_3$OCHO  have
 low upper level energies which only sample the freshly evaporated species,
 and  then can not probe the innermost and hottest structure of the hot cores.
 Nitrogen and oxygen separation is clearly seen in 29 hot cores in which
 C$_2$H$_5$CN and oxygen-bearing molecules peak at different spatial
 positions (see Figure 2 and the figures in the supplementary material).

\subsection {Chemistry}

 COMs are suggested to be particularly important in both astrophysics and
astrochemistry, but also to be linked to origin of life. However
no firm conclusions about their formation are reached, even for
commonly detected CH$_3$OCHO and C$_2$H$_5$CN. Various chemical
models are proposed to interpret their origination (Charnley et
al. 1992, 1995; Caselli et al. 1993; Millar et al. 1991; Garrod et
al. 2006, 2008; Taquet et al. 2012, 2015, 2016; Rodgers \&
Charnley 2001, 2003; Pols et al. 2018; Choudhury et al. 2015). In
summary, main chemical routes are gas-phase and grain-surface
chemical reactions as well as interaction between gas-phase and
grain-surface molecules. Large sample observations of their
relative abundances and spatial distributions will be able to
provide clues in investigating their forming pathways.

  Higher gas-phase abundances of CH$_3$OH are derived from
(sub)millimeter wavelength observations toward massive star formation regions,
which is attributed to icy CH$_3$OH evaporated from grain mantles
leading to large abundance by various chemical models. From Figure
3, column densities of CH$_3$OH increase as rotation temperatures
rise, providing additional evidence that CH$_3$OH originates from
grain surface and  is then released into gas phase when gas
temperature is higher than 100 K. Our large sample observations
give positive correlation between CH$_3$OH and CH$_3$OCHO which
appears to indicate that CH$_3$OCHO is formed in gas phase and is
related to the evaporated CH$_3$OH as proposed by Charnley et al.
(1992, 1995). However, gas phase mechanisms by molecule-ion
reactions  have very low efficiency in producing CH$_3$OCHO (Horn
et al. 2004; Geppert et al. 2006). The relationship between
the observed column densities and rotation temperatures of
CH$_3$OCHO show the same trend as of CH$_3$OH that has also been repeorted by previous observations (Law et al. 2021), 
which may hint that CH$_3$OCHO also originates from grain surface and then is
evaporated into phase as temperature increases. The warm up model
developed by Garrod et al. (2006) suggested that CH$_3$OCHO is
mainly synthesized on grain surface in lower temperatures and gas
phase reactions play a dominant  role in forming CH$_3$OCHO when the
gas temperature is above 200 K. In this scheme, formation of
CH$_3$OCHO  is attributed to both grain surface and gas phase
reactions. In our case, the observations of CH$_3$OCHO transitions
at 3 mm band have low upper level energies and should sample
freshly evaporated species. The rotation temperatures of
CH$_3$OCHO in most cores are lower than 200 K. We then favor
CH$_3$OCHO observed by us at lower frequency band is synthesized
on grain surface. Future higher angular resolution observations of
high energy level lines can detect inner and even hotter
components of the hot cores, and  can test if the gas phase
reactions play an important role in forming CH$_3$OCHO.

C$_2$H$_5$CN can be formed by ion-molecule reactions in gas phase,
but the processes are quite slow. Grain surface reactions through
hydrogenation of HC$_3$N is a more efficient mechanism (Blake et
al. 1987; Caselli et al. 1993; Charnley et al. 2004). Contrary to
the two complex oxygen-bearing molecules, the column densities of
C$_2$H$_5$CN did not show positive correlation with its rotation
temperatures. A possible explanation is that C$_2$H$_5$CN tends to
have higher gas temperature than the oxygen-bearing COMs while our
lower energy level lines can only trace lower excitation
temperatures in the outer envelopes of the hot cores (\"Oberg et al.
2013). Large amount of C$_2$H$_5$CN are not fully evaporated from
grain surface. Surprisingly the column densities of C$_2$H$_5$CN
correlate well with those of CH$_3$OH which is inconsistent with
previous observations. Previous observations have shown that
nitrogen-bearing molecules have strong correlation with other
nitrogen-bearing molecules and  no correlation with oxygen-bearing
COMs (Bisschop et al. 2007; Bergner et al. 2017; Suzuki et al.
2018). As argued before, our observations may only sample freshly
evaporated species which have not involved in subsequent gas phase
reactions yet. The observed correlation between C$_2$H$_5$CN and
CH$_3$OH may reflect initial chemical environments or physical
differences (Van't Hoff et al. 2020).

 The spatial separation between C$_2$H$_5$CN and the two oxygen-bearing
molecules can be seen from Figure 2 and the figures in the supplementary material. But no obvious 
temperature differentiation is observed as in
Orion KL, W3(OH) complex and G34.26+0.15 (Blake et al. 1987;
Crockett et al. 2014; Wyrowski et al. 1999; Qin et al. 2015;
Mookerjea et al. 2007), due to our targeted lines having low upper
level energies. Another reason is that in most cases our
beam may cover part of both nitrogen and nitrogen cores. Nitrogen
and oxygen differentiation is a long-standing problem. Different
chemical models give different explanations (Caselli et al. 1993;
Garrod et al. 2008; Rodgers \& Charnley 2003; Van't Hoff et al.
2020). Higher resolution observations of highly excited lines from
various COMS toward a large sample of hot cores are needed to
interpret the observed differentiation.

\section{Conclusions}

We have performed a systematic hot core survey employing
C$_2$H$_5$CN, CH$_3$OCHO and CH$_3$OH commonly observed in hot
cores, based on the data obtained by the ATOMS project. We
summarize the main results in the following. \\

1. The ATOMS 3 mm band were set up to cover multiple C$_2$H$_5$CN, CH$_3$OCHO
and CH$_3$OH transitions, which were then used for deriving their
rotation temperatures and column densities. We identify a dense
core as a hot core if more than 3 transitions of the COMs are
identified and have rotation temperature above 100 K. In total, 60
hot cores are identified, out of which 45 are newly detected, thus representing
the largest hot core sample with similar angular
resolution and spectral coverage. \\

2. The observations have shown that
line intensities of  C$_2$H$_5$CN  are stronger than those of
CH$_3$OCHO lines in 19 cores while the opposite is observed in 5
cores. There are total absence of C$_2$H$_5$CN line emission in 4 cores
while CH$_3$OCHO lines are not detected in 9 cores. Probably the different spectral features 
reflect the differences in physical and chemical environments. \\

3. Line images of the three molecules show compact source
structure concentrated on small regions for the hot cores. There
are 28 hot cores associated with UC H{\sc ii} regions. Based on
relative positions between UC H{\sc ii} regions and CH$_3$OH
emission peaks, we classify the detected cores into two
categories: externally heated and internally heated hot cores. 24
hot cores are externally heated by radiation from adjacent UC
H{\sc ii} regions while the others are heated internally. \\

4. Our large sample shows that column densities of CH$_3$OCHO and
CH$_3$OH correlate well with their rotation temperatures, and
column densities of the two molecules have a positive correlation.
The results support that CH$_3$OCHO and CH$_3$OH  originate from the grain surface chemistry. \\

5. Nitrogen and oxygen differentiation is observed in 29 hot cores
with nitrogen-bearing and oxygen-bearing molecules peaking at
different positions, but no clear differences in temperature and
column density are seen from our data. \\

In summary, our ALMA 12 m array observations provide the largest robust hot core
sample with similar angular resolution and spectral coverage. 24
hot cores are considered to be heated by external energy sources,
and therefore hot cores may be not considered at evolutionary
sequence of massive star formation. The targeted lines of ATOMS have upper level energies of
C$_2$H$_5$CN and CH$_3$OCHO less than 139 K, hence the hottest
components are not observed.  The heating mechanisms as well as nitrogen and oxygen
differentiation in rotation temperature are needed to be verified by future
 observations of highly excited lines of various COMs. The future observations
 can also test whether gas phase chemical reactions play an important role in higher
temperatures, as suggested by chemical models.

\section*{Acknowledgements}

This work has been supported by the National Key R\&D Program of
China (No. 2017YFA0402701), by the National Natural Science
Foundation of China (grant Nos. 12033005, 12073061, 12122307, 12103045, 11973013, and  11873086), and by the international partnership program of Chinese
Academy of Sciences through grant No.114231KYSB20200009,  Shanghai Pujiang Program 20PJ1415500, and the science research grants from the China Manned 
Space Project with No. CMS-CSST-2021-B06. This research was carried
out in part at the Jet Propulsion Laboratory, which is operated by
the California Institute of Technology under contract with NASA. J.-E. Lee was supported by the National Research Foundation of Korea (NRF) grant 
funded by the Korea government (MSIT; grant No. 2021R1A2C1011718). C.W.L. is supported by the Basic Science Research Program through the National Research Foundation 
of Korea (NRF) funded by the Ministry of Education, Science and Technology (NRF-2019R1A2C1010851). This
work is sponsored in part by the Chinese Academy of Sciences (CAS), through a grant to the
CAS South America Center for Astronomy (CASSACA) in Santiago, Chile. G.G. and L.B. acknowledge support by the ANID BASAL project FB210003. This paper makes use
of the following ALMA data: ADS/JAO.ALMA\#2019.1.00685.S.
ALMA is a partnership of ESO (representing its member states),
NSF (USA), and NINS (Japan), together with NRC (Canada),
MOST and ASIAA (Taiwan), and KASI (Republic of Korea), in cooperation with the Republic of Chile. The Joint ALMA
Observatory is operated by ESO, AUI/NRAO, and NAOJ.

\section*{Data Availability}

The data underlying this article are available in the ALMA archive.



\bibliographystyle{mnras}
\bibliography{example} 

%


\begin{landscape}
\begin{table}

   \centering
  \caption{Physical parameters of the hot cores}
    \begin{tabular}{lccccccccccccccccccr} 
       \hline\hline
    Source &RA   & DEC&$\theta_{\rm source}$& & C$_2$H$_5$CN &&& &CH$_3$OCHO&&&~~~~~~~~~~~~~~~~~~~~~~CH$_3$OH&\\

        \cline{5-7}\cline{9-11}\cline{13-14}
        & h~m~s   & ${\circ}$~$\prime$~$\prime\prime$ &$\prime\prime$&T$_{\rm rot}$ (K) &N (cm$^{-2}$)&f&&T$_{\rm rot}$ (K)&N (cm$^{-2}$)&f&&T$_{\rm rot}$ (K)&N (cm$^{-2}$)\\

     \hline
I08303--4303$^S$ & 08:32:08.68 & --43:13:45.78 &2.2& 104$\pm$16 & (7.1$\pm$1.0)$\times$10$^{14}$ & (2.8$\pm$0.4)$\times$10$^{-3}$ &  & 101$\pm$9 & (1.5$\pm$0.2)$\times$10$^{16}$ & (6.1$\pm$0.1)$\times$10$^{-2}$ &  & 101 & 2.5$\times$10$^{17}$ \\
I08470--4243$^S$ & 08:48:47.79 & --42:54:27.90 &1.8& 125$\pm$18 & (1.3$\pm$0.2)$\times$10$^{15}$ & (1.5$\pm$0.3)$\times$10$^{-3}$ &  & 200$\pm$30 & (2.1$\pm$0.2)$\times$10$^{17}$ & (2.5$\pm$0.1)$\times$10$^{-1}$ &  & 202$\pm$32 & (8.5$\pm$1.1)$\times$10$^{17}$ \\
I09018--4816 & 09:03:33.46 & --48:28:01.69 & 3.4& &  &  &  &  &  &  &  & 160$\pm$8 & (1.9$\pm$0.1)$\times$10$^{17}$ \\
I11298--6155$^E$ & 11:32:05.59 & --62:12:25.62 &3.1&  &  &  &  & 124$\pm$6 & (1.3$\pm$0.3)$\times$10$^{16}$ & (8.3$\pm$5.3)$\times$10$^{-2}$ &  & 124 & 1.6$\times$10$^{17}$ \\
I12326--6245$^{E,S}$ & 12:35:35.09 & --63:02:31.91 &1.3& 130$\pm$32 & (3.1$\pm$0.2)$\times$10$^{15}$ & (8.4$\pm$0.5)$\times$10$^{-3}$ &  & 137$\pm$33 & (2.3$\pm$0.2)$\times$10$^{16}$ & (6.2$\pm$0.6)$\times$10$^{-2}$ &  & 137 & 3.7$\times$10$^{17}$ \\
I13079--6218 & 13:11:13.75 & --62:34:41.55 &2.4& 100$\pm$13 & (2.1$\pm$0.2)$\times$10$^{16}$ & (1.2$\pm$0.2)$\times$10$^{-2}$ &  & 110$\pm$9 & (1.1$\pm$0.3)$\times$10$^{17}$ & (8.5$\pm$2)$\times$10$^{-2}$ &  & 110 & 1.3$\times$10$^{18}$ \\
I13134--6242 & 13:16:43.2 & --62:58:32.3 &2.2& 139$\pm$4 & (8.2$\pm$0.5)$\times$10$^{15}$ & (7.5$\pm$0.5)$\times$10$^{-3}$ &  & 160$\pm$5 & (2.1$\pm$0.1)$\times$10$^{17}$ & (1.9$\pm$0.1)$\times$10$^{-1}$ &  & 160 & 1.1$\times$10$^{18}$ \\
I13140--6226$^S$ & 13:17:15.49 & --62:42:24.42 &5.5& 139$\pm$4 & (5.9$\pm$0.1)$\times$10$^{14}$ & (7.4$\pm$0.2)$\times$10$^{-3}$ &  & 120$\pm$17 & (6.7$\pm$0.1)$\times$10$^{15}$ & (8.4$\pm$0.1)$\times$10$^{-2}$ &  & 120 & 8.1$\times$10$^{16}$ \\
I13471--6120$^{E,S}$ & 13:50:41.81 &--61:35:10.67 & 1& 134$\pm$8 & (2.2$\pm$0.5)$\times$10$^{15}$ & (1.7$\pm$0.4)$\times$10$^{-3}$ &  & 126$\pm$10 & (1.1$\pm$0.2)$\times$10$^{17}$ & (8.5$\pm$1.3)$\times$10$^{-2}$ &  & 126 & 1.3$\times$10$^{18}$ \\
I13484--6100 & 13:51:58.31 & --61:15:41.5 &2.5& 131$\pm$18 & (1.6$\pm$0.6)$\times$10$^{15}$ & (1.5$\pm$0.5)$\times$10$^{-2}$ &  &  &  &  &  & 126 & 1.1$\times$10$^{17}$ \\
I14498--5856$^S$ & 14:53:42.68 & --59:08:52.89 &2.8& 144$\pm$13 & (8.6$\pm$0.1)$\times$10$^{14}$ & (1.9$\pm$0.1)$\times$10$^{-3}$ &  & 106$\pm$3 & (1.5$\pm$0.1)$\times$10$^{16}$ & (3.3$\pm$0.3)$\times$10$^{-2}$ &  & 135$\pm$8 & (4.5$\pm$0.4)$\times$10$^{17}$ \\
I15254--5621$^E$ & 15:29:19.39 & --56:31:22.34 &0.8& 148$\pm$3 & (4.2$\pm$0.1)$\times$10$^{15}$ & (2.3$\pm$0.2)$\times$10$^{-3}$ &  & 146$\pm$14 & (9.2$\pm$0.2)$\times$10$^{16}$ & (5.2$\pm$0.4)$\times$10$^{-2}$ &  & 165$\pm$13 & (1.8$\pm$0.1)$\times$10$^{18}$ \\
I15437--5343 & 15:47:32.73 & --53:52:38.8 &2.3& 128$\pm$20 & (1.3$\pm$0.1)$\times$10$^{15}$ & (1.7$\pm$0.1)$\times$10$^{-3}$ &  & 106$\pm$11 & (2.1$\pm$0.1)$\times$10$^{16}$ & (2.8$\pm$0.2)$\times$10$^{-2}$ &  & 170$\pm$28 & (7.5$\pm$0.4)$\times$10$^{17}$ \\
I15520--5234$^{E,S}$ & 15:55:48.47 & --52:43:06.75 &2.9& 118$\pm$13 & (1.6$\pm$0.1)$\times$10$^{15}$ & (3.6$\pm$0.3)$\times$10$^{-3}$ &  & 160$\pm$30 & (3.7$\pm$0.1)$\times$10$^{16}$ & (8.4$\pm$0.4)$\times$10$^{-2}$ &  & 168$\pm$28 & (4.4$\pm$0.1)$\times$10$^{17}$ \\
I16060--5146$^{E,S}$ & 16:09:52.64 & --51:54:54.49 &1.9& 110$\pm$5 & (3$\pm$0.2)$\times$10$^{15}$ & (2$\pm$0.1)$\times$10$^{-2}$ &  & 110$\pm$22 & (2.5$\pm$0.2)$\times$10$^{16}$ & (1.7$\pm$0.2)$\times$10$^{-1}$ &  & 110 & 1.5$\times$10$^{17}$ \\
I16065--5158$^{E,S}$ & 16:10:19.99 & --52:06:07.25 &3.1& 125$\pm$15 & (2$\pm$0.1)$\times$10$^{16}$ & (1.5$\pm$0.2)$\times$10$^{-2}$ &  & 150$\pm$43 & (8.9$\pm$0.1)$\times$10$^{16}$ & (6.8$\pm$0.1)$\times$10$^{-2}$ &  & 174$\pm$14 & (1.3$\pm$0.1)$\times$10$^{18}$ \\
I16071--5142$^{E}$ & 16:10:59.59 & --51:50:23.37 &4.2& 161$\pm$6 & (1.1$\pm$0.1)$\times$10$^{16}$ & (1.1$\pm$0.1)$\times$10$^{-2}$ &  & 150$\pm$20 & (1.1$\pm$0.1)$\times$10$^{17}$ & (1.1$\pm$0.1)$\times$10$^{-1}$ &  & 190$\pm$16 & (9.8$\pm$0.2)$\times$10$^{17}$ \\
I16076--5134$^{E,S}$ & 16:11:26.59 & --51:41:57.84 &4.3& 110$\pm$18 & (3.4$\pm$0.2)$\times$10$^{15}$ & (2$\pm$0.2)$\times$10$^{-2}$ &  & 121$\pm$32 & (2.5$\pm$0.3)$\times$10$^{16}$ & (1.5$\pm$0.2)$\times$10$^{-1}$ &  & 130$\pm$14 & (1.7$\pm$0.1)$\times$10$^{17}$ \\
I16164--5046$^{E}$ & 16:20:11.08 & --50:53:14.75 &2.8& 133$\pm$20 & (7.4$\pm$0.1)$\times$10$^{15}$ & (8.1$\pm$0.2)$\times$10$^{-3}$ &  & 157$\pm$19 & (1.2$\pm$0.2)$\times$10$^{17}$ & (1.3$\pm$0.2)$\times$10$^{-1}$ &  & 187$\pm$74 & (9.1$\pm$0.1)$\times$10$^{17}$ \\
I16172--5028$^{E,S}$ & 16:21:02.97 & --50:35:12.6 &1.8& 110$\pm$32 & (1.9$\pm$0.3)$\times$10$^{15}$ & (3.3$\pm$0.5)$\times$10$^{-3}$ &  &  &  &  &  & 204$\pm$18 & (5.7$\pm$0.1)$\times$10$^{17}$ \\
I16272--4837c1 & 16:30:58.77 & --48:43:53.57 &2.2& 220$\pm$8 & (4.2$\pm$0.1)$\times$10$^{16}$ & (1.3$\pm$0.1)$\times$10$^{-2}$ &  & 230$\pm$37 & (8.9$\pm$0.2)$\times$10$^{17}$ & (2.8$\pm$0.1)$\times$10$^{-1}$ &  & 240$\pm$21 & (3.2$\pm$0.1)$\times$10$^{18}$ \\
I16272--4837c2 & 16:30:58.68 & --48:43:51.32 &2& 120$\pm$3 & (3.1$\pm$0.1)$\times$10$^{15}$ & (6.9$\pm$0.1)$\times$10$^{-3}$ &  & 150$\pm$9 & (3.9$\pm$0.1)$\times$10$^{16}$ & (8.7$\pm$0.1)$\times$10$^{-2}$ &  & 180$\pm$6 & (4.5$\pm$0.1)$\times$10$^{17}$ \\
I16272--4837c3 & 16:30:57.29 & --48:43:39.87 &2.1& 103$\pm$11 & (2.1$\pm$0.1)$\times$10$^{15}$ & (6.9$\pm$0.1)$\times$10$^{-3}$ &  & 123$\pm$12 & (3.8$\pm$0.1)$\times$10$^{15}$ & (8.7$\pm$0.1)$\times$10$^{-2}$ &  & 123$\pm$6 & (2.1$\pm$0.1)$\times$10$^{17}$ \\
I16318--4724$^E$ & 16:35:33.96 & --47:31:11.59 &2.2& 130$\pm$5 & (2.7$\pm$0.1)$\times$10$^{16}$ & (9.6$\pm$0.8)$\times$10$^{-3}$ &  & 148$\pm$15 & (1.9$\pm$0.5)$\times$10$^{17}$ & (6.8$\pm$0.2)$\times$10$^{-2}$ &  & 192$\pm$18 & (2.8$\pm$0.2)$\times$10$^{18}$ \\
I16344--4658$^{E}$ & 16:38:09.49 & --47:04:59.73 &1.7& 150$\pm$4 & (5.5$\pm$0.1)$\times$10$^{15}$ & (1.1$\pm$0.1)$\times$10$^{-2}$ &  & 160$\pm$15 & (5.2$\pm$0.1)$\times$10$^{16}$ & (1.1$\pm$0.1)$\times$10$^{-1}$ &  & 160$\pm$12 & (4.8$\pm$0.1)$\times$10$^{17}$ \\
I16348--4654$^{E,S}$ & 16:38:29.65 & --47:00:35.67 &0.8& 270$\pm$13 & (2$\pm$0.1)$\times$10$^{17}$ & (1.3$\pm$0.1)$\times$10$^{-2}$ &  & 213$\pm$12 & (2$\pm$0.1)$\times$10$^{18}$ & (1.3$\pm$0.1)$\times$10$^{-1}$ &  & 220$\pm$12 & (1.5$\pm$0.1)$\times$10$^{19}$ \\
I16351--4722$^{E,S}$ & 16:38:50.50 & --47:28:00.68 &1.5& 121$\pm$10 & (5.7$\pm$0.1)$\times$10$^{15}$ & (7.6$\pm$0.1)$\times$10$^{-3}$ &  & 150$\pm$31 & (6$\pm$0.3)$\times$10$^{16}$ & (8$\pm$0.4)$\times$10$^{-2}$ &  & 150 & 7.5$\times$10$^{17}$ \\
I16458--4512$^{S}$ & 16:49:30.04 & --45:17:44.58 &1.1&  &  &  &  & 120$\pm$4 & (2.1$\pm$0.1)$\times$10$^{16}$ & (1.4$\pm$0.1)$\times$10$^{-1}$ &  & 120 & 1.5$\times$10$^{17}$ \\
I16484--4603 & 16:52:04.66 & --46:08:33.85 &1.3& 120$\pm$12 & (1.7$\pm$0.3)$\times$10$^{15}$ & (1$\pm$0.3)$\times$10$^{-3}$ &  & 127$\pm$5 & (6.1$\pm$0.3)$\times$10$^{16}$ & (3.6$\pm$0.3)$\times$10$^{-2}$ &  & 151$\pm$4 & (1.7$\pm$0.1)$\times$10$^{18}$ \\
I16547--4247$^{S}$ & 16:58:17.18 &--42:52:07.57 &2.5&  114$\pm$8 & (3.1$\pm$0.1)$\times$10$^{15}$ & (7.8$\pm$0.3)$\times$10$^{-3}$ &  & 150$\pm$20 & (2.2$\pm$0.2)$\times$10$^{16}$ & (5.5$\pm$0.4)$\times$10$^{-2}$ &  & 170$\pm$5 & (4$\pm$0.1)$\times$10$^{17}$ \\
I17008--4040 & 17:04:22.91 & --40:44:22.91 &1.7& 161$\pm$7 & (1.6$\pm$0.1)$\times$10$^{16}$ & (5$\pm$0.1)$\times$10$^{-3}$ &  & 150$\pm$14 & (2.8$\pm$0.1)$\times$10$^{17}$ & (8.8$\pm$0.1)$\times$10$^{-2}$ &  & 180$\pm$12 & (3.2$\pm$0.1)$\times$10$^{18}$ \\
I17016--4124c1$^{E}$ & 17:05:10.97 & --41:29:06.95 &0.9& 153$\pm$6 & (3.6$\pm$0.1)$\times$10$^{16}$ & (7.8$\pm$0.1)$\times$10$^{-3}$ &  & 143$\pm$14 & (4.9$\pm$0.1)$\times$10$^{17}$ & (1.1$\pm$0.1)$\times$10$^{-1}$ &  & 193$\pm$14 & (4.6$\pm$0.1)$\times$10$^{18}$ \\
I17016--4124c2 & 17:05:11.20 & --41:29:07.05 &1.8& 101$\pm$12 & (5$\pm$0.2)$\times$10$^{15}$ & (1.7$\pm$0.1)$\times$10$^{-3}$ &  & 165$\pm$32 & (2$\pm$0.3)$\times$10$^{17}$ & (6.6$\pm$0.1)$\times$10$^{-2}$ &  & 180$\pm$11 & (3.1$\pm$0.1)$\times$10$^{18}$ \\
I17158--3901c1 & 17:19:20.43 & --39:03:51.58 &2.9& 152$\pm$9 & (1$\pm$0.1)$\times$10$^{15}$ & (3.2$\pm$0.4)$\times$10$^{-3}$ &  & 151$\pm$12 & (2.2$\pm$0.2)$\times$10$^{16}$ & (7.1$\pm$0.5)$\times$10$^{-2}$ &  & 150$\pm$2 & (3.1$\pm$0.2)$\times$10$^{17}$ \\
I17158--3901c2$^{S}$ & 17:19:20.47 & --39:03:49.20 & 2.8&152$\pm$3 & (1.5$\pm$0.1)$\times$10$^{15}$ & (1.4$\pm$0.1)$\times$10$^{-3}$ &  &  &  &  &  & 150 & 1.1$\times$10$^{17}$ \\
I17175--3544$^{E}$ & 17:20:53.42 & --35:46:57.72 &3.4& 174$\pm$9 & (5.4$\pm$0.2)$\times$10$^{16}$ & (8.6$\pm$0.4)$\times$10$^{-3}$ &  & 183$\pm$8 & (1.4$\pm$0.1)$\times$10$^{18}$ & (2.2$\pm$0.1)$\times$10$^{-1}$ &  & 182$\pm$6 & (6.3$\pm$0.1)$\times$10$^{18}$ \\
I17220--3609$^{E,S}$ & 17:25:25.22 & --36:12:45.34 &1.6& 183$\pm$14 & (1.6$\pm$0.1)$\times$10$^{16}$ & (5.7$\pm$0.3)$\times$10$^{-3}$ &  & 173$\pm$22 & (3.8$\pm$0.1)$\times$10$^{17}$ & (1.4$\pm$0.1)$\times$10$^{-1}$ &  & 176$\pm$6 & (2.8$\pm$0.1)$\times$10$^{18}$ \\
I17233--3606$^{E,S}$ & 17:26:42.46 & --36:09:17.85 &3.9& 160$\pm$4 & (3.9$\pm$0.1)$\times$10$^{16}$ & (4.1$\pm$0.2)$\times$10$^{-2}$ &  & 100$\pm$14 & (8.9$\pm$0.5)$\times$10$^{16}$ & (9.3$\pm$0.6)$\times$10$^{-2}$ &  & 165$\pm$8 & (9.6$\pm$0.4)$\times$10$^{17}$ \\
I17441--2822$^{E}$ & 17:47:20.17 & --28:23:04.74 &1.3& 160$\pm$18 & (2.5$\pm$0.1)$\times$10$^{16}$ & (1.8$\pm$0.1)$\times$10$^{-2}$ &  & 144$\pm$7 & (9.6$\pm$0.2)$\times$10$^{16}$ & (6.9$\pm$0.2)$\times$10$^{-2}$ &  & 207$\pm$8 & (1.4$\pm$0.1)$\times$10$^{18}$ \\
I18032--2032c1 & 18:06:14.92 & --20:31:43.22 &0.9&  &  &  &  & 204$\pm$17 & (1.2$\pm$0.1)$\times$10$^{17}$ & (1.1$\pm$0.1)$\times$10$^{-1}$ &  & 204 & 1.1$\times$10$^{18}$ \\
I18032--2032c2 & 18:06:14.88 & --20:31:39.59 &2.4& 127$\pm$5 & (2$\pm$0.1)$\times$10$^{16}$ & (4$\pm$0.3)$\times$10$^{-2}$ &  & 130$\pm$12 & (7$\pm$0.2)$\times$10$^{16}$ & (1.5$\pm$0.4)$\times$10$^{-1}$ &  & 146$\pm$5 & (4.7$\pm$0.2)$\times$10$^{17}$ \\
I18032--2032c3 & 18:06:14.80 & --20:31:37.26 &1.1& 110$\pm$3 & (2.4$\pm$0.1)$\times$10$^{15}$ & (4$\pm$0.3)$\times$10$^{-2}$ &  &  &  &  &  & 110 & 1.8$\times$10$^{17}$ \\
I18032--2032c4 & 18:06:14.66 & --20:31:31.57 &2.1& 130$\pm$16 & (8$\pm$0.1)$\times$10$^{15}$ & (3.8$\pm$0.1)$\times$10$^{-3}$ &  & 140$\pm$13 & (1.7$\pm$0.2)$\times$10$^{17}$ & (8.1$\pm$0.3)$\times$10$^{-2}$ &  & 157$\pm$8 & (2.1$\pm$0.1)$\times$10$^{18}$ \\
I18056--1952$^{S}$ & 18:08:38.23 & --19:51:50.31 &1.1& 285$\pm$9 & (1.9$\pm$0.1)$\times$10$^{18}$ & (6.3$\pm$0.1)$\times$10$^{-2}$ &  & 133$\pm$5 & (2.1$\pm$0.1)$\times$10$^{18}$ & (7$\pm$0.3)$\times$10$^{-2}$ &  & 226$\pm$9 & (3$\pm$0.1)$\times$10$^{19}$ \\
I18089--1732 & 18:11:51.45 & --17:31:28.96 &1.7& 277$\pm$9 & (4.1$\pm$0.1)$\times$10$^{16}$ & (1.6$\pm$0.1)$\times$10$^{-2}$ &  & 128$\pm$14 & (2.3$\pm$0.5)$\times$10$^{17}$ & (8.8$\pm$0.2)$\times$10$^{-2}$ &  & 200$\pm$7 & (2.6$\pm$0.1)$\times$10$^{18}$ \\
I18117--1753$^{S}$ & 18:14:39.51 & --17:52:00.08 &1.6& 180$\pm$28 & (9.5$\pm$0.1)$\times$10$^{15}$ & (5.9$\pm$0.2)$\times$10$^{-3}$ &  & 160$\pm$25 & (1.3$\pm$0.2)$\times$10$^{17}$ & (8.1$\pm$1.1)$\times$10$^{-2}$ &  & 163$\pm$5 & (1.6$\pm$0.1)$\times$10$^{18}$ \\
I18159--1648c1 & 18:18:54.66 & --16:47:50.28 &1.7& 157$\pm$26 & (7.6$\pm$0.2)$\times$10$^{15}$ & (7.6$\pm$0.1)$\times$10$^{-3}$ &  & 161$\pm$18 & (2.4$\pm$0.1)$\times$10$^{17}$ & (2.4$\pm$0.4)$\times$10$^{-1}$ &  & 228$\pm$12 & (1.1$\pm$0.2)$\times$10$^{18}$ \\
I18159--1648c2 & 18:18:54.34 & --16:47:49.97 &1.9& 104$\pm$20 & (1.1$\pm$0.2)$\times$10$^{15}$ & (1.8$\pm$0.3)$\times$10$^{-3}$ &  & 110$\pm$22 & (2.9$\pm$0.3)$\times$10$^{16}$ & (4.8$\pm$0.4)$\times$10$^{-2}$ &  & 165$\pm$11 & (6$\pm$0.1)$\times$10$^{17}$ \\

     \hline

    \end{tabular}

\end{table}
\end{landscape}

\begin{landscape}
\begin{table}

   \centering
  \contcaption{}
    \begin{tabular}{lccccccccccccccccccr} 
       \hline\hline
    Source &RA   & DEC&$\theta_{\rm source}$& & C$_2$H$_5$CN &&& &CH$_3$OCHO&&&~~~~~~~~~~~~CH$_3$OH\\

        \cline{5-7}\cline{9-11}\cline{13-14}
        & h~m~s   & ${\circ}$~$\prime$~$\prime\prime$&$\prime\prime$&T$_{\rm rot}$ (K)&N (cm$^{-2}$)&f&&T$_{\rm rot}$ (K)&N (cm$^{-2}$)&f&&T$_{\rm rot}$ (K)&N (cm$^{-2}$)\\

     \hline
I18182--1433$^{S}$ & 18:21:09.05 & --14:31:47.88 &2.1&120$\pm$10 & (1.1$\pm$0.1)$\times$10$^{15}$ & (5.2$\pm$0.5)$\times$10$^{-3}$ &  & 122$\pm$4 & (2.4$\pm$0.2)$\times$10$^{16}$ & (1.1$\pm$0.1)$\times$10$^{-1}$ &  & 120 & 2.1$\times$10$^{17}$ \\
I18236--1205$^{S}$ & 18:26:25.79 & --12:03:53.08 &1.4& 110$\pm$8 & (1.8$\pm$0.1)$\times$10$^{15}$ & (1.1$\pm$0.1)$\times$10$^{-2}$ &  &  &  &  &  & 138$\pm$20 & (1.7$\pm$0.1)$\times$10$^{17}$ \\
I18290--0924$^{S}$ & 18:31:44.13 & --09:22:12.25 &1.4& 120$\pm$4 & (3.3$\pm$0.9)$\times$10$^{15}$ & (5.8$\pm$1.6)$\times$10$^{-3}$ &  &  &  &  &  & 120 & 5.7$\times$10$^{17}$ \\
I18316--0602 & 18:34:20.91 & --05:59:42.0 & 2.9&186$\pm$35 & (2.8$\pm$0.1)$\times$10$^{15}$ & (3.2$\pm$0.1)$\times$10$^{-3}$ &  & 108$\pm$204 & (2.1$\pm$0.15)$\times$10$^{16}$ & (2.4$\pm$0.2)$\times$10$^{-2}$ &  & 167$\pm$7 & (8.7$\pm$0.3)$\times$10$^{17}$ \\
I18411--0338$^{S}$ & 18:43:46.23 & --03:35:29.77 &1& 134$\pm$4 & (9.6$\pm$0.5)$\times$10$^{15}$ & (1.1$\pm$0.1)$\times$10$^{-2}$ &  & 160$\pm$15 & (7$\pm$0.1)$\times$10$^{16}$ & (7.7$\pm$0.5)$\times$10$^{-2}$ &  & 160$\pm$14 & (9.1$\pm$0.6)$\times$10$^{17}$ \\
I18469--0132$^{E,S}$ & 18:49:33.05 & --01:29:03.34 &0.9& 124$\pm$8 & (5.1$\pm$0.1)$\times$10$^{15}$ & (3.6$\pm$0.2)$\times$10$^{-2}$ &  & 136$\pm$15 & (9.4$\pm$0.2)$\times$10$^{16}$ & (6.7$\pm$0.4)$\times$10$^{-2}$ &  & 145$\pm$9 & (1.4$\pm$0.1)$\times$10$^{18}$ \\
I18507+0110$^{E,S}$ & 18:53:18.56 & +01:14:58.23 &1.3& 172$\pm$6 & (9$\pm$0.2)$\times$10$^{16}$ & (8.2$\pm$0.2)$\times$10$^{-3}$ &  & 200$\pm$11 & (1.6$\pm$0.2)$\times$10$^{18}$ & (1.5$\pm$0.2)$\times$10$^{-1}$ &  & 240$\pm$16 & (1.1$\pm$0.1)$\times$10$^{19}$ \\
I18507+0121 & 18:53:18.01 & +01:25:25.56 &1.7& 160$\pm$4 & (3.7$\pm$0.1)$\times$10$^{16}$ & (8.8$\pm$0.3)$\times$10$^{-3}$ &  & 200$\pm$12 & (6.1$\pm$0.1)$\times$10$^{17}$ & (1.5$\pm$0.1)$\times$10$^{-1}$ &  & 220$\pm$8 & (4.2$\pm$0.1)$\times$10$^{18}$ \\
I18517+0437$^{S}$ & 18:54:14.24 & +04:41:40.65 &2.3& 190$\pm$30 & (8$\pm$1)$\times$10$^{14}$ & (8.4$\pm$1.1)$\times$10$^{-4}$ &  & 140$\pm$23 & (2.8$\pm$0.3)$\times$10$^{16}$ & (2.9$\pm$0.4)$\times$10$^{-2}$ &  & 164$\pm$8 & (9.5$\pm$0.3)$\times$10$^{17}$ \\
I19078+0901c1$^{E,S}$ & 19:10:13.16 & +09:06:12.49 &1.7& 140$\pm$22 & (2.5$\pm$0.2)$\times$10$^{15}$ & (1.5$\pm$0.1)$\times$10$^{-2}$ &  &  &  &  &  & 140 & 1.7$\times$10$^{17}$ \\
I19078+0901c2 & 19:10:14.13 & +09:06:24.67 &1.8& 120$\pm$12 & (3.2$\pm$0.2)$\times$10$^{15}$ & (1.5$\pm$0.1)$\times$10$^{-2}$ &  & 115$\pm$6 & (4.6$\pm$0.3)$\times$10$^{16}$ &  &  & 170$\pm$8 & (8.7$\pm$0.3)$\times$10$^{17}$ \\
I19095+0930$^{E,S}$ & 19:11:53.99 & +09:35:50.27 &0.5& 160$\pm$11 & (7.5$\pm$0.2)$\times$10$^{15}$ & (1.8$\pm$0.1)$\times$10$^{-3}$ &  & 121$\pm$3 & (1.3$\pm$0.1)$\times$10$^{17}$ & (3.1$\pm$0.1)$\times$10$^{-2}$ &  & 156$\pm$15 & (4.2$\pm$0.1)$\times$10$^{18}$ \\

     \hline
    \end{tabular}
    Notes: The superscript E in column 1 indicates that the hot core is externally heated, while S denotes that there is spatial separation between nitrogen- and oxygen-bearing molecules. $\theta_{\rm source}$  is the source size derived from the deconvolved continuum size on assuming $\pi\theta^{2}$=$\pi$ab, where a and b are major and minor axes of the continuum source from Gaussian fitting. f denotes the relative abundance of the molecular species (C$_2$H$_5$CN or CH$_3$OCHO) with respect to CH$_3$OH.

\end{table}
\end{landscape}
\section*{Author affiliations}
$^{1}$Department of Astronomy,  Yunnan University, Kunming 650091, People's Republic of China\\
$^{2}$Shanghai Astronomical Observatory, Chinese Academy of Sciences, 80 Nandan Road, Shanghai 200030, People's Republic of China\\
$^{3}$Jet Propulsion Laboratory, California Institute of Technology, 4800 Oak Grove Drive, Pasadena, CA 91109, USA\\
$^{4}$National Astronomical Observatories, Chinese Academy of Sciences, Beijing 100101, People's Republic of China\\
$^{5}$University of Chinese Academy of Sciences, Beijing 100049, People's Republic of China\\
$^{6}$NAOC-UKZN Computational Astrophysics Centre, University of KwaZulu-Natal,Durban 4000, South Africa\\
$^{7}$Center for Astrophysics, Harvard \& Smithsonian, 60 Garden Street,Cambridge, MA 02138, USA\\
$^{8}$Department of Astronomy, Peking University, 100871 Beijing, People's Republic of China\\
$^{9}$Departamento de Astronom'ia, Universidad de Chile, Las Condes, 7591245 Santiago, Chile\\
$^{10}$Department of Physics, University of Helsinki, PO Box 64, FI-00014 Helsinki, Finland\\
$^{11}$Korea Astronomy and Space Science Institute, 776 Daedeokdaero, Yuseong-gu, Daejeon 34055, Republic of Korea\\
$^{12}$University of Science and Technology, Korea (UST), 217 Gajeong-ro,Yuseong-gu, Daejeon 34113, Republic of Korea\\
$^{13}$School of Physics and Astronomy, Sun Yat-sen University, 2 Daxue Road, Zhuhai, Guangdong 519082, People's Republic of China\\
$^{14}$Yunnan Observatories, Chinese Academy of Sciences, 396 Yangfangwang, Guandu District, Kunming, 650216, People's Republic of China\\
$^{15}$Chinese Academy of Sciences South America Center for Astronomy, National Astronomical Observatories, CAS, Beijing 100101, China \\
$^{16}$Institute of Astronomy and Astrophysics, Academia Sinica, 11F of Astronomy-Mathematics Building, AS/NTU No. 1, Section 4, Roosevelt Road., Taipei 10617, Taiwan\\
$^{17}$School of Space Research, Kyung Hee University, Yongin-Si, Gyeonggi-Do 17104, Republic of Korea\\
$^{18}$Kavli Institute for Astronomy and Astrophysics, Peking University, 5 Yiheyuan Road, Haidian District, Beijing 100871, People's Republic of China\\
$^{19}$Department of Physics, Anhui Normal University, Wuhu, Anhui 241002, People's Republic of China\\
$^{20}$Institute of Astrophysics, School of Physics and Electronical Science, Chuxiong Normal University, Chuxiong, 675000, People's Republic of China\\
$^{21}$Institute of Astronimy and Astrophysics, Anqing  Normal University, Anqing, 246133, People's Republic of China\\
$^{22}$College of Science, Yunnan Agricultural University, Kunming 650201, People's Republic of China\\
$^{24}$Physical Research Laboratory, Navrangpura, Ahmedabad-380 009, India \\
$^{25}$Indian Institute of Space Science and Technology, Thiruvananthapuram 695 547, India

\clearpage
\appendix
\section{Supplementary material}
\subsection{Spectra in SPW 7}

\begin{figure*}
\includegraphics[width=17cm,height=20cm]{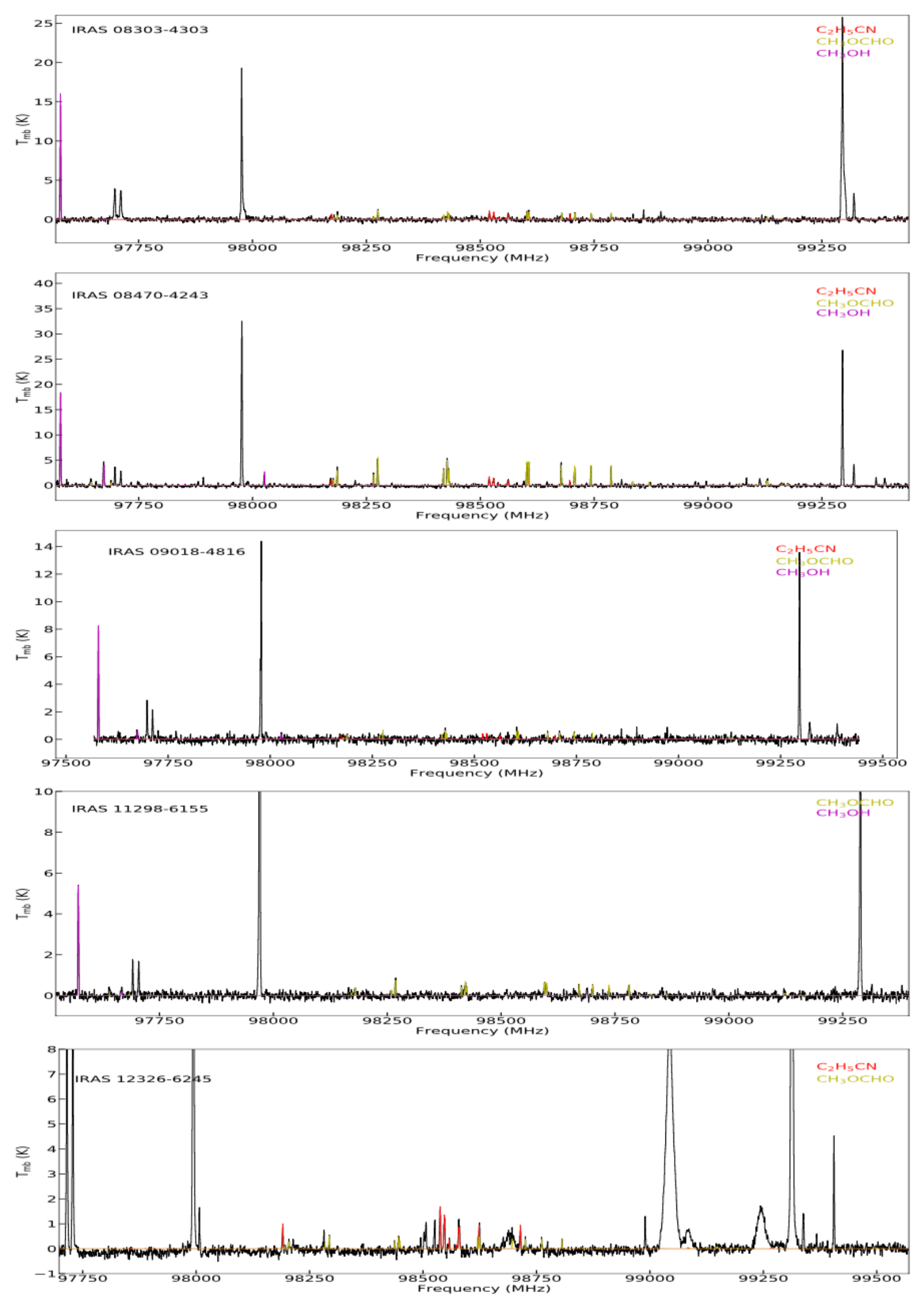}
\caption{Spectra in SPW 7  for the 60 hot cores. The observed
spectra are shown in black curves and the XCLASS modelled spectra
are coded in color.}
\end{figure*}

\begin{figure*}
\includegraphics[width=17cm,height=20cm]{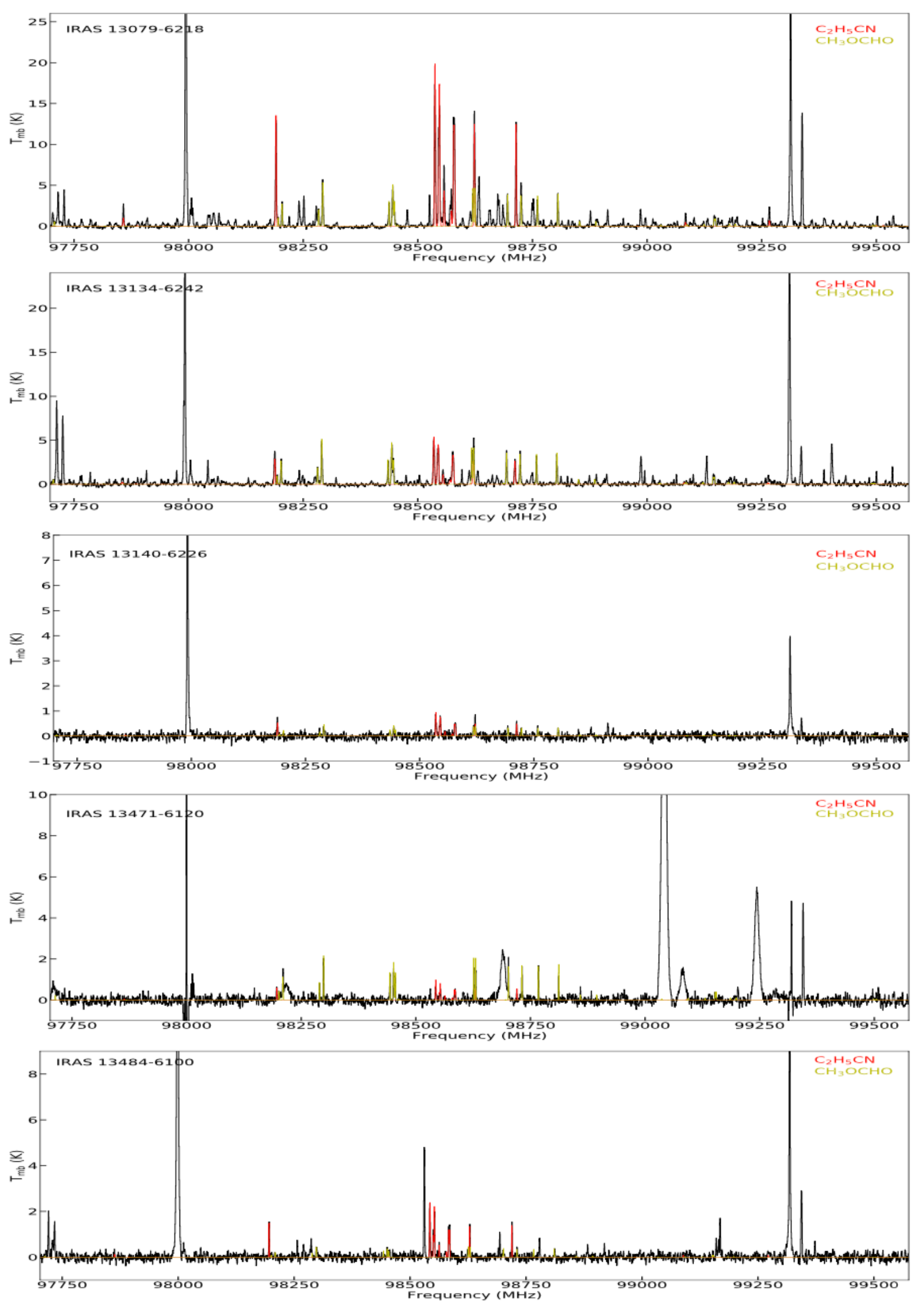}
\contcaption{}
\end{figure*}

\begin{figure*}
\includegraphics[width=17cm,height=20cm]{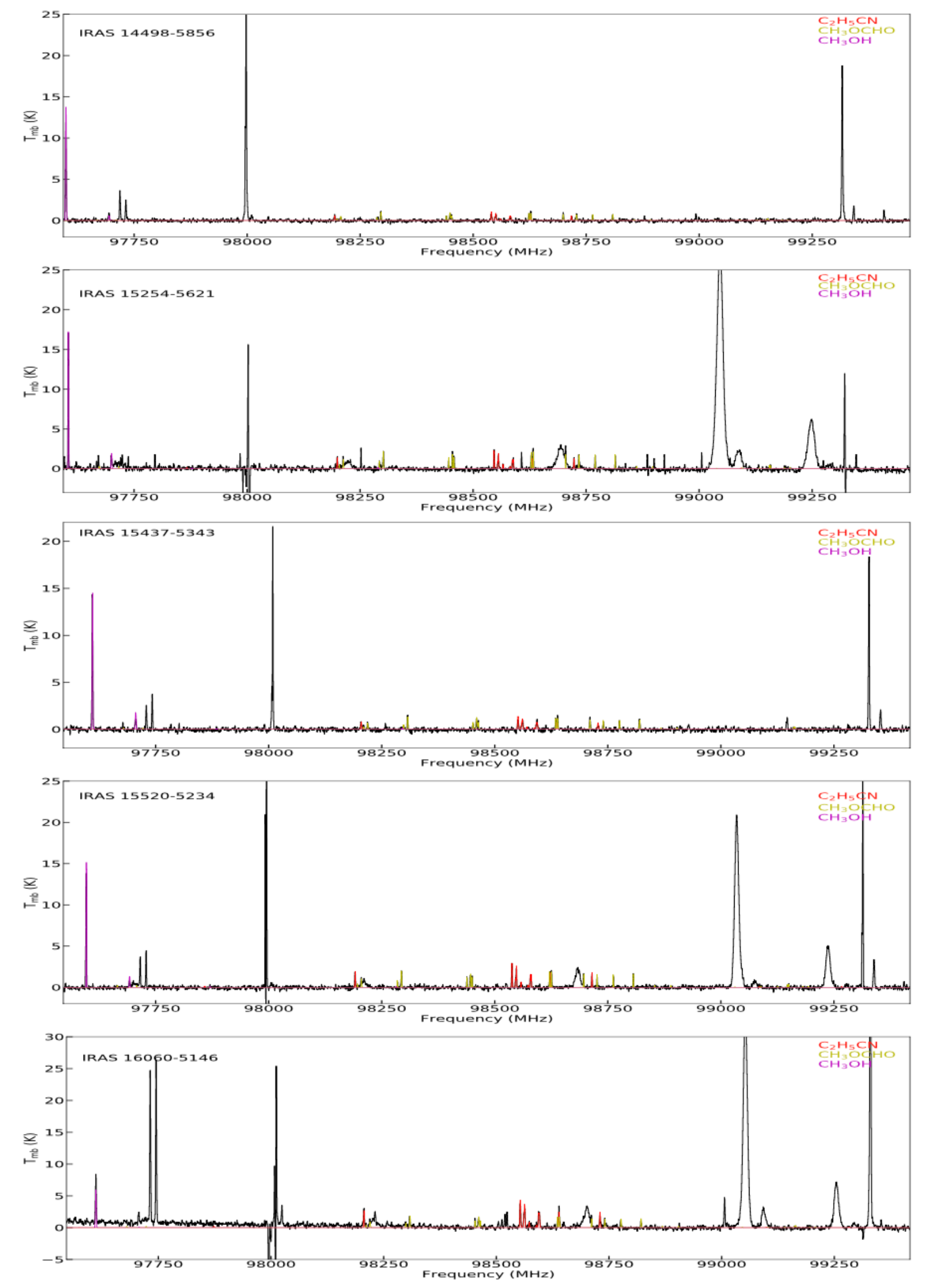}
\contcaption{}
\end{figure*}

\begin{figure*}
\includegraphics[width=17cm,height=20cm]{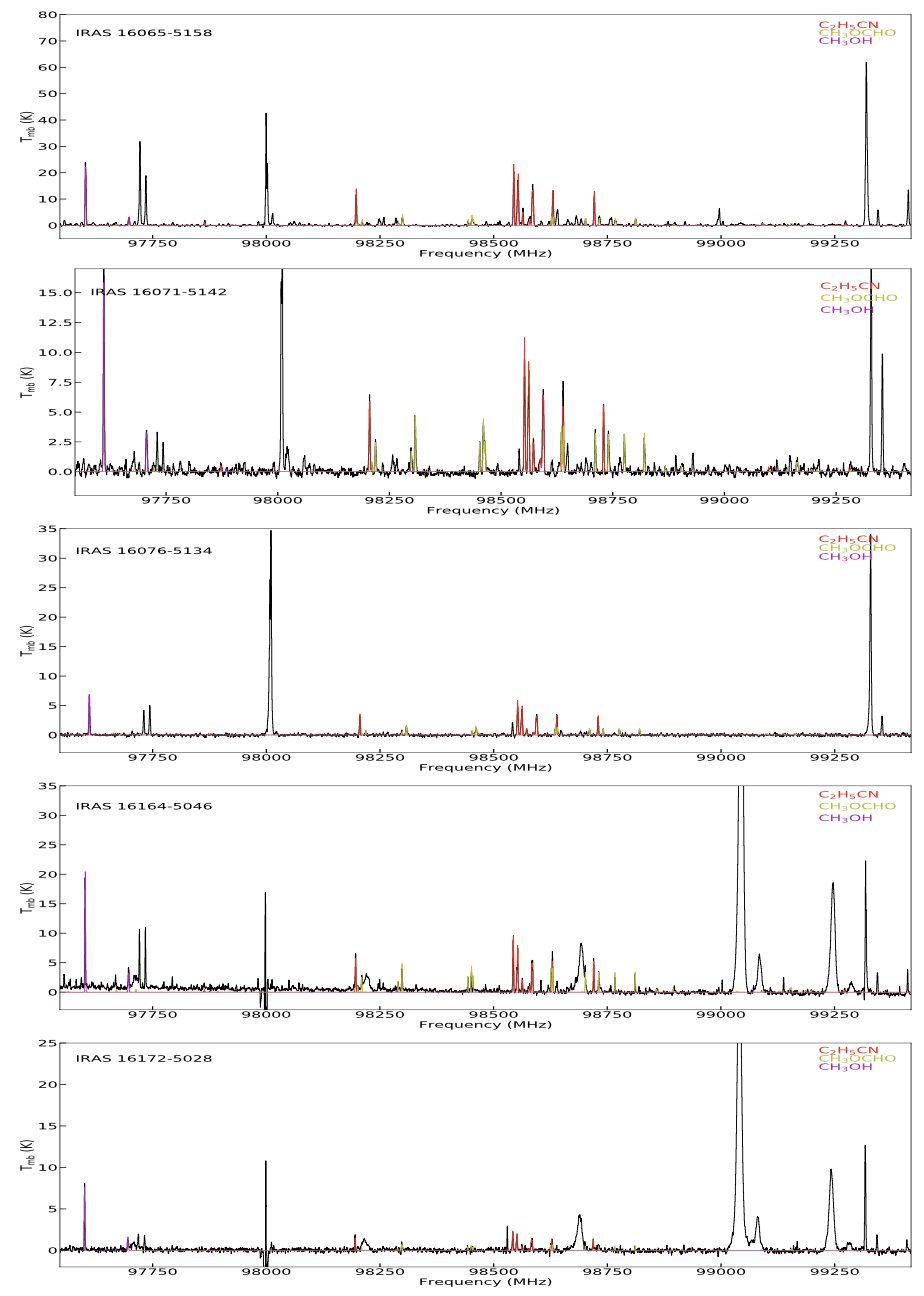}
\contcaption{}
\end{figure*}

\begin{figure*}
\includegraphics[width=17cm,height=20cm]{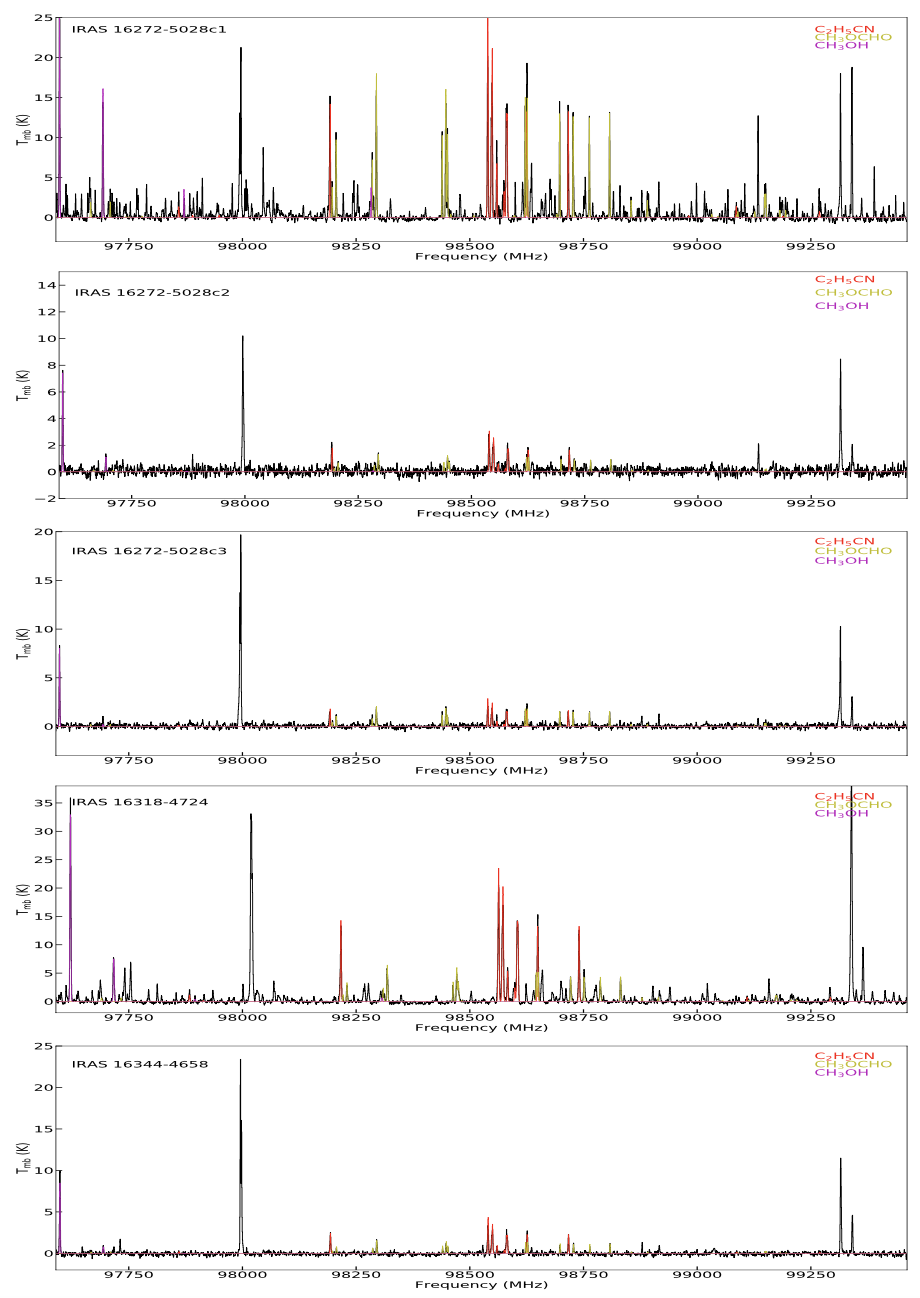}
\contcaption{}
\end{figure*}

\begin{figure*}
\includegraphics[width=17cm,height=20cm]{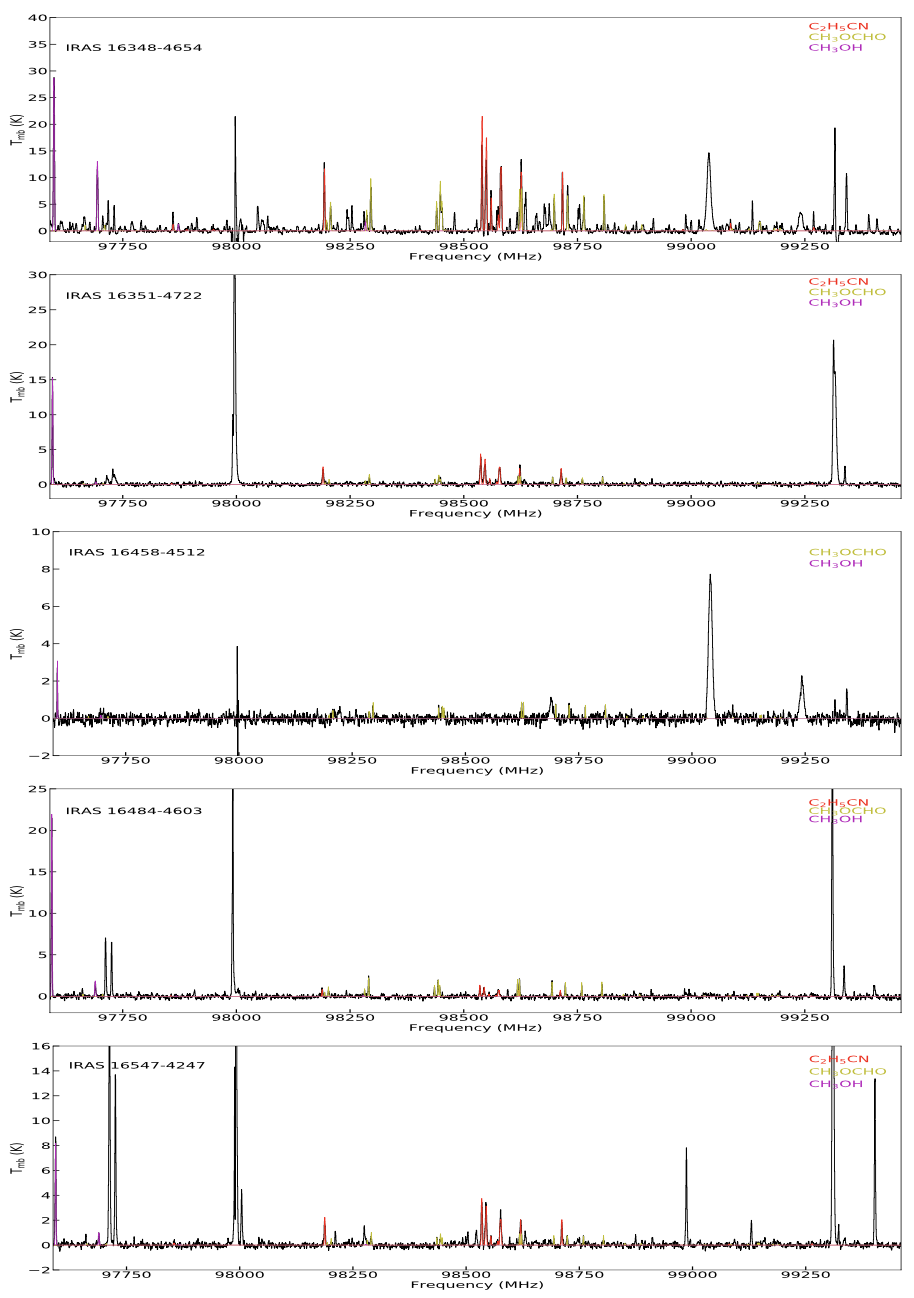}
\contcaption{}
\end{figure*}

\begin{figure*}
\includegraphics[width=17cm,height=20cm]{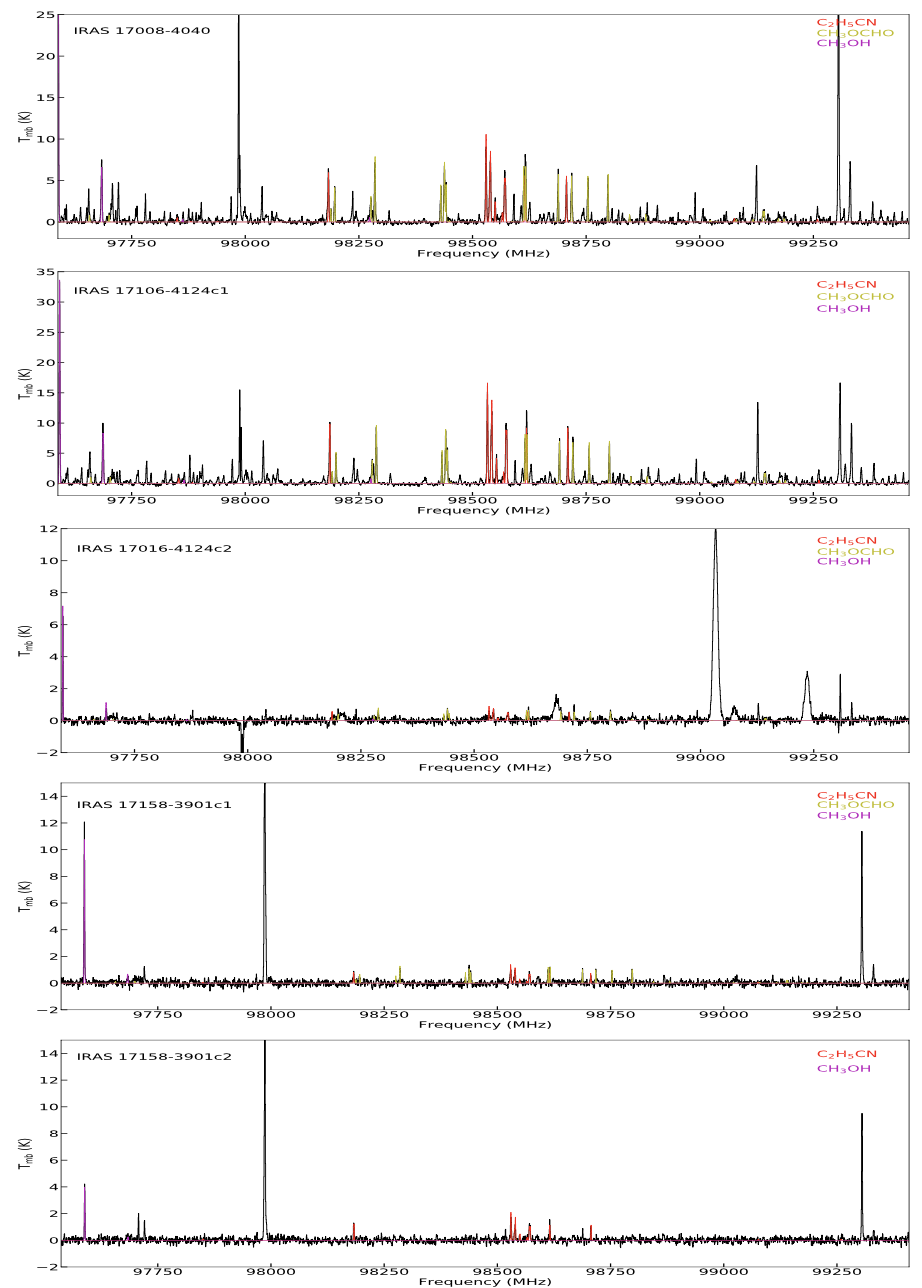}
\contcaption{}
\end{figure*}

\begin{figure*}
\includegraphics[width=17cm,height=20cm]{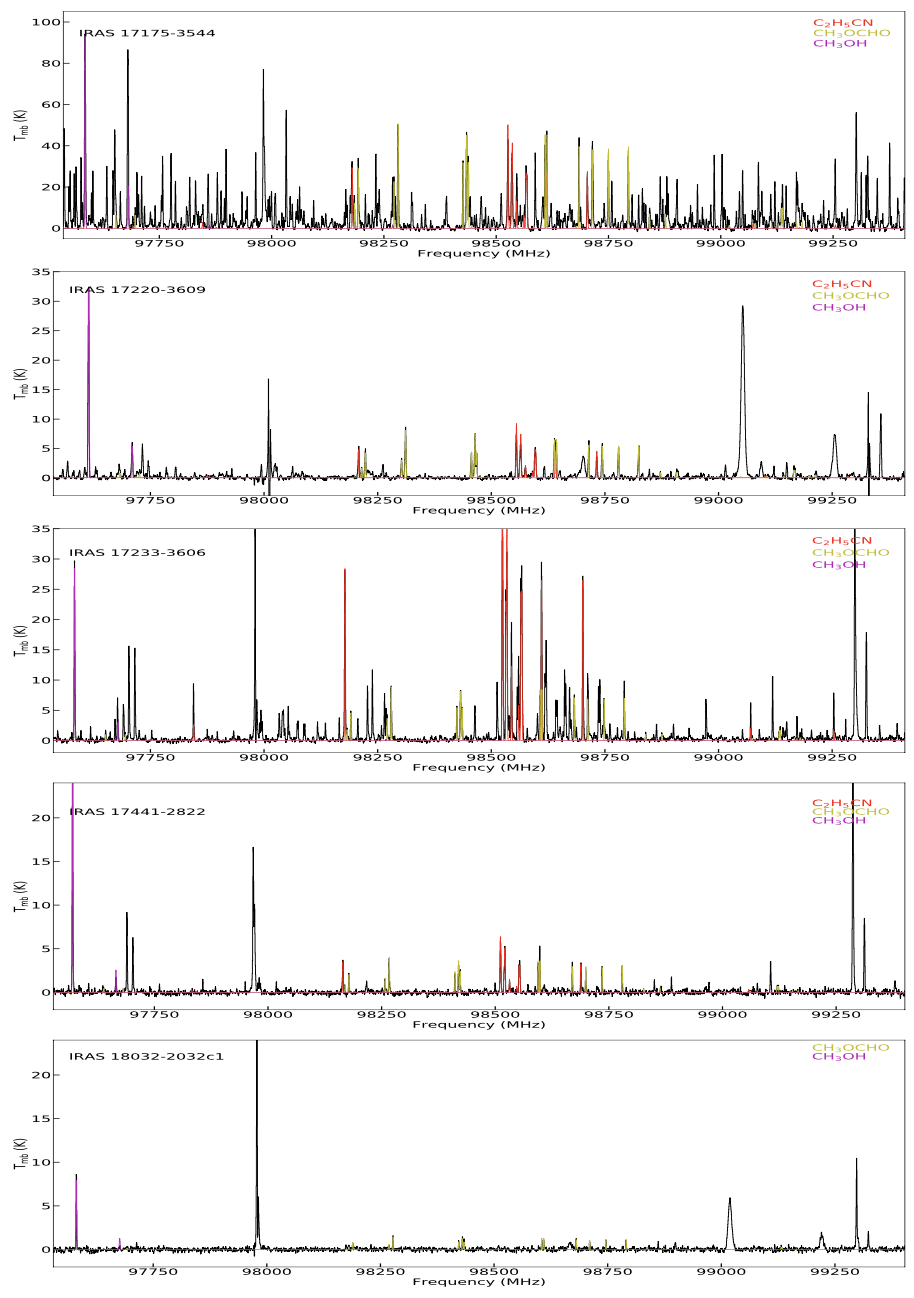}
\contcaption{}
\end{figure*}

\begin{figure*}
\includegraphics[width=17cm,height=20cm]{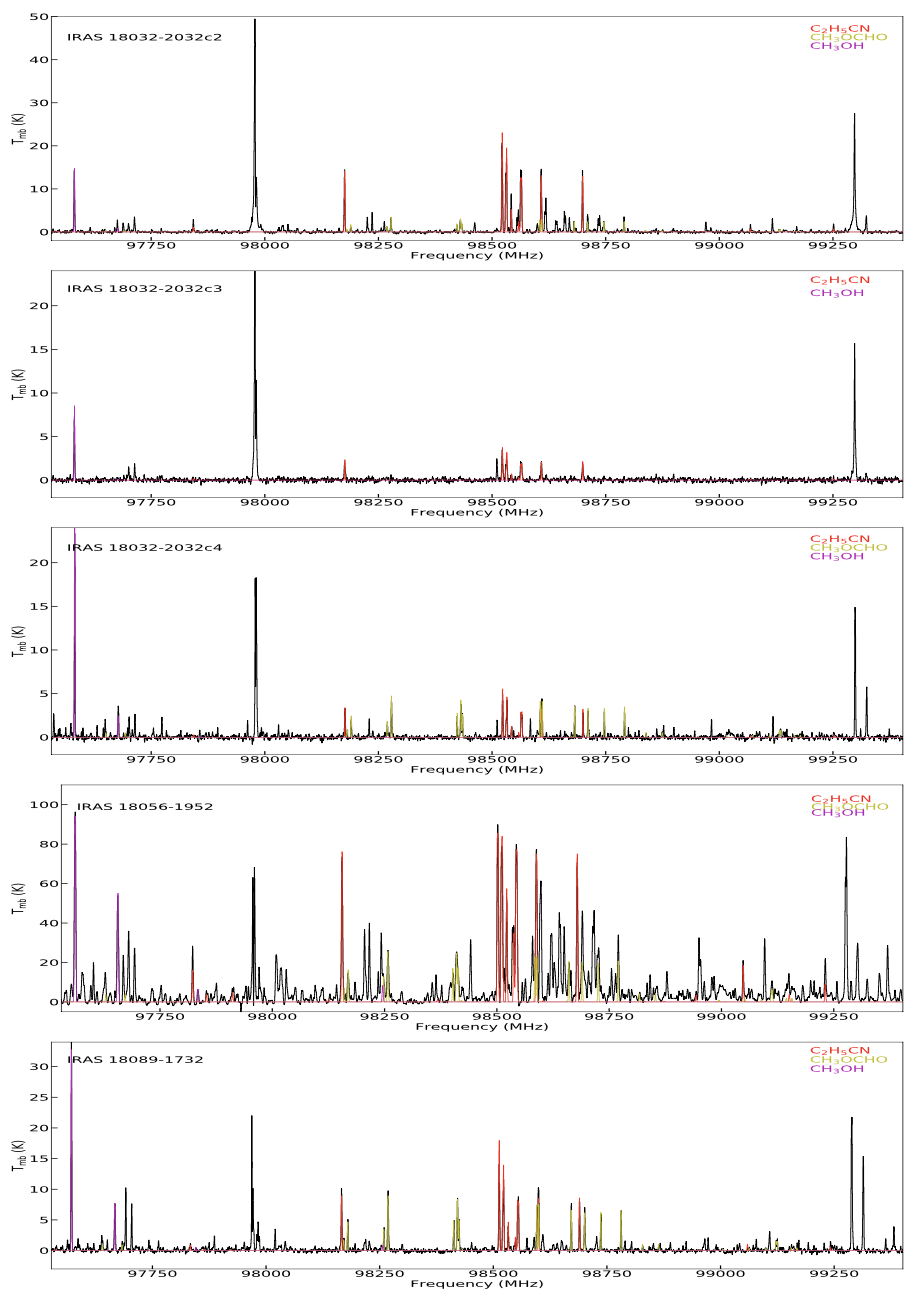}
\contcaption{}
\end{figure*}

\begin{figure*}
\includegraphics[width=17cm,height=20cm]{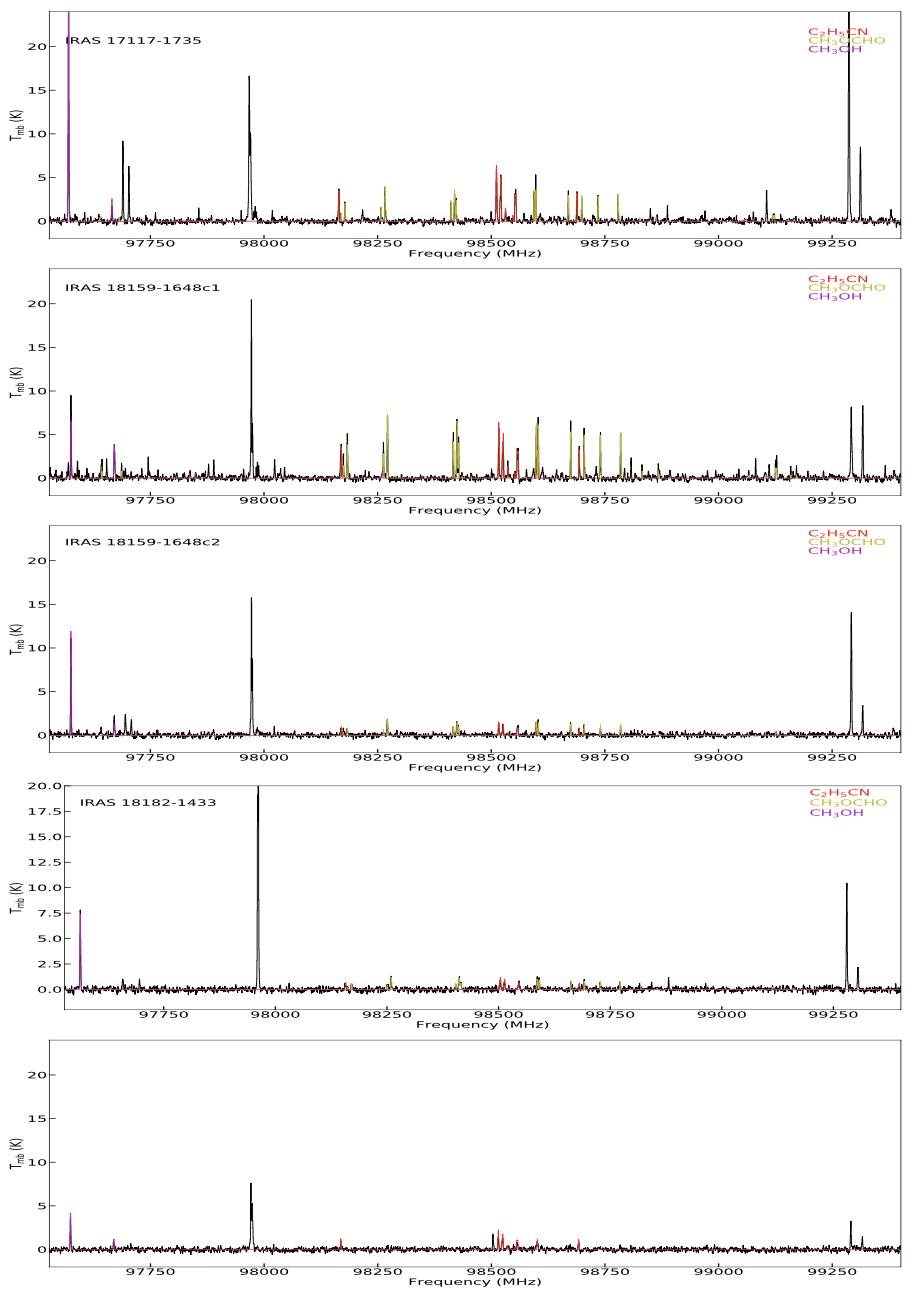}
\contcaption{}
\end{figure*}

\begin{figure*}
\includegraphics[width=17cm,height=20cm]{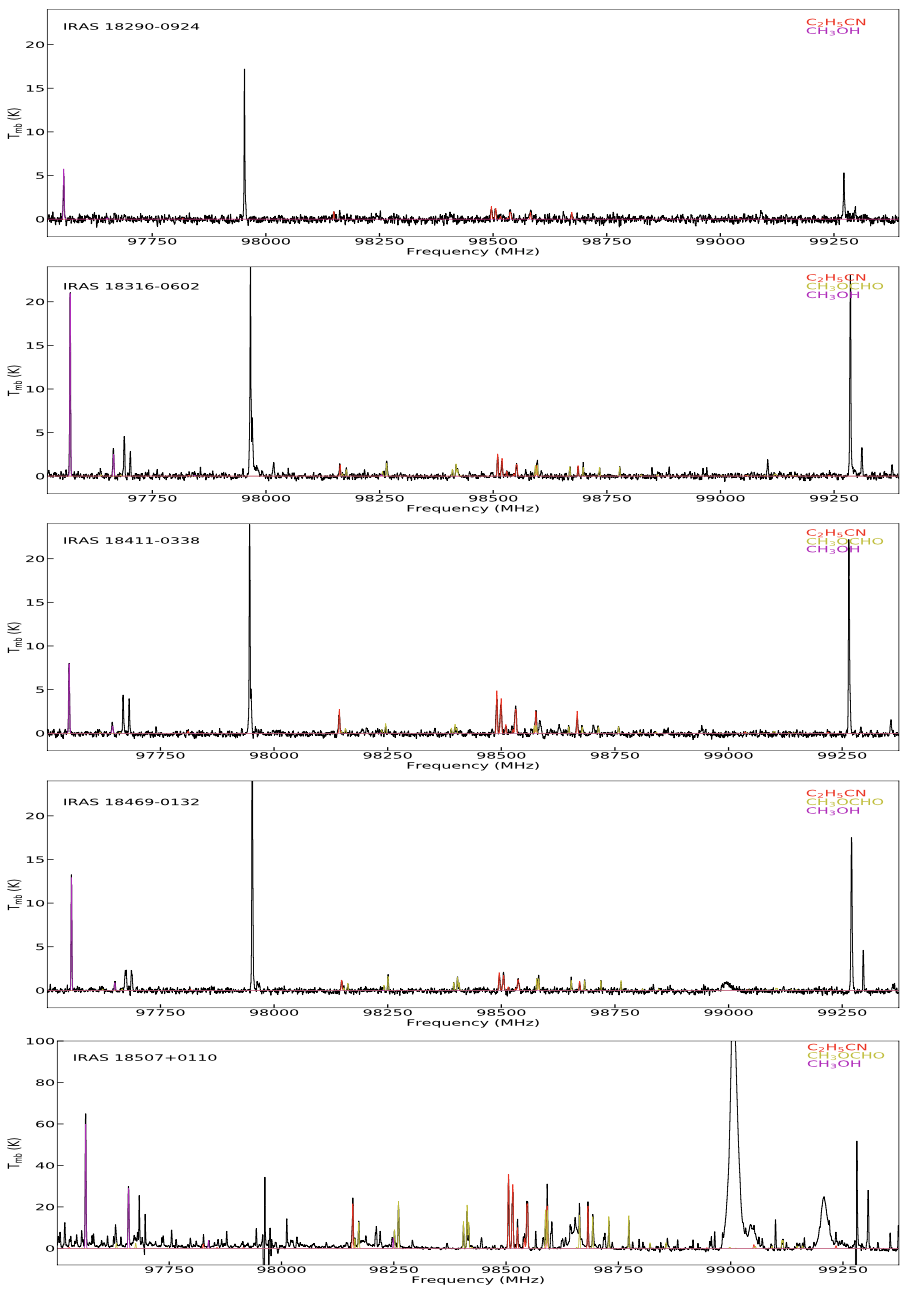}
\contcaption{}
\end{figure*}

\begin{figure*}
\includegraphics[width=17cm,height=20cm]{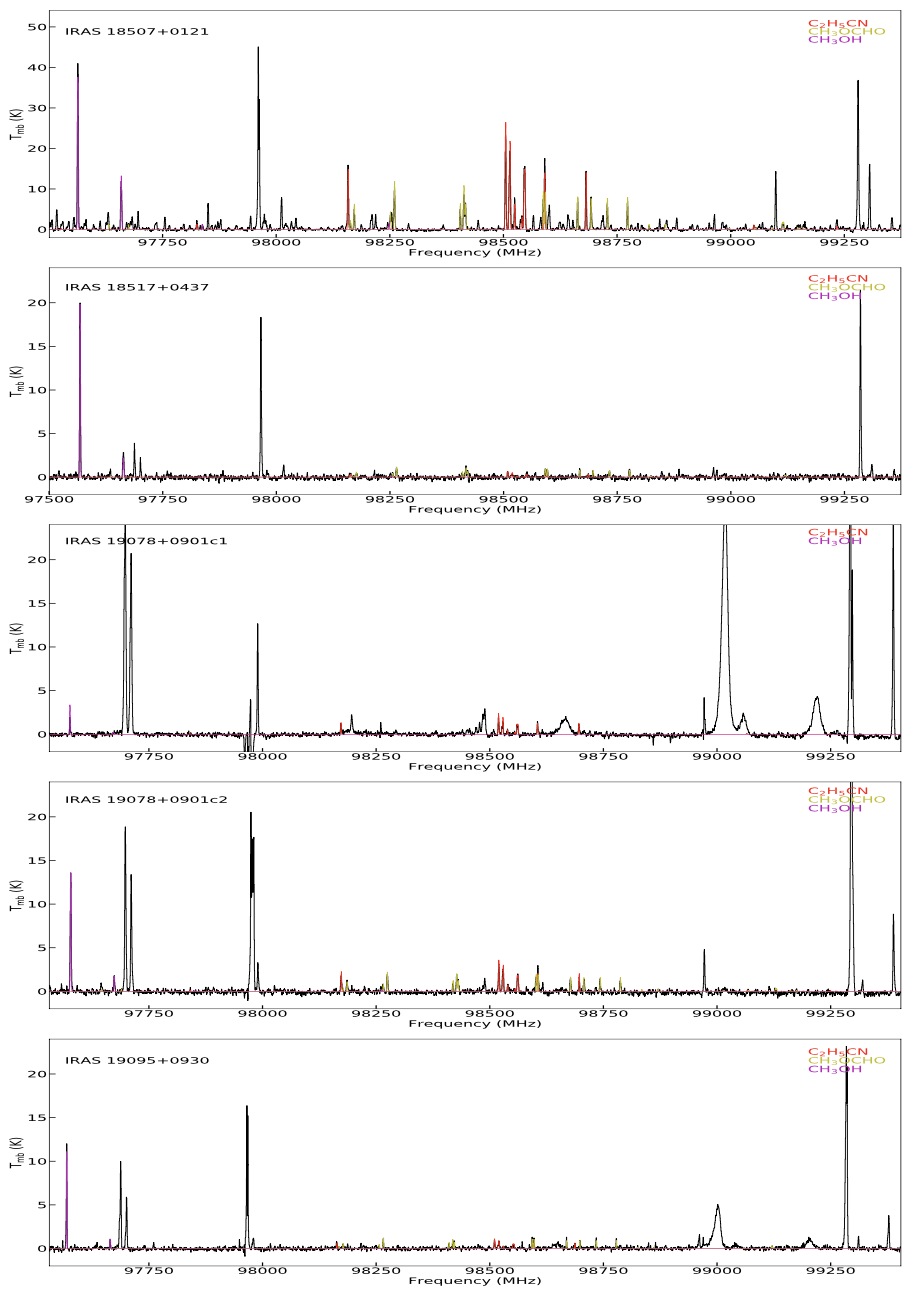}
\contcaption{}
\end{figure*}

\clearpage
\subsection{Spectra in SPW 8}

\begin{figure*}
\includegraphics[width=17cm,height=20cm]{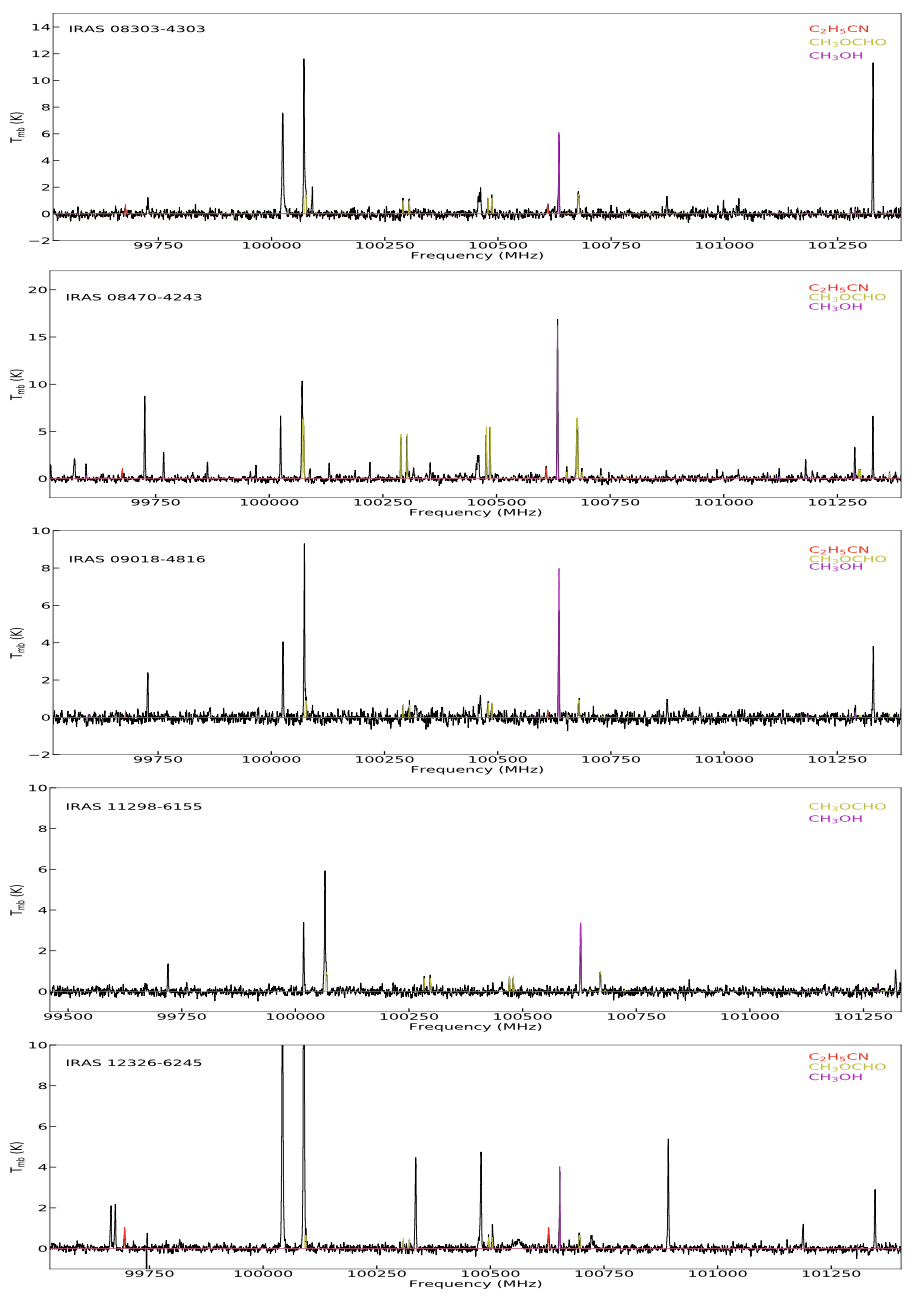}
\caption{Spectra in SPW 8  for the 60 hot cores. The observed
spectra are shown in black curves and the XCLASS modelled spectra
are coded in color.}
\end{figure*}

\begin{figure*}
\includegraphics[width=17cm,height=20cm]{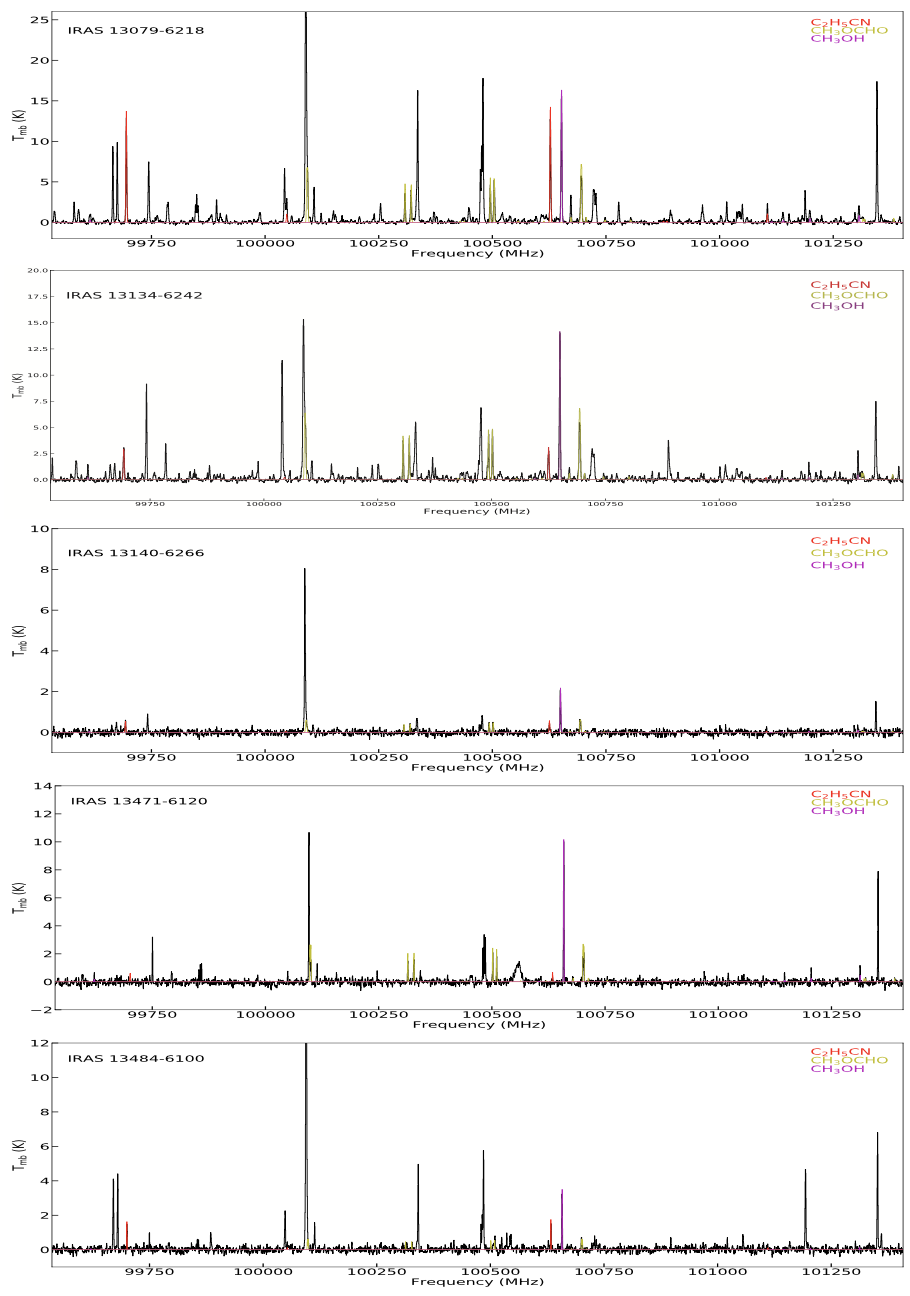}
\contcaption{}
\end{figure*}

\begin{figure*}
\includegraphics[width=17cm,height=20cm]{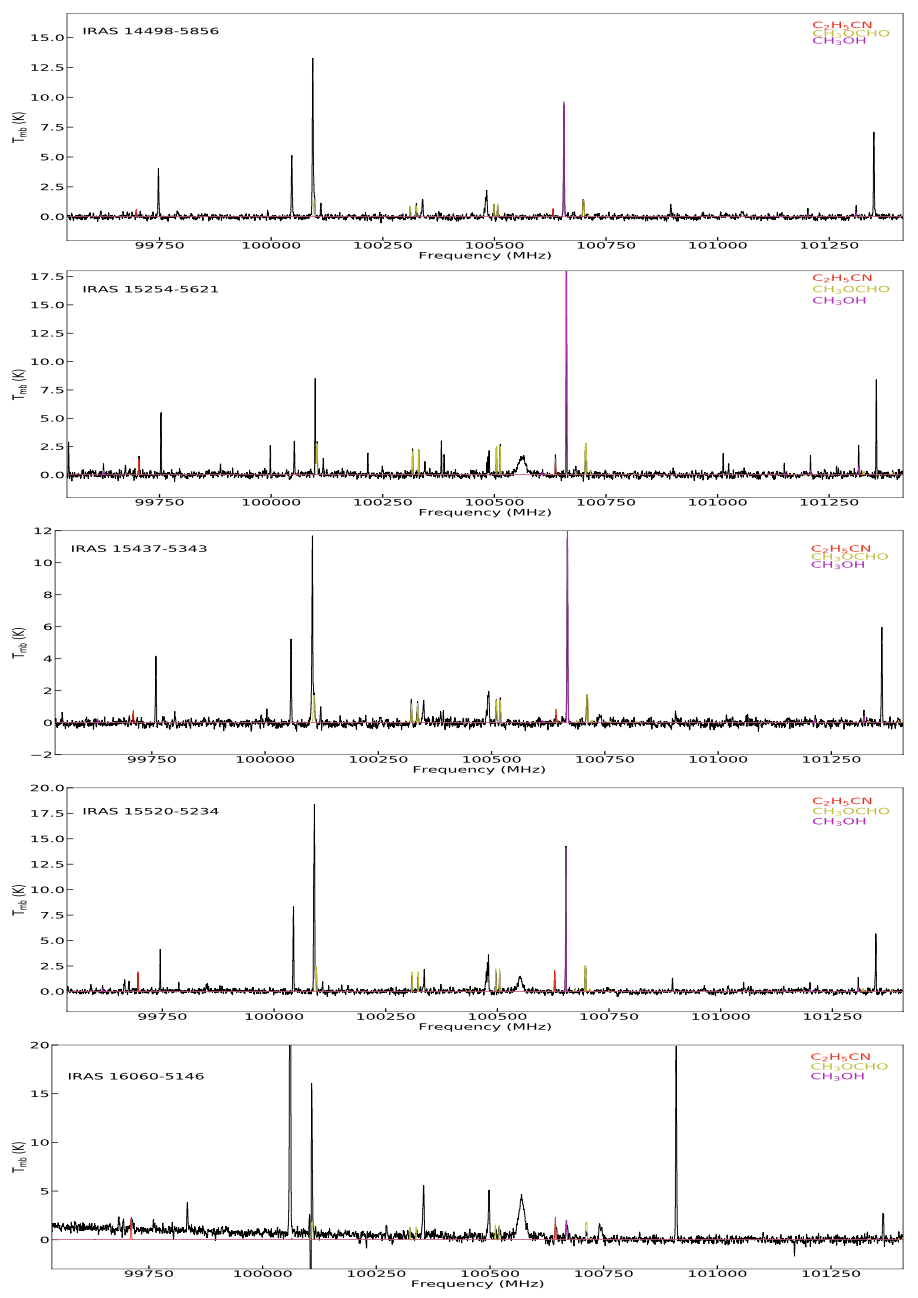}
\contcaption{}
\end{figure*}

\begin{figure*}
\includegraphics[width=17cm,height=20cm]{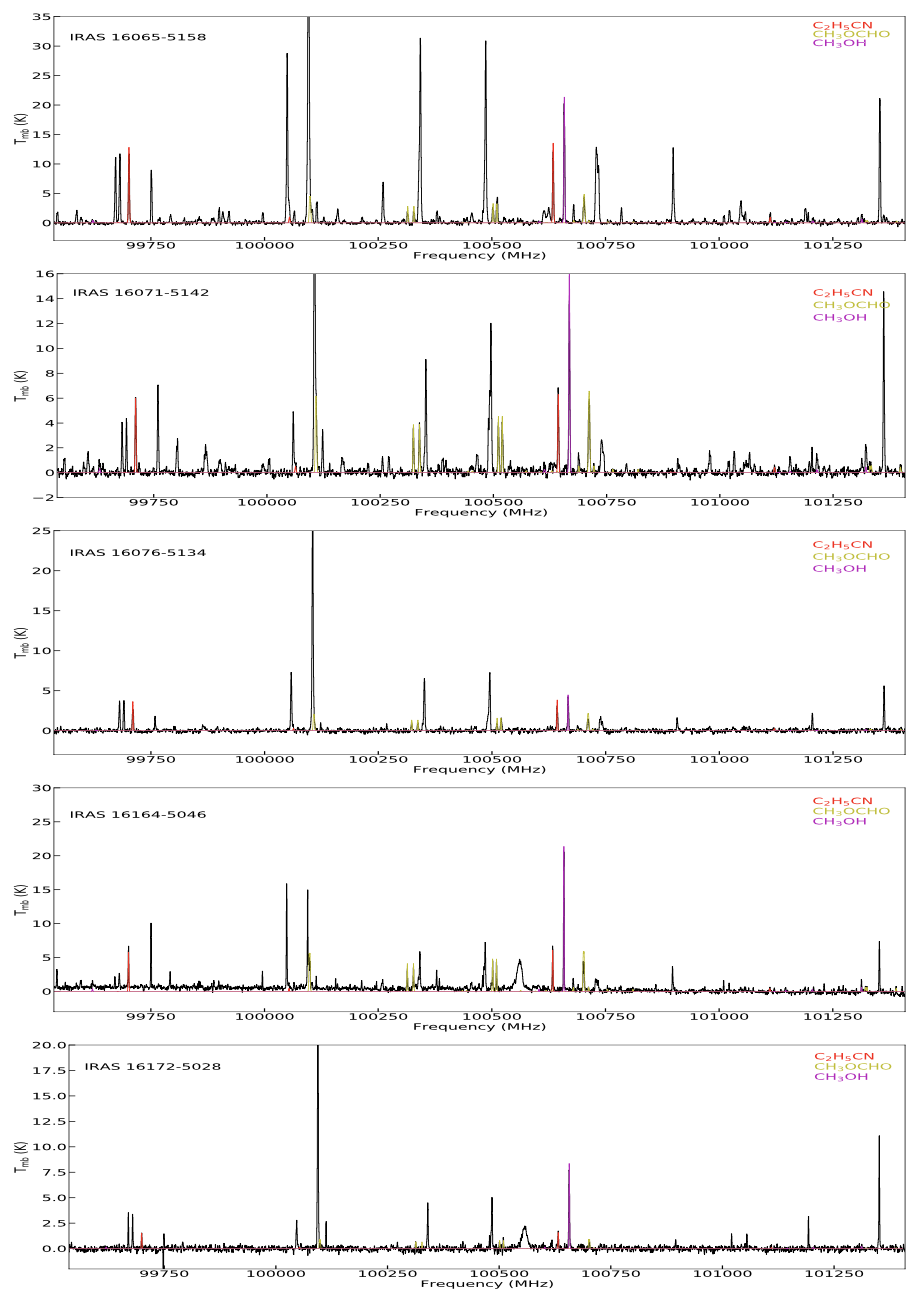}
\contcaption{}
\end{figure*}

\begin{figure*}
\includegraphics[width=17cm,height=20cm]{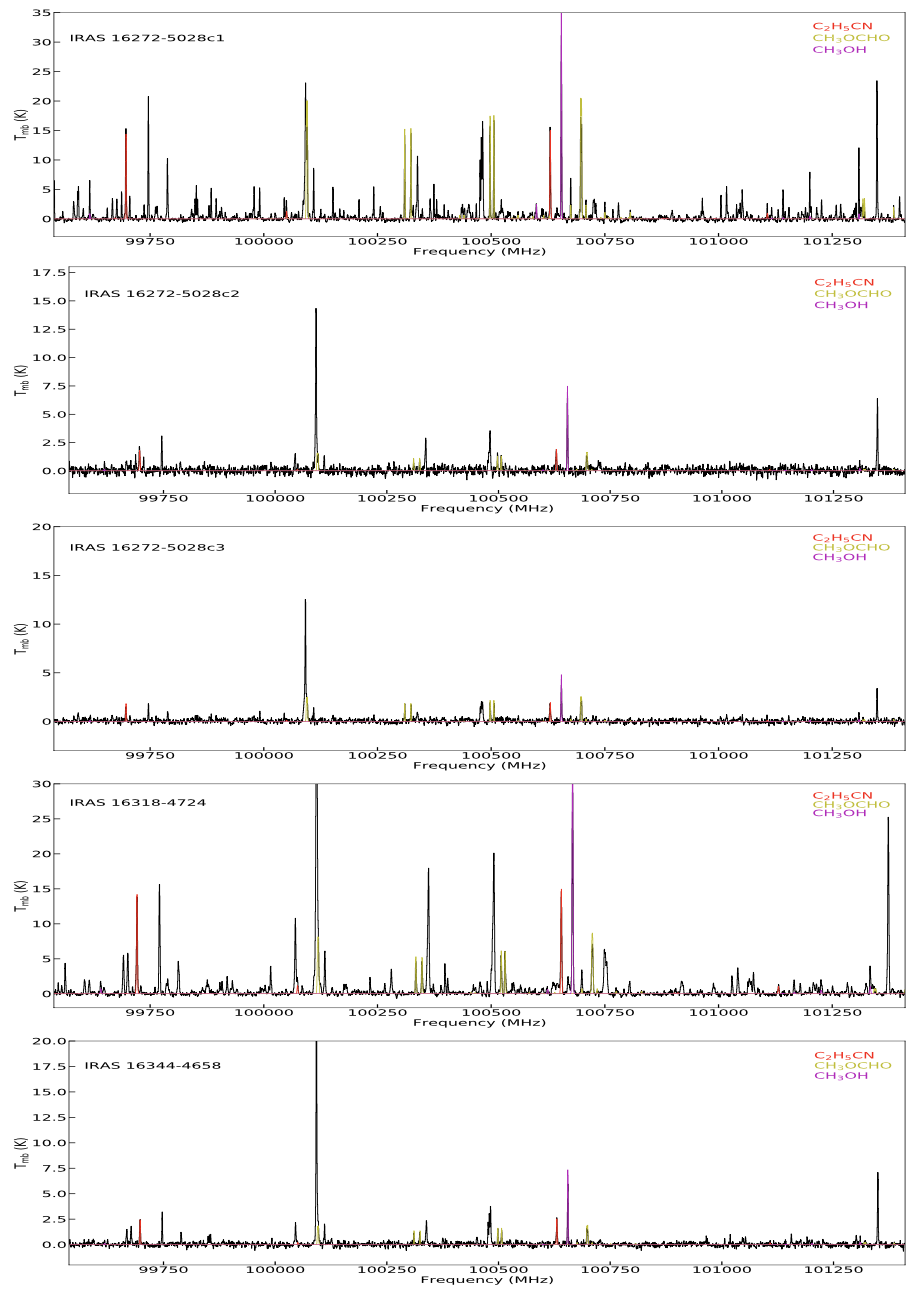}
\contcaption{}
\end{figure*}

\begin{figure*}
\includegraphics[width=17cm,height=20cm]{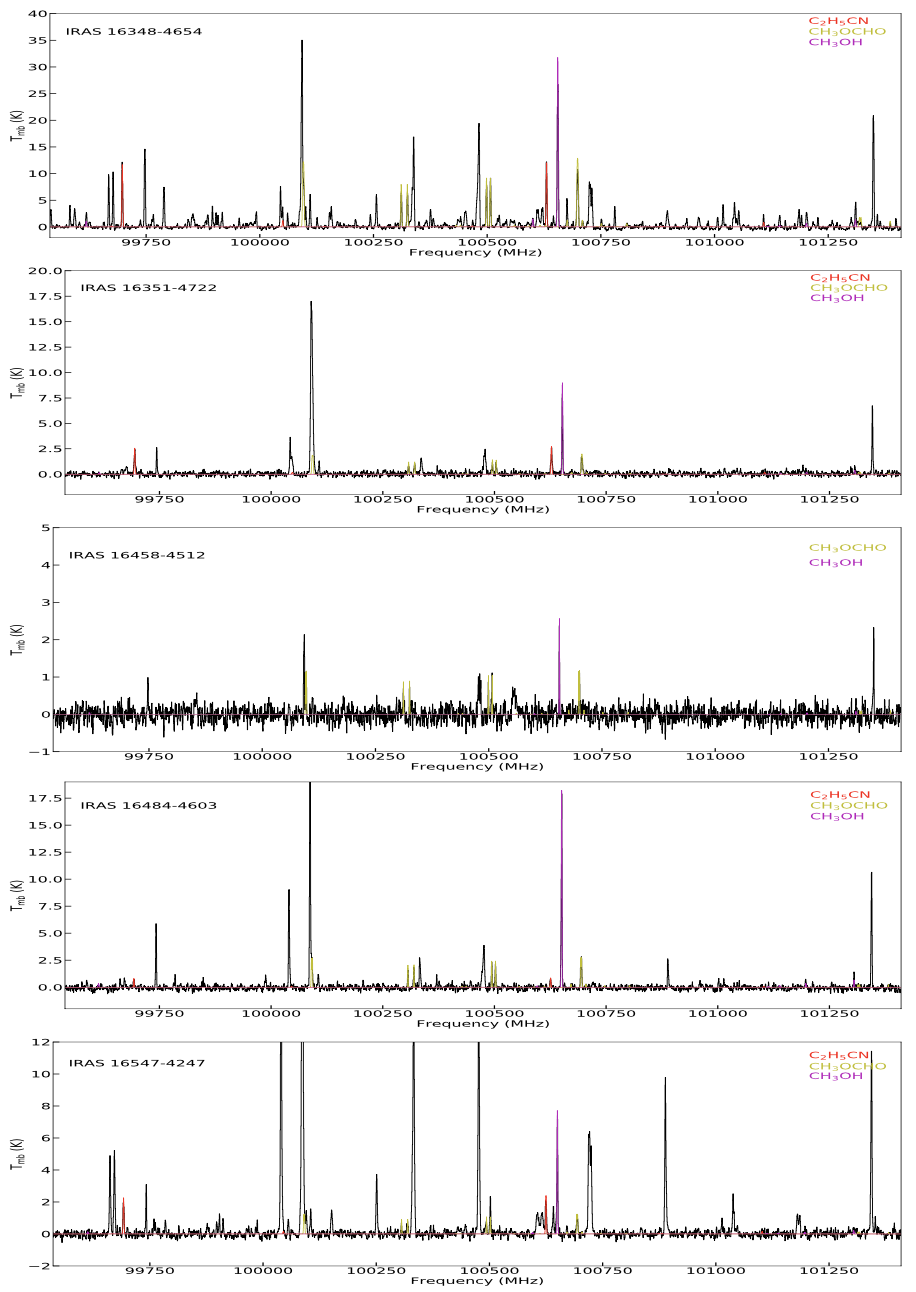}
\contcaption{}
\end{figure*}

\begin{figure*}
\includegraphics[width=17cm,height=20cm]{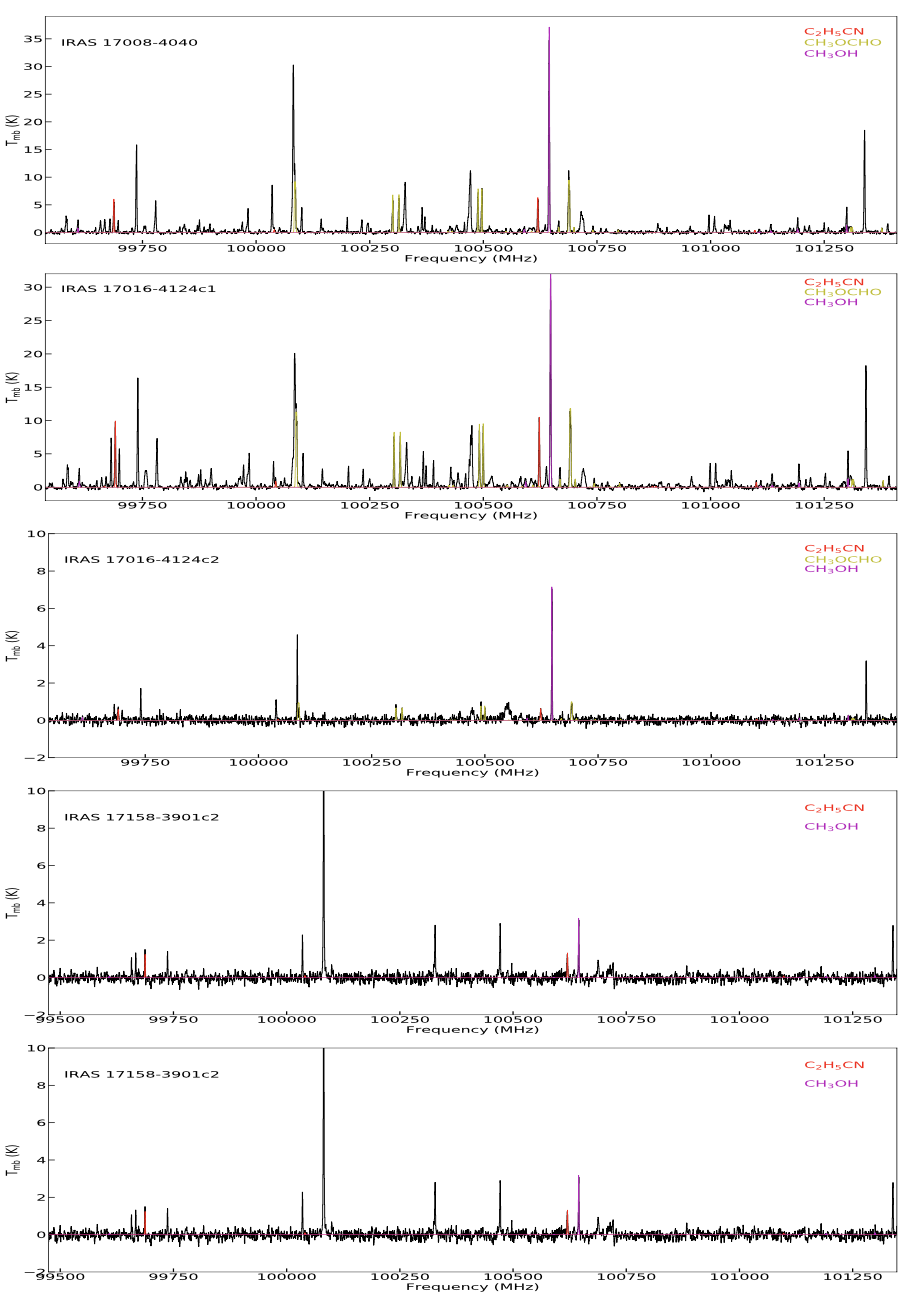}
\contcaption{}
\end{figure*}

\begin{figure*}
\includegraphics[width=17cm,height=20cm]{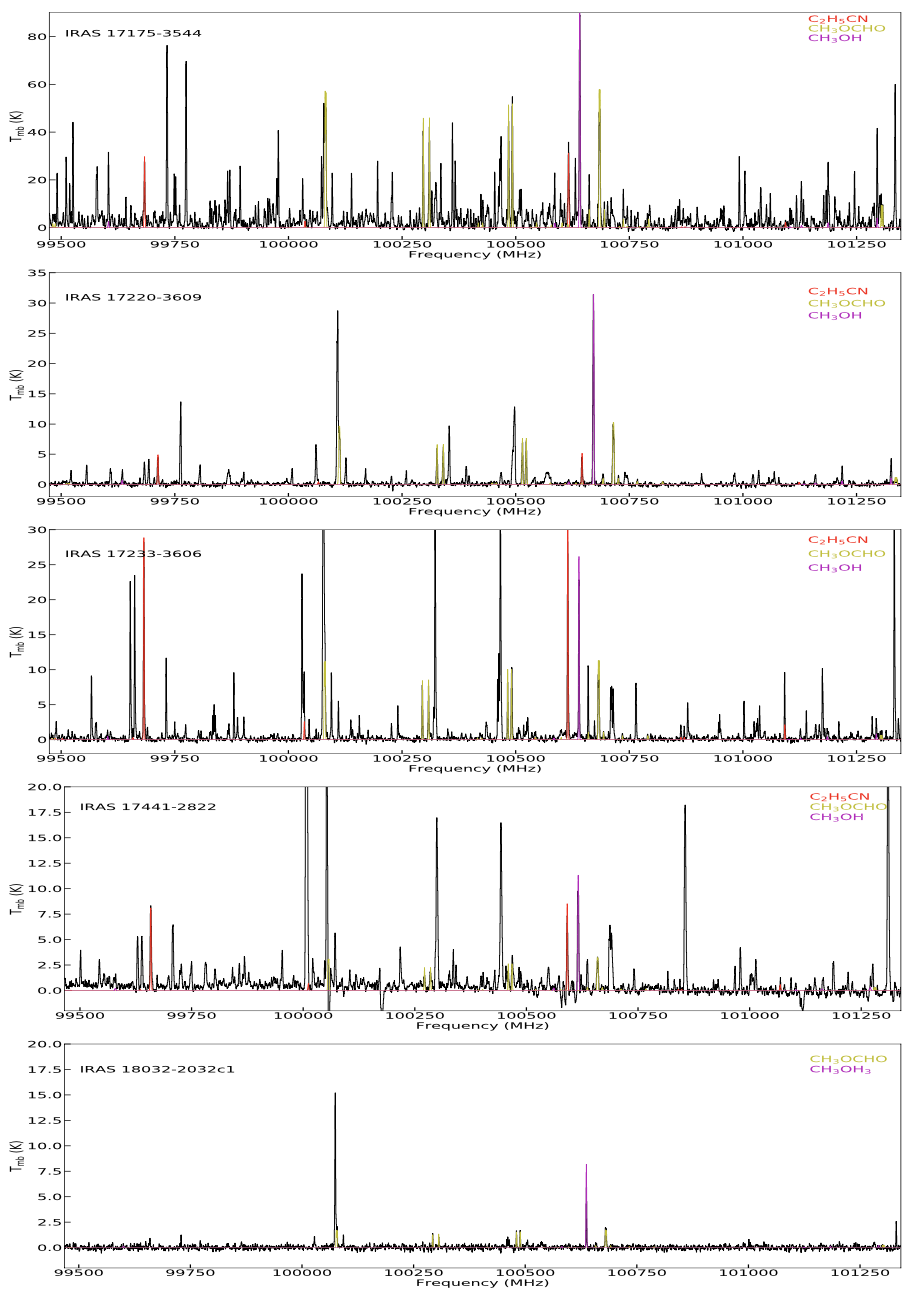}
\contcaption{}
\end{figure*}

\begin{figure*}
\includegraphics[width=17cm,height=20cm]{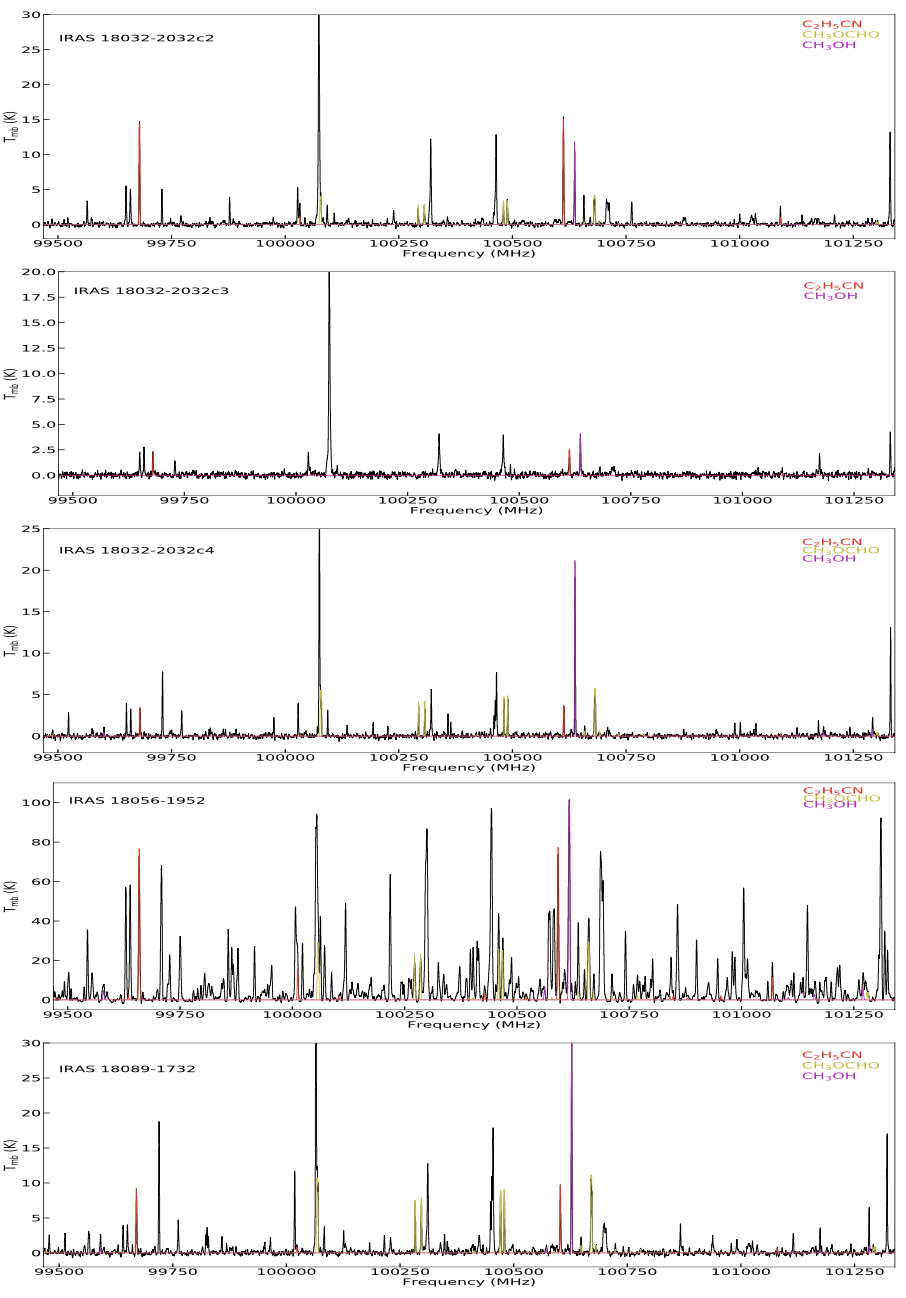}
\contcaption{}
\end{figure*}

\begin{figure*}
\includegraphics[width=17cm,height=20cm]{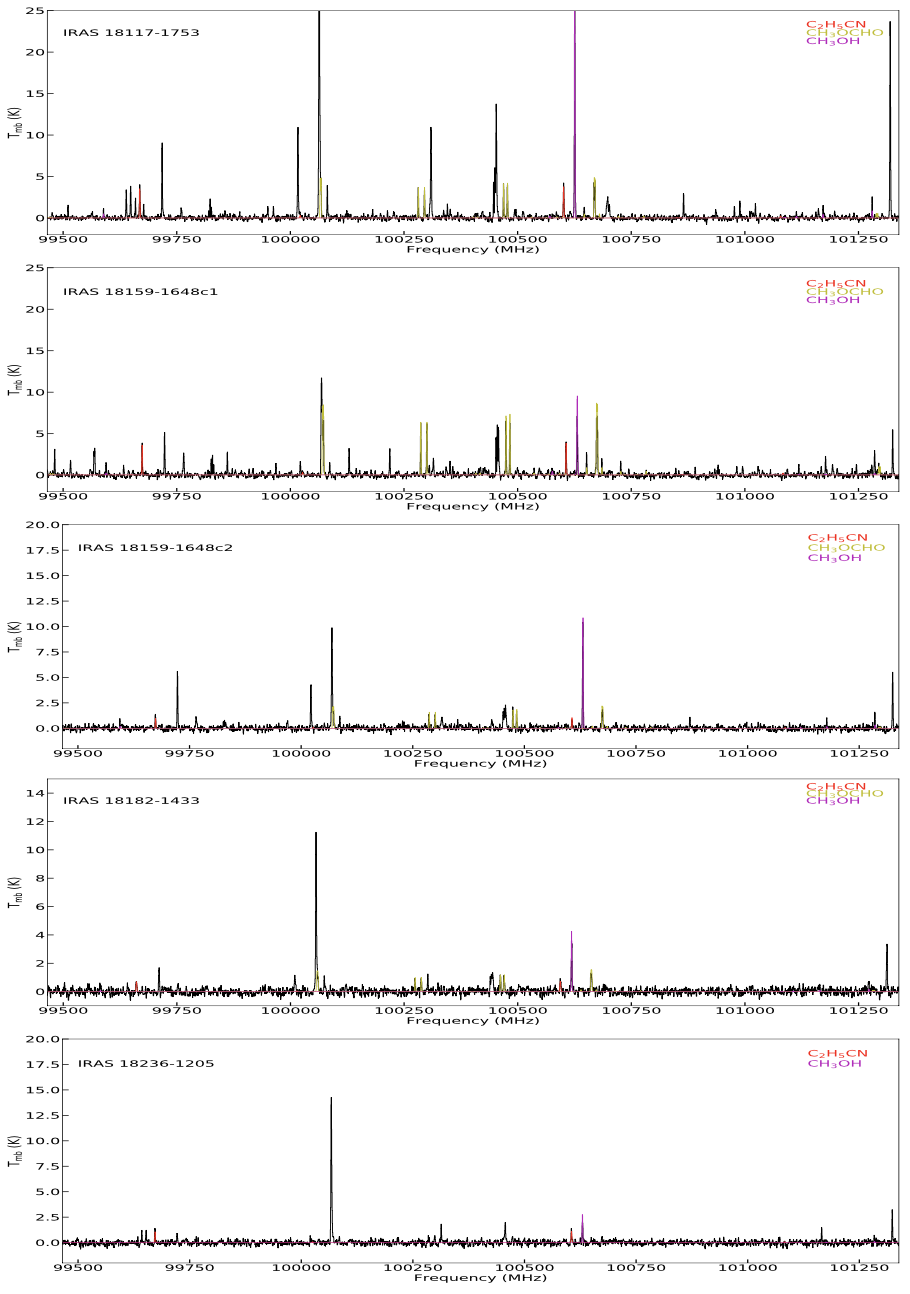}
\contcaption{}
\end{figure*}

\begin{figure*}
\includegraphics[width=17cm,height=20cm]{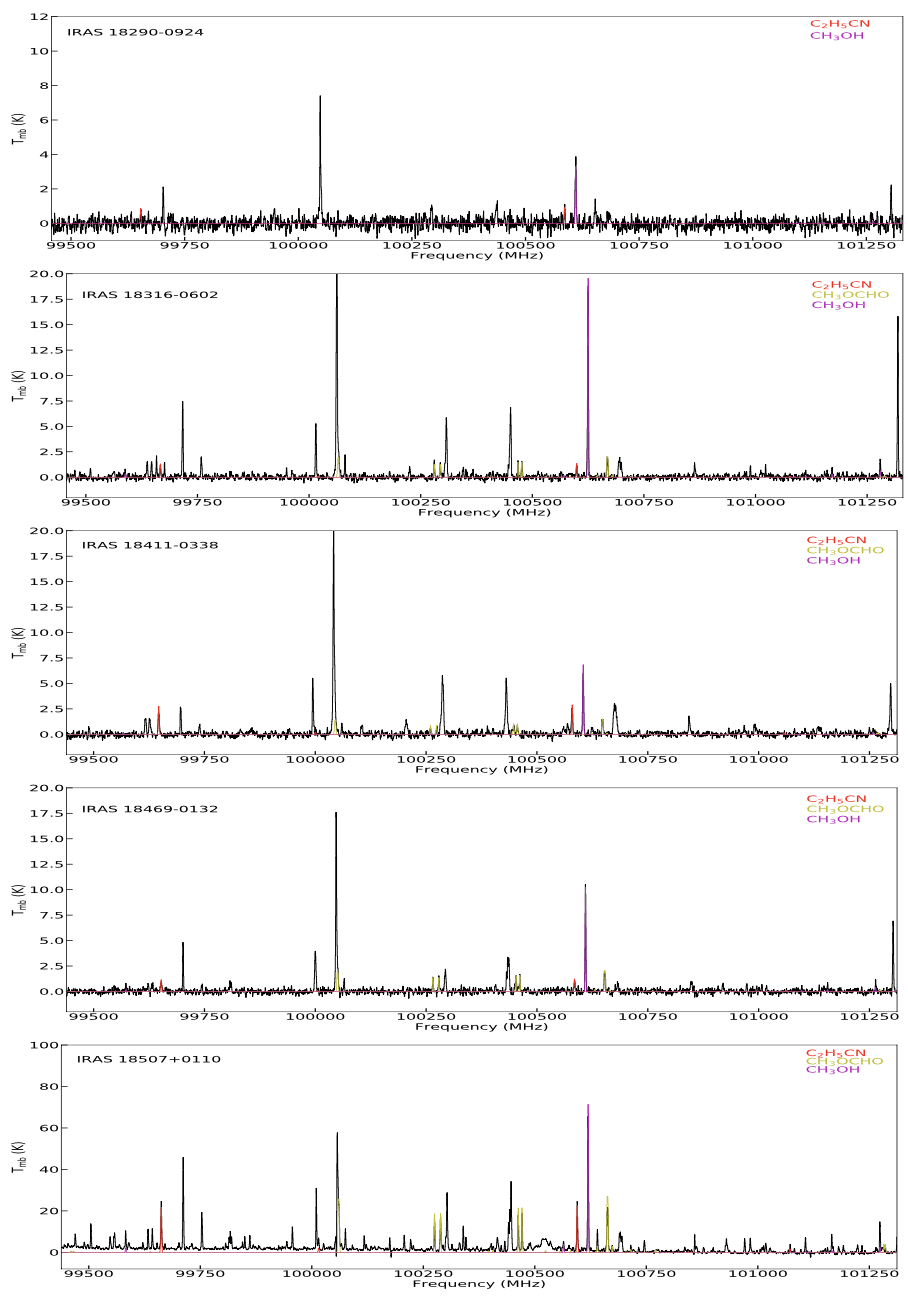}
\contcaption{}
\end{figure*}

\begin{figure*}
\includegraphics[width=17cm,height=20cm]{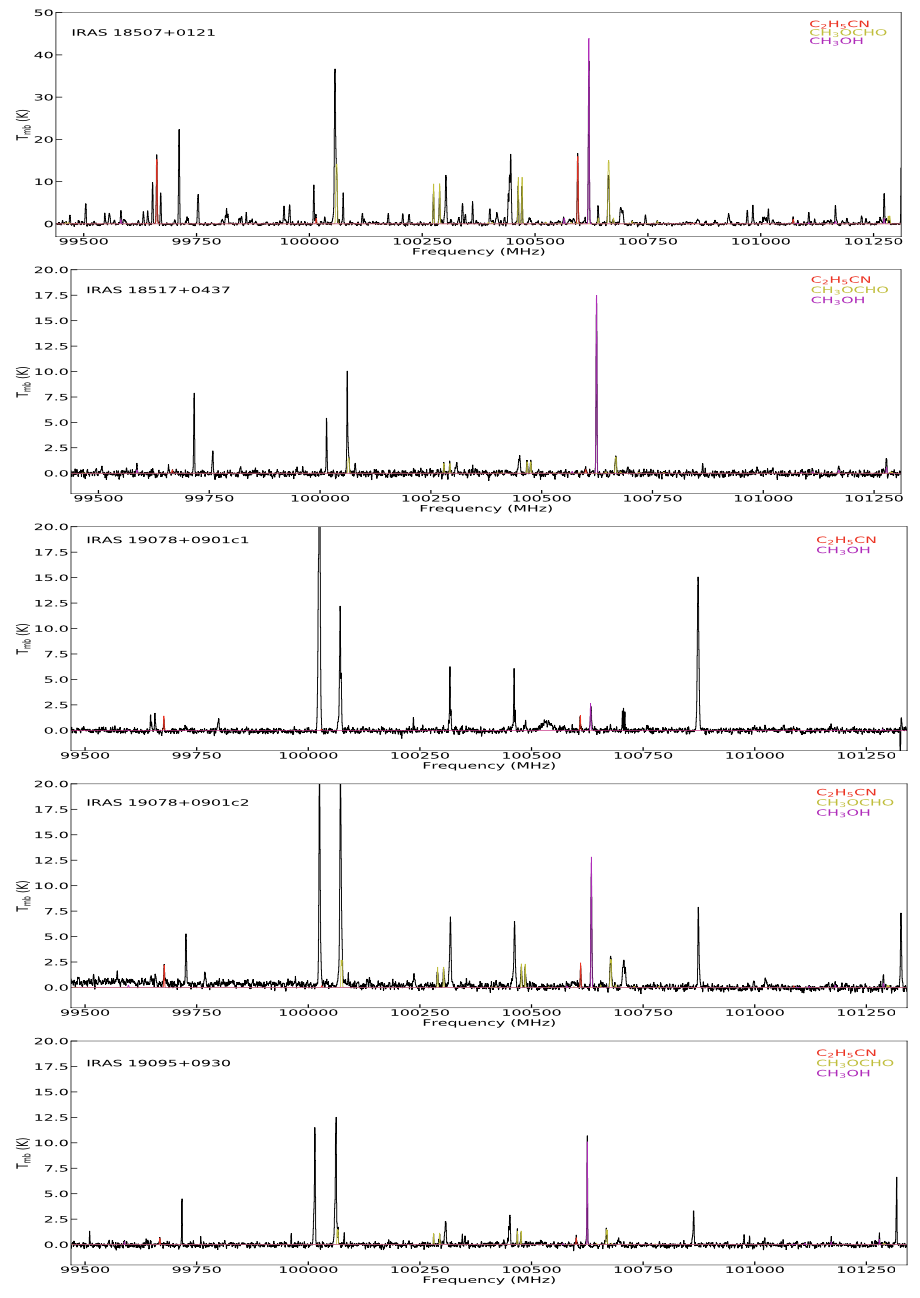}
\contcaption{}
\end{figure*}

\clearpage

\subsection{Images of the continuum and organic molecular lines for externally heated hot cores}

\begin{figure*}[tbh]
\centering
\includegraphics[width=11cm,height=22cm]{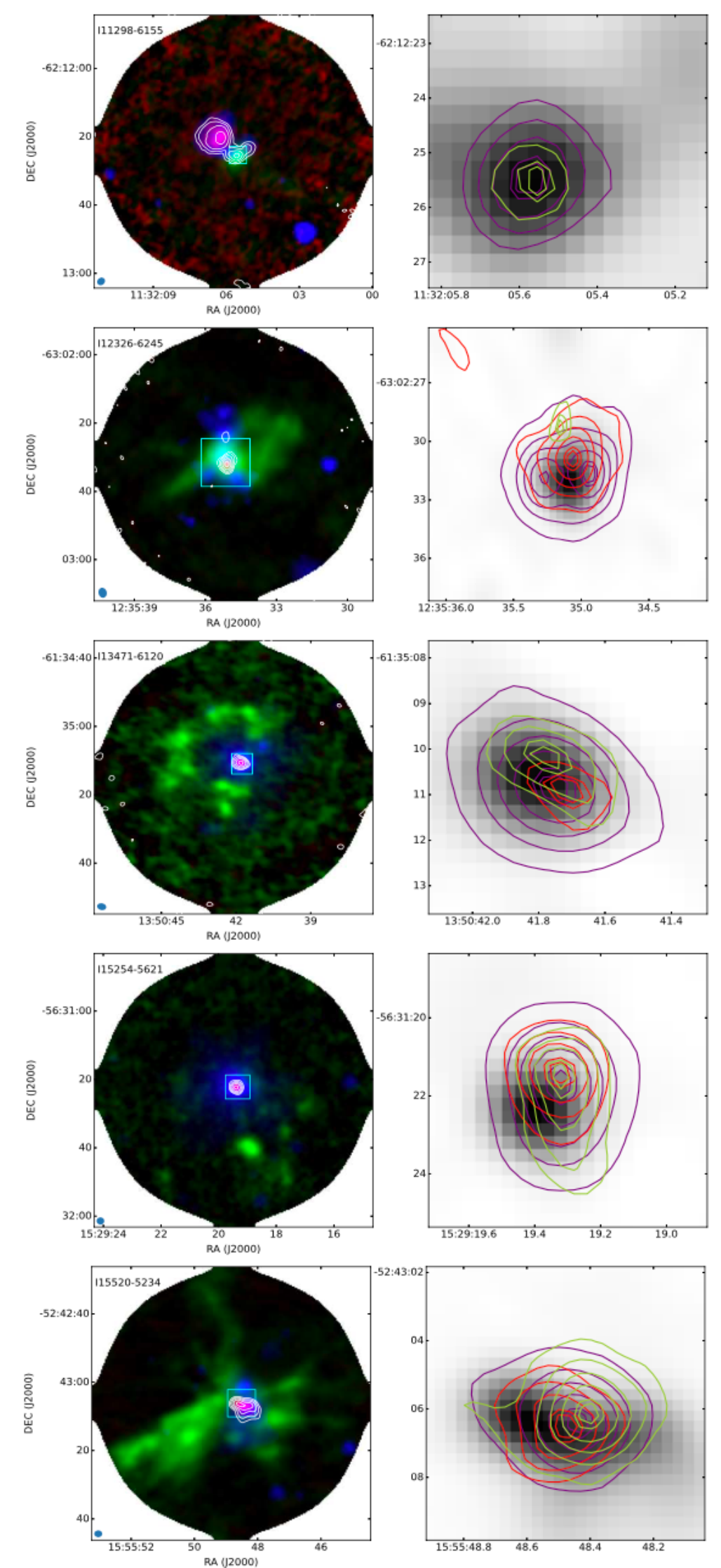}
\caption{Images of the continuum and organic molecular lines. In
the left panels, the background shows the three-color image
composed by H40$\alpha$ (red), SiO (green) and Spitzer 4.5
$\micron$ (blue), and the white contours represent the 3 mm
continuum; the green rectangles mark the imaging regions of the right panels. In the right panel, the background shows the 3 mm continuum.
The red, cygn and yellow contours represent the integrated
intensities of C$_2$H$_5$CN, CH$_3$OH and CH$_3$OCHO,
respectively. The contour levels are 10 to 90 percent (stepped by
20 percent) of the peak values. The innermost contour has a level
of 95 percent of the peak value.}
\end{figure*}

\addtocounter{figure}{-1}
\begin{figure*}
\centering
\includegraphics[width=12cm,height=25cm]{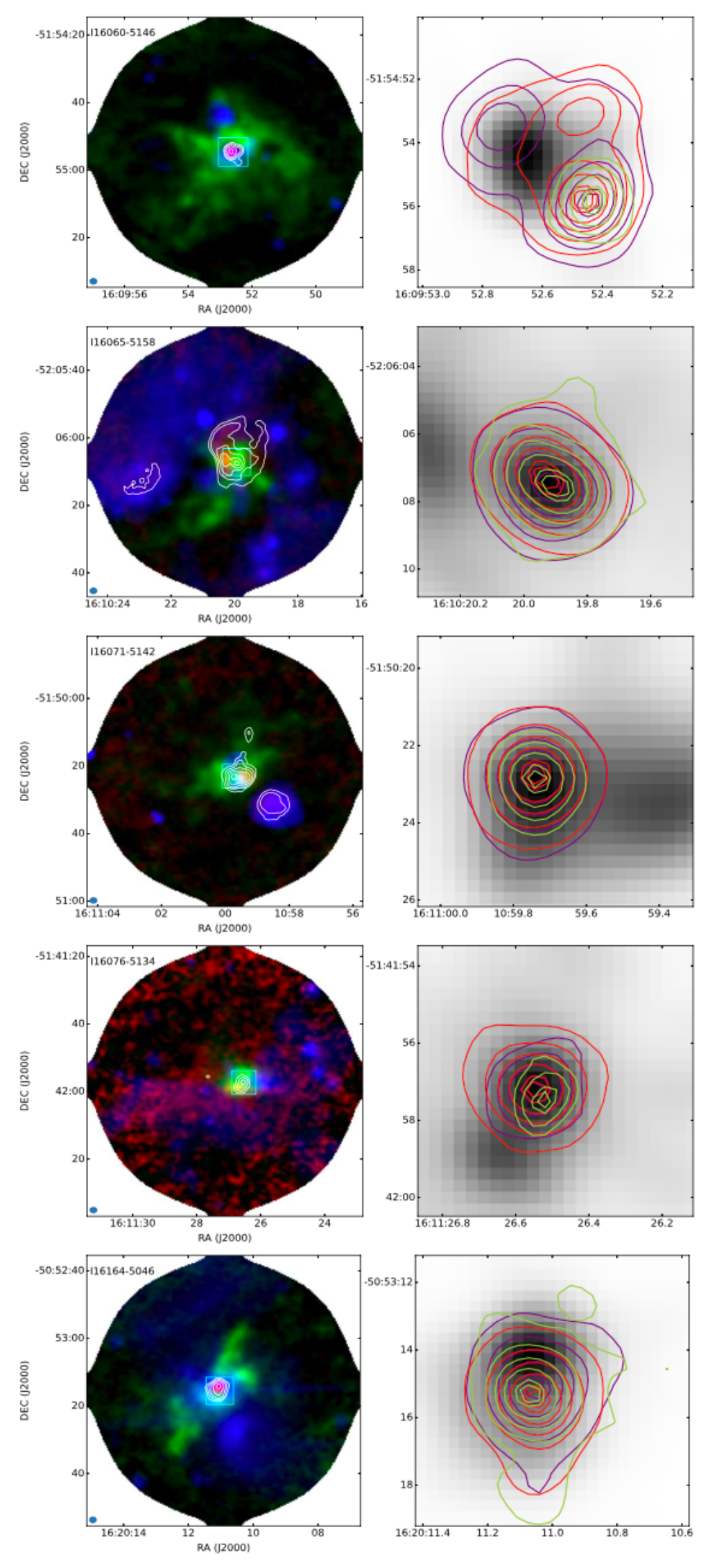}
\caption{\it -- continued}
\end{figure*}

\addtocounter{figure}{-1}
\begin{figure*}
\centering
\includegraphics[width=12cm,height=25cm]{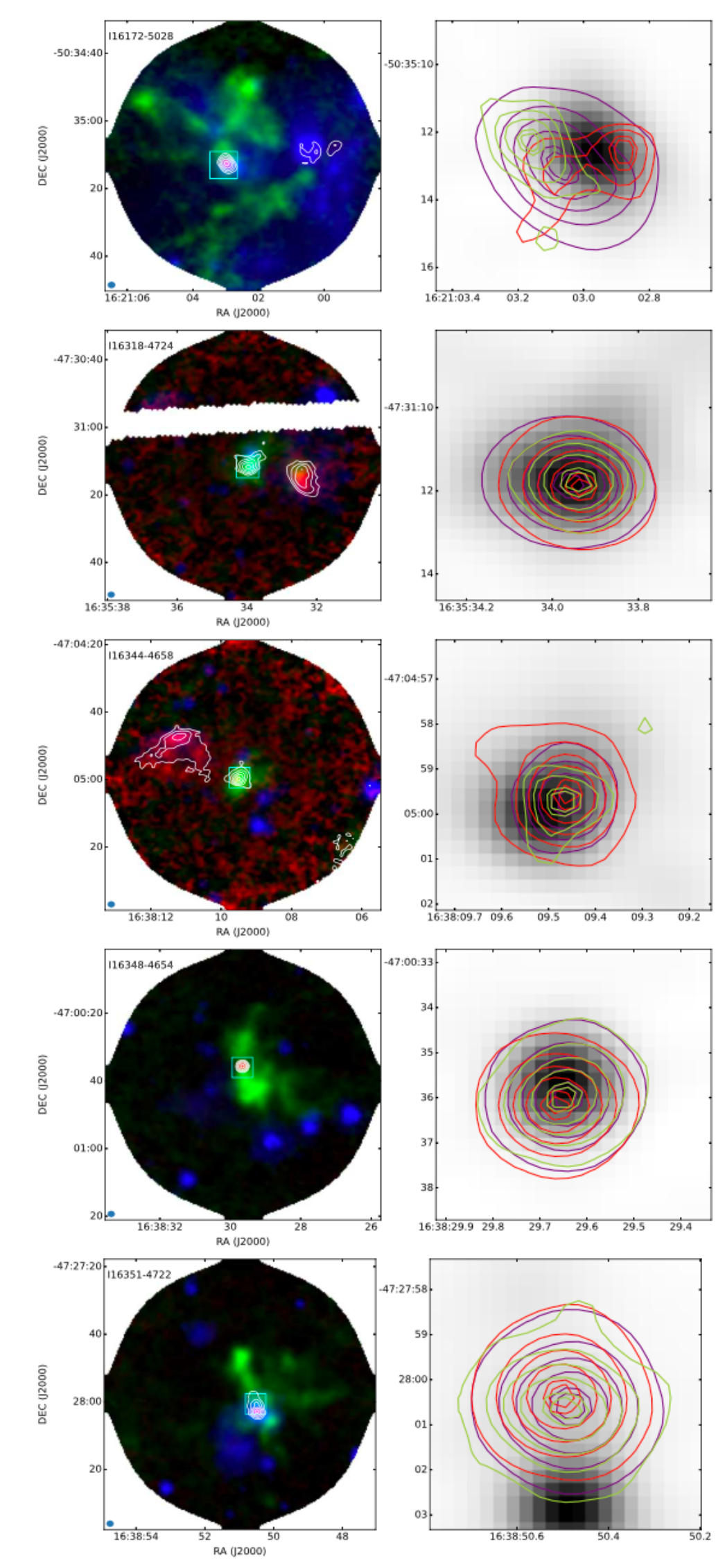}
\caption{\it -- continued}
\end{figure*}

\addtocounter{figure}{-1}
\begin{figure*}
\includegraphics[width=12cm,height=25cm]{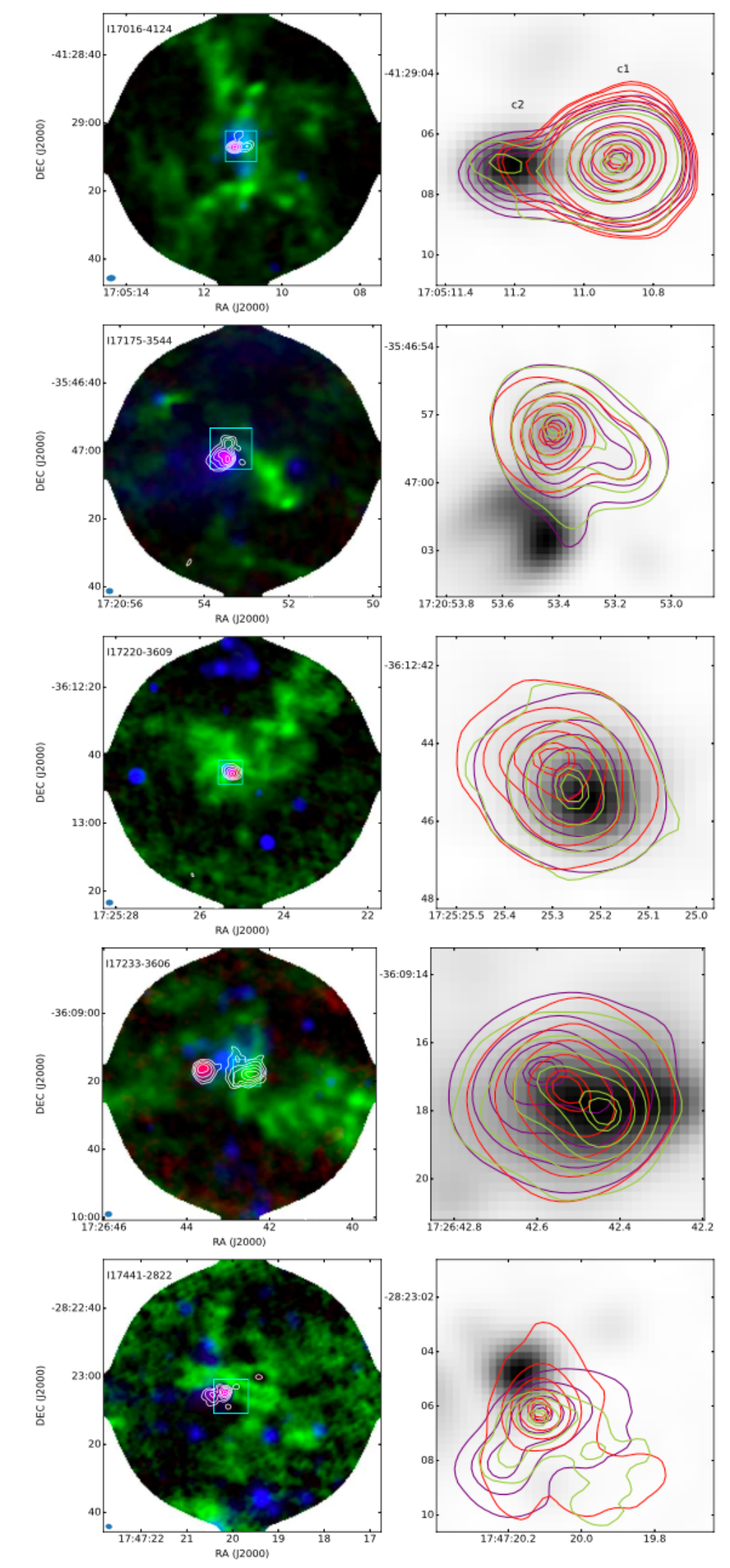}
\caption{\it -- continued}
\end{figure*}

\addtocounter{figure}{-1}
\begin{figure*}
\includegraphics[width=12cm,height=25cm]{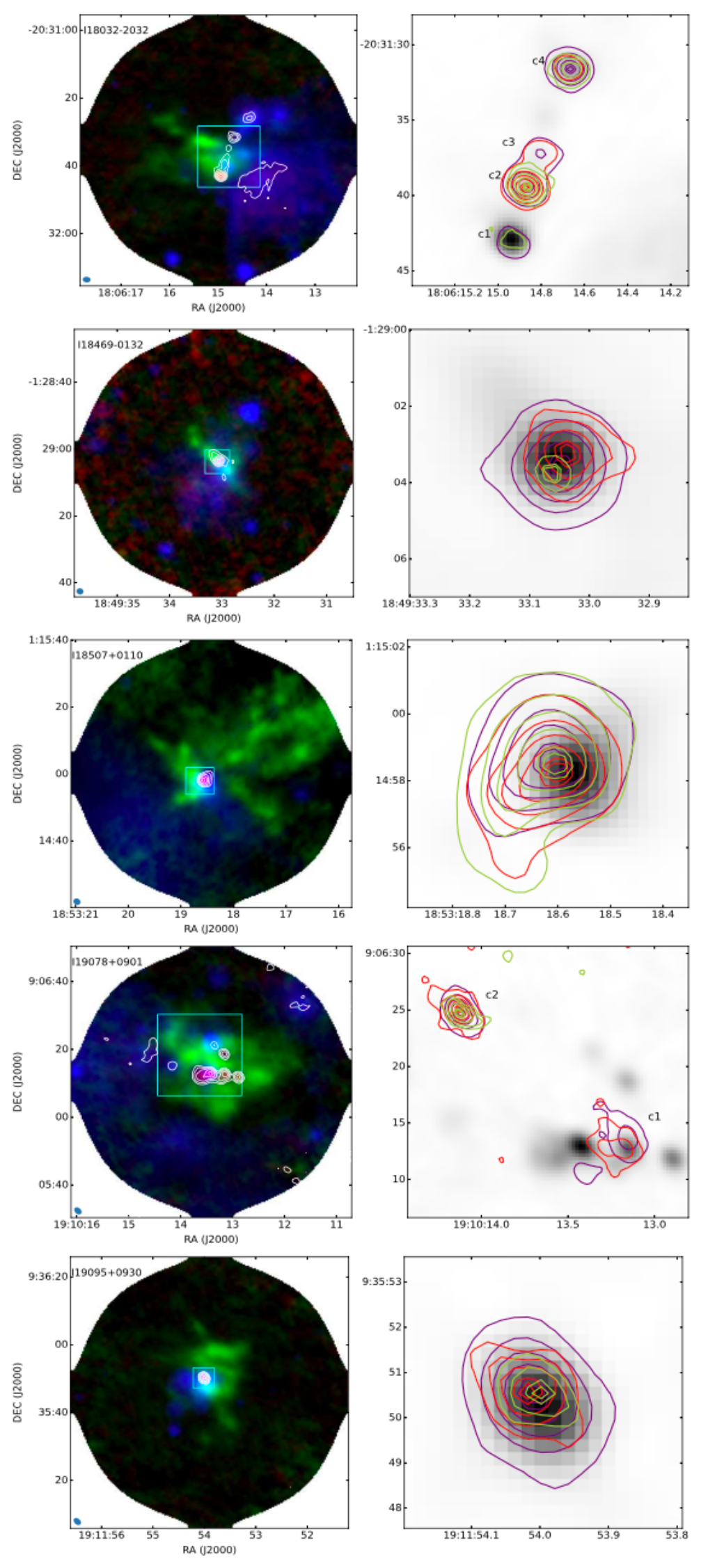}
\caption{\it -- continued}
\end{figure*}

\clearpage
\subsection{Images of the continuum and organic molecular lines for
internally heated hot cores}

\begin{figure*}
\includegraphics[width=11cm,height=22cm]{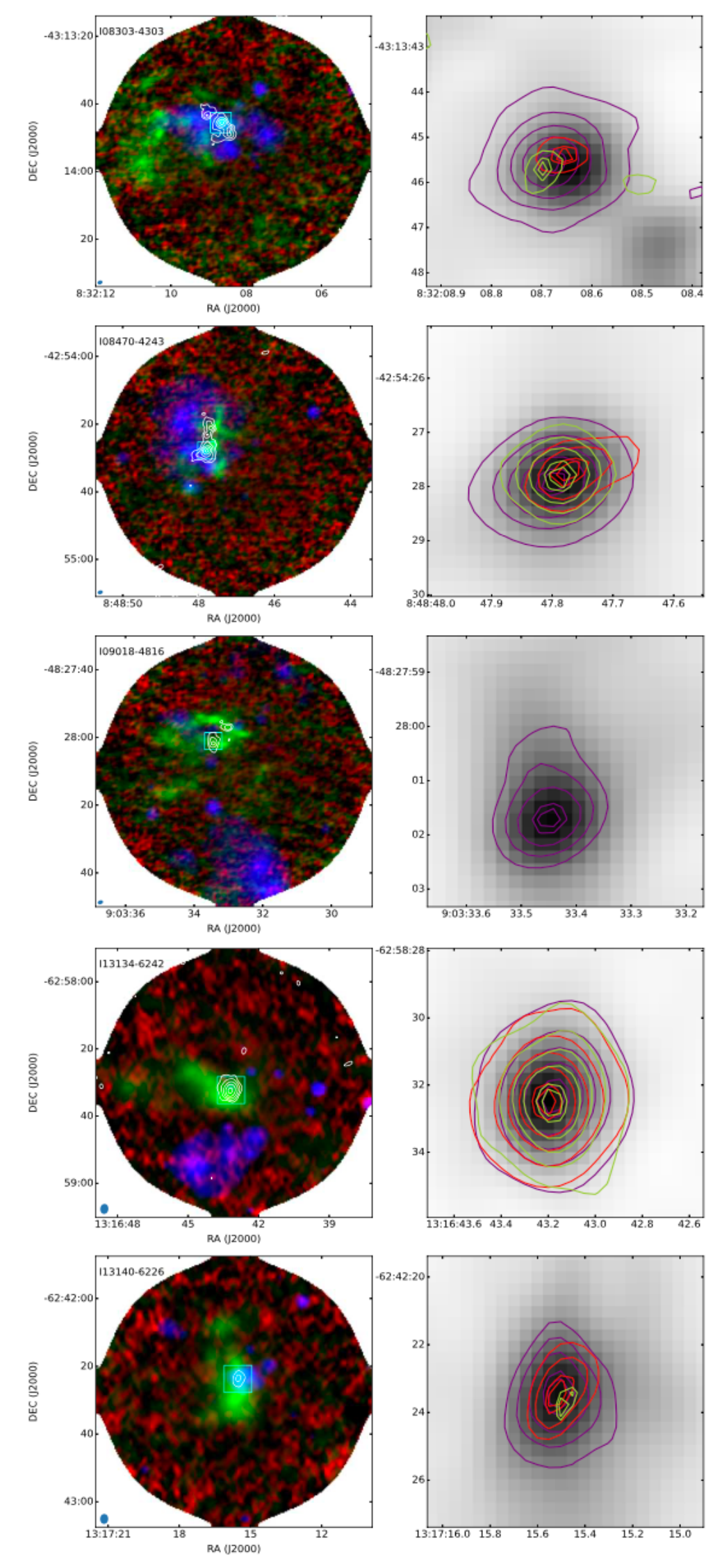}
\caption{Images of the continuum and organic molecular lines. In
the left panels, the background shows the three-color image
composed by H40$\alpha$ (red), SiO (green) and Spitzer 4.5
$\micron$ (blue), and the white contours represent the 3 mm
continuum; the green rectangles mark the imaging regions of the right panels. In the right panel, the background shows the 3 mm continuum.
The red, cygn and yellow contours represent the integrated
intensities of C$_2$H$_5$CN, CH$_3$OH and CH$_3$OCHO,
respectively. The contour levels are 10 to 90 percent (stepped by
20 percent) of the peak values. The innermost contour has a level
of 95 percent of the peak value.}
\end{figure*}

\addtocounter{figure}{-1}
\begin{figure*}
\includegraphics[width=12cm,height=25cm]{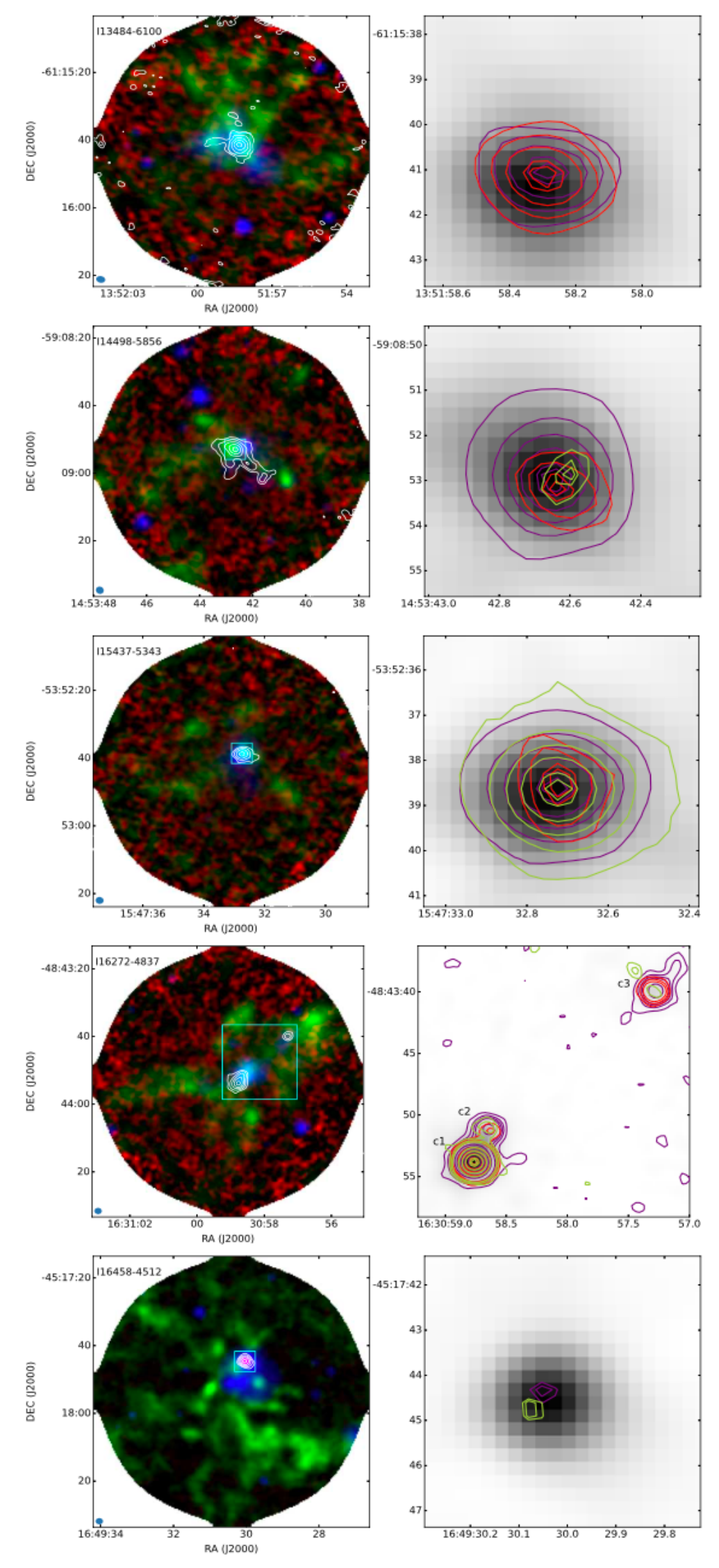}
\caption{\it -- continued}
\end{figure*}

\addtocounter{figure}{-1}
\begin{figure*}
\includegraphics[width=12cm,height=25cm]{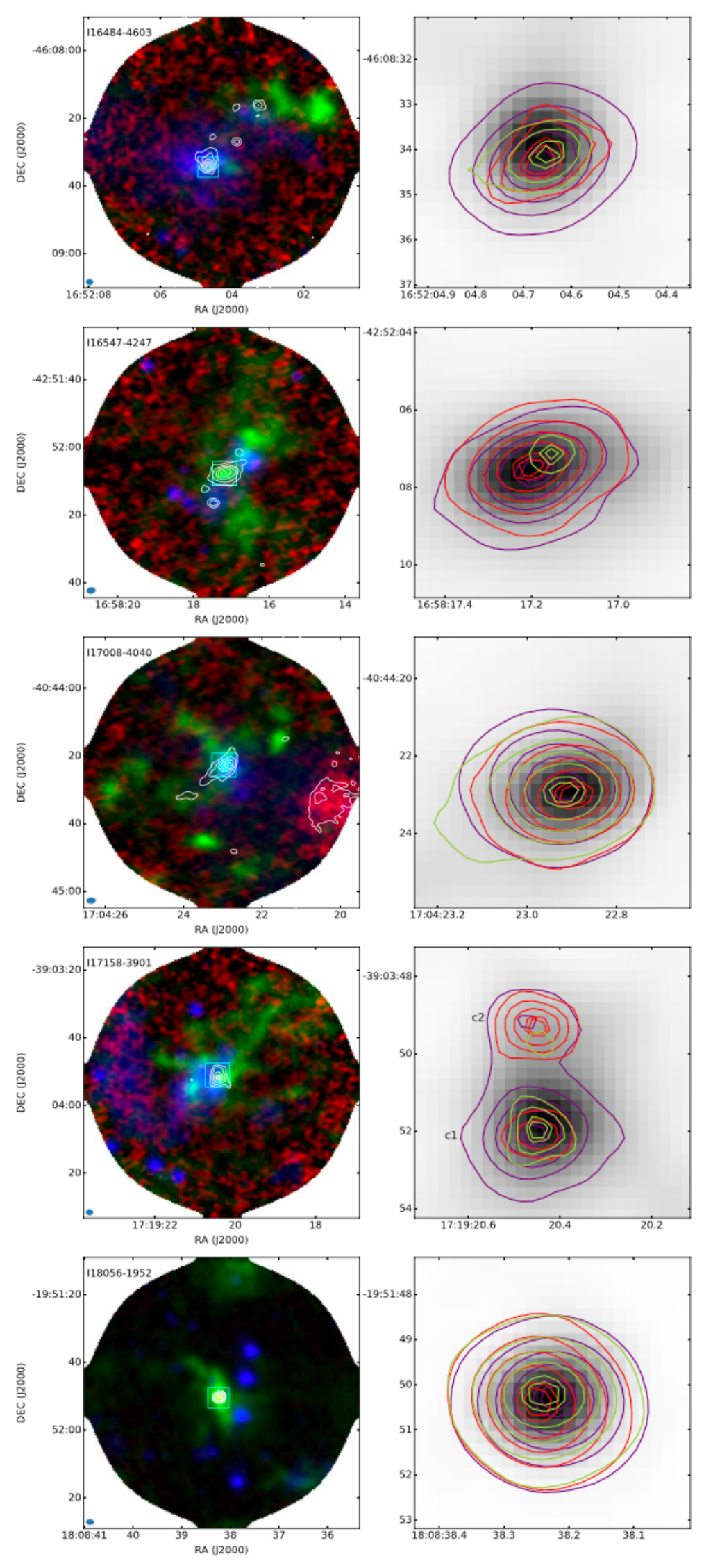}
\caption{\it -- continued}
\end{figure*}

\addtocounter{figure}{-1}
\begin{figure*}
\includegraphics[width=12cm,height=25cm]{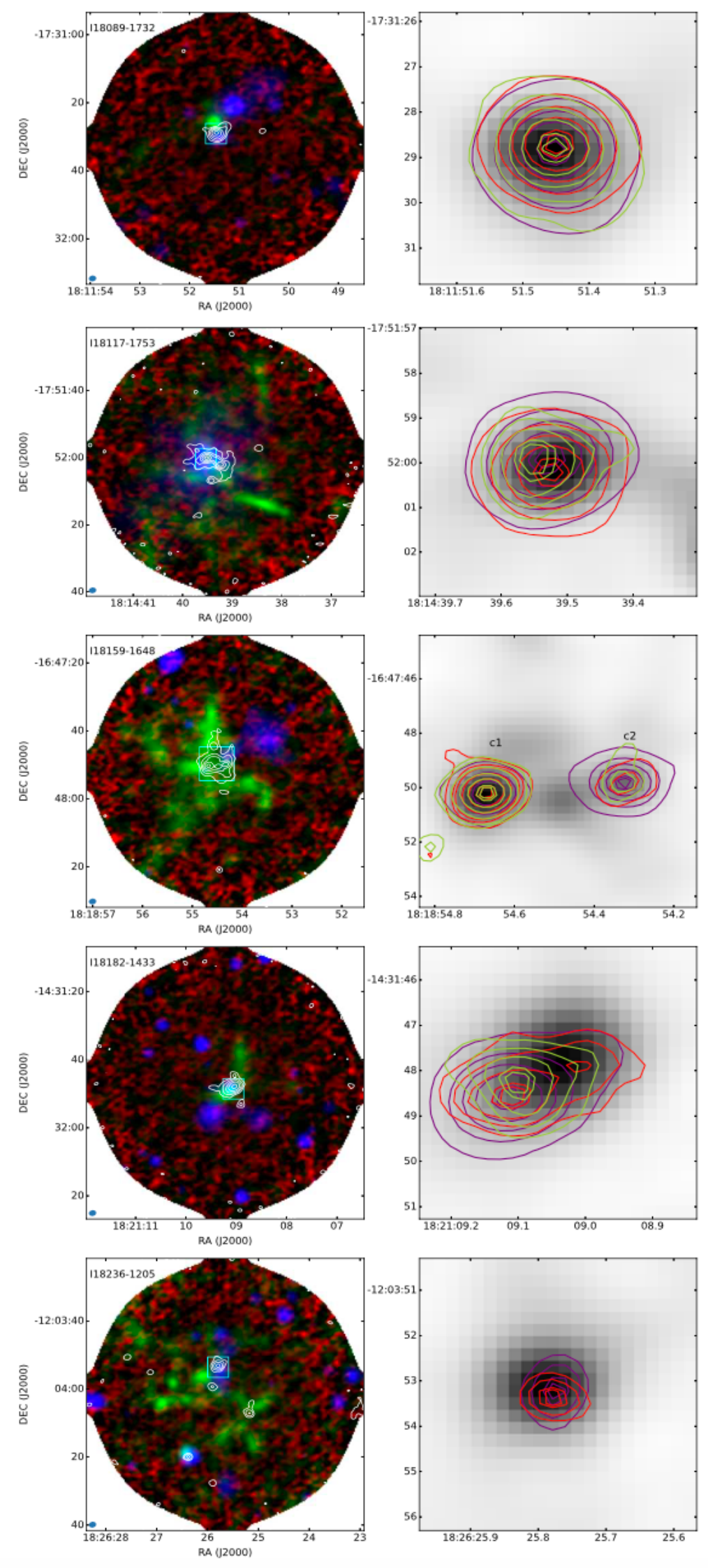}
\caption{\it -- continued}
\end{figure*}

\addtocounter{figure}{-1}
\begin{figure*}
\includegraphics[width=12cm,height=25cm]{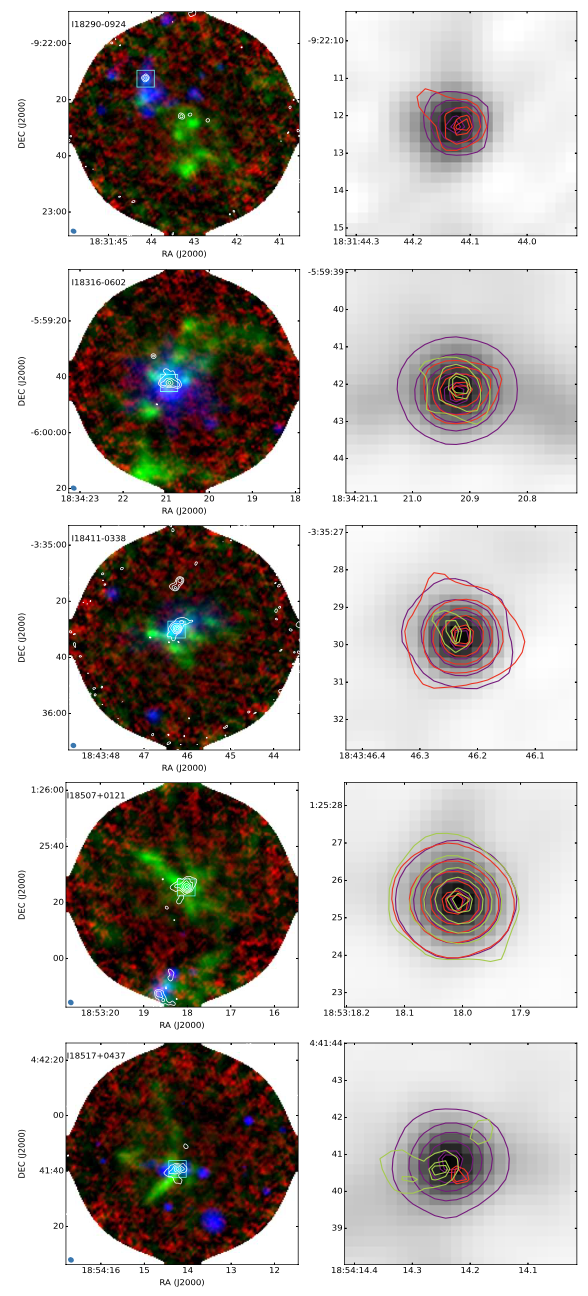}
\caption{\it -- continued}
\end{figure*}


\bsp    
\label{lastpage}
\end{document}